# Gravity-driven coatings on curved substrates: a differential geometry approach

**Pier Giuseppe Ledda[1] †, M. Pezzulla[2], E. Jambon-Puillet[3], P-T Brun[3] and F. Gallaire[1]**

[1]Laboratory of Fluid Mechanics and Instabilities, École Polytechnique Fédérale de Lausanne, CH-1015 Lausanne, Switzerland

[2]Slender Structures Lab, Department of Mechanical and Production Engineering, Århus University, Inge Lehmanns Gade 10, 8000 Århus C, Denmark

[3]Department of Chemical and Biological Engineering, Princeton University,Princeton, New-Jersey 08540, USA

Drainage and spreading processes in thin liquid films have received considerable attention in the past decades. Yet, our understanding of three-dimensional cases remains sparse, with only a few studies focusing on flat and axisymmetric substrates. Here, we exploit differential geometry to understand the drainage and spreading of thin films on curved substrates, under the assumption of negligible surface tension and hydrostatic gravity effects. We develop a solution for the drainage on a local maximum of a generic substrate. We then investigate the role of geometry in defining the spatial thickness distribution via an asymptotic expansion in the vicinity of the maximum. Spheroids with a much larger (respectively smaller) height than the equatorial radius are characterized by an increasing (respectively decreasing) coating thickness when moving away from the pole. These thickness variations result from a competition between the variations of the substrate's slope and mean curvature. The coating of a torus presents larger thicknesses and a faster spreading on the inner region than on the outer region, owing to the different curvatures in these two regions. In the case of an ellipsoid with three different axes, spatial modulations in the drainage solution are observed as a consequence of a faster drainage along the short principal axis, faithfully reproduced by a three-dimensional asymptotic solution. Leveraging the conservation of mass, an analytical solution for the average spreading front is obtained. The solutions are in agreement with numerical simulations and experimental measurements obtained from the coating of a curing polymer on diverse substrates.

## 1. Introduction

Coating flows are found in many environmental, chemical and engineering processes (Weinstein & Ruschak 2004), such as spin coating (Scriven 1988; Schwartz & Roy 2004) and dip coating (Landau & Levich 1942). Additionally, coating assisted fabrication methodologies recently showed potential in the fabrication of curved spherical shells (Lee *et al.* 2016) and inflatable soft tentacles (Jones *et al.* 2021). The plethora of observed coating patterns motivated a great deal of studies aimed at understanding the underlying physical mechanisms (Weinstein & Ruschak 2004). Typical examples include inertia-driven Kapitza waves (Kapitza 1948; Kapitza & Kapitza 1965), Marangoni effects due to gradients in surface tension (Oron 2000; Scheid 2013; Hosoi & Bush 2001; Xue *et al.* 2020) the formation of drop s (Rayleigh 1882; Taylor 1950; Chandrasekhar 2013; Fermigier *et al.* 1992) and rivulets

† Email address for correspondence: pier.ledda@epfl.ch

(Lerisson *et al.* 2019, 2020; Ledda *et al.* 2020; Ledda & Gallaire 2021) below horizontal and inclined substrates, respectively. Such formation of elongated structures along the streamwise direction is also typical of contact-line driven instabilities, often called *fingering*, and occurs when a fluid spreads on a dry substrate (Oron *et al.* 1997; Kondic 2003; Weinstein & Ruschak 2004; Craster & Matar 2009). Such patterns are identified as the physical origin for several geological structures such as stalactites (Short *et al.* 2005; Camporeale & Ridolfi 2012) and flutings in limestone caves (Camporeale 2015; Bertagni & Camporeale 2017; Ledda *et al.* 2021) and due to solidification and melting of water (Camporeale 2015), while physical or chemical erosion leads to scallops (Meakin & Jamtveit 2010) or linear karren (Bertagni & Camporeale 2021) patterns. Gravity currents, widely encountered in environmental fluid dynamics, are flows driven by gravity differences typically imputed to the presence of one phase heavier than the other which spreads on a substrate. Examples typically involve complex rheologies (Balmforth *et al.* 2000, 2002, 2006) and include oil spreading on the sea (Hoult 1972), lava (Balmforth *et al.* 2000) and pyroclastic flows due to a volcano eruption, dust storms, avalanches (Simpson 1982; Huppert 1986; Balmforth & Kerswell 2005; Huppert 2006), slurry and sheet flows (Ancey 2007).

The analysis of spreading of currents requires the knowledge of the position, velocity and thickness of the advancing front. If the inertia of the flow is negligible, the dominant balance to describe the *viscous* gravity current is given by viscosity and buoyancy. With the aim of comparing their results with those of Keulegan (1957), Huppert & Simpson (1980) investigated the two-dimensional viscous gravity current on a horizontal substrate, driven by hydrostatic gravity effects. By combining a lubrication approximation with the volume conservation, the authors determined a self-similar solution for the thickness and spreading front, recovering the result of Smith (1969) in the case of the release of an initial amount of fluid. The general problem for different initial and boundary conditions, such as continuous feeding (Didden & Maxworthy 1982; Huppert 1982*b*), was investigated by Gratton & Minotti (1990) via a phase-plane formalism. When the substrate is inclined, a gravity component parallel to the substrate is introduced, which often dominates the dynamics. Huppert (1982*a*) highlighted that the lubrication solution at the leading order presents a discontinuity at the front, as long as surface tension and hydrostatic pressure gradients along the film are neglected. The thickness distribution far from the front is recovered by only considering the drainage along the in-plane directions of the substrate, referred here as *drainage* solution (Huppert 1982*a*). For the inclined plane case, the thickness solution far from the front reads $h \propto x^{1/2}t^{-1/2}$ and is formally analogous to the result of Jeffreys (1930). The mathematical derivation may be more involved when the substrate is curved, e.g. in the case of the release of an initial volume of fluid on the outerside of a cone (Acheson 1990), a cylinder or a sphere (Takagi & Huppert 2010; Lee *et al.* 2016; Balestra *et al.* 2019), or in the inside of a downward-pointing cone or a sphere (Xue & Stone 2021; Lin *et al.* 2021).

From now on, we focus on the diverging spreading on a curved substrate away from the pole, while gravity points downwards (Takagi & Huppert 2010; Lee *et al.* 2016; Balestra *et al.* 2019), see figure 1. In this case, the drainage solution fairly reproduces the experimental observations since the hydrostatic pressure gradient due to the gravity component orthogonal to the substrate does not induce any instability of the thin film free surface. Takagi & Huppert (2010) studied the drainage and spreading on a cylinder and a sphere, in the vicinity of the pole. The drainage thickness scales as $h \sim t^{-1/2}$ both for the cylinder and the sphere. More refined drainage solutions were obtained by Balestra *et al.* (2019) and Lee *et al.* (2016) for the cylinder and the sphere, respectively. In both cases, an increase of the thickness moving from the pole to the equator is observed. However, as highlighted by the numerical simulations of Duruk *et al.* (2021) for the coating of a spheroid, for some values of its aspect ratio, the thickness decreases moving from the pole to the equator.

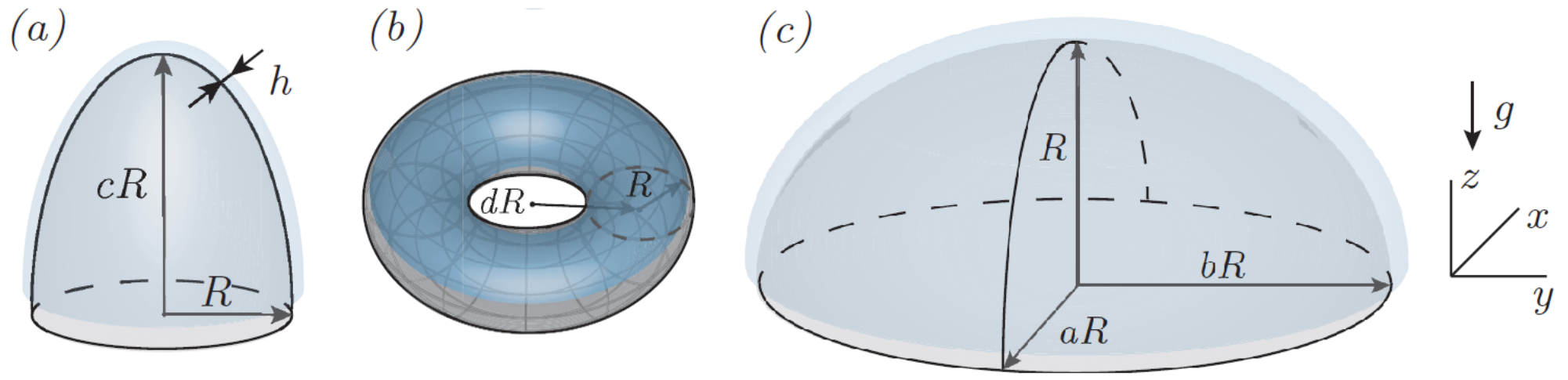

Figure 1: Different coated substrates considered in this work: (*a*) spheroid, (*b*) torus, (*c*) ellipsoid.

The latter example shows the effect of the substrate geometry in the resulting thickness distribution which stems from drainage induced by gravity. We therefore aim at exploring the role of the substrate in this process, which still needs to be systematically studied, even in the simple case of axisymmetric substrates. When the symmetry of the substrate is broken, spatial non-uniformities may also modify the picture previously described and require further investigation. Despite the abundance of studies on spreading in different conditions, the problem of three-dimensional drainage and spreading has been the object of limited studies on flat substrates (Lister 1992; Xue & Stone 2020), to the best of our knowledge. The role of the substrate in inducing three-dimensional drainage still needs to be assessed.

A lubrication model for generic substrates was developed in Roy *et al.* (2002) and Howell (2003) by considering a generic orthogonal local coordinate system. The same result was obtained by Thiffeault & Kamhawi (2006) via classical differential geometry where the equations are written in the natural, local (*general*) coordinates system, not necessarily orthogonal. General coordinates define a local coordinate system, with the advantage of deriving general equations that can be used for any geometry and without the need of defining principal directions. The literature about the topic is extremely vast; for our purposes, the essential tools can be found in Deserno (2004) and Irgens (2019).

The lubrication equation in general coordinates offers the yet unexplored opportunities to systematically study the three-dimensional drainage and spreading on complex substrates through analytical solutions. In this work, we develop analytical solutions and approximations for the drainage and spreading problem on several substrates, with the aim of identifying relevant features of coatings on curved substrates. In the spirit of Huppert (1982*a*), we consider the case in which the tangential gravity components dominate the film thickness dynamics and we neglect the hydrostatic pressure and surface tension effects, keeping only the leading order terms given by the drainage gravity components. This approach is suitable to obtain simple analytical expressions to shed light on the leading effect of the substrate on the thickness distribution. The paper is organized as follows. In Section 2, we introduce the coating problem of a generic substrate and the differential geometry tools necessary to understand the flow configuration. We then obtain a general solution in the vicinity of a local maximum of a diverging substrate. The following sections focus on how geometry influences the drainage around local maxima using the geometries reported in figure 1. Section 3 is devoted to the study of the drainage and spreading on a spheroid, where we show that depending on the aspect ratio, the film can either get thicker or thinner as we move away from the pole. Subsequently, Section 4 studies the problem of non-symmetric drainage and spreading on a torus. We conclude by studying the spatially non-uniform drainage and spreading solution on ellipsoids, in Section 5. Eventually, the analytical and numerical results are compared to experimental measurements.

## 2. The coating problem of a generic substrate

### 2.1. *Problem definition and metric terms in general coordinates*

In this section, we introduce the essential differential geometry tools to solve the problem of the coating on a generic substrate. For a complete description of differential geometry and general coordinates, we refer to Deserno (2004). The derivation of the lubrication equation for generic curved substrates can be found in Roy *et al.* (2002), Thiffeault & Kamhawi (2006) and Wray *et al.* (2017). The geometry is sketched in figure 2. We consider a generic substrate $h^0$, on which lies a fluid film of thickness $h$, and introduce a Cartesian reference frame $(x, y, z)$. The substrate is identified by the position vector $\boldsymbol{X}\left(x^{\{1\}}, x^{\{2\}}\right)$, where $(x^{\{1\}}, x^{\{2\}})$ denote the local coordinates used to parameterize the surface (e.g. the zenith and the azimuth for spherical coordinates, the radial coordinate and the azimuth for a cone). The flow equations are solved in the local and natural reference frame of the substrate. We introduce the local coordinate vectors parallel to the substrate $\mathbf{e}_i = \partial_i \boldsymbol{X}$, $i = 1, 2$ (not necessarily orthonormal), and the normal coordinate vector $\mathbf{e}_3 = \frac{\mathbf{e}_1 \times \mathbf{e}_2}{|\mathbf{e}_1 \times \mathbf{e}_2|}$. From the knowledge of the local coordinate vectors, we introduce the $2 \times 2$ symmetric metric tensor $\mathbb{G}_{ij}$ and the square root of the determinant of the metric on the substrate $w$, which is related to the area element on the surface $\mathrm{d}A$ through $dA = w \mathrm{d}x^{\{1\}} \mathrm{d}x^{\{2\}}$ :

$$\mathbb{G}_{ij} = \mathbf{e}_i \cdot \mathbf{e}_j = \mathbb{G}_{ji}, \quad w = \left(\det \mathbb{G}_{ij}\right)^{1/2}. \tag{2.1}$$

The metric tensor defines the generic line element $\mathrm{d}s$ as $\mathrm{d}s^2 = \mathbb{G}_{11}\left(\mathrm{d}x^{\{1\}}\right)^2 + 2\mathbb{G}_{12}\mathrm{d}x^{\{1\}}\mathrm{d}x^{\{2\}} + \mathbb{G}_{22}\left(\mathrm{d}x^{\{2\}}\right)^2$. Therefore, the dimensions of each component depend on the considered parameterization $(x^{\{1\}}, x^{\{2\}})$, so that each part that composes $\mathrm{d}s^2$ has the dimensions of the square of a length. We also introduce the second fundamental form and the curvature tensor, which respectively read:

$$\mathbb{S}_{ij} = \partial_i \mathbf{e}_j \cdot \mathbf{e}_3, \quad \mathbb{K}_i^{\{j\}} = \mathbb{S}_{ik}\mathbb{G}^{\{kj\}}, \tag{2.2}$$

where $\mathbb{G}^{\{ij\}}$ is the inverse metric tensor, i.e. $\mathbb{G}^{\{ij\}} = \mathbb{G}_{ij}^{-1}$. The mean $\mathcal{K}$ and the Gaussian $\mathcal{G}$ curvatures read $\mathcal{K} = \operatorname{tr}\mathbb{K}$ and $\mathcal{G} = \det\mathbb{K}$, respectively. Following Einstein's notation for the summation, a generic vector $\mathbf{f}$ can be written in terms of its covariant and contravariant base, i.e. $\mathbf{f} = f^{\{i\}}\mathbf{e}_i = f_i \mathbf{e}^{\{i\}}$, where $\mathbf{e}^{\{i\}}$ is the covector defined as $\mathbf{e}^{\{i\}} \cdot \mathbf{e}_j = \delta_{ij}$. The two contravariant components, parallel to the substrate, of the gravity vector read $g_t^{\{i\}} = \mathbf{g} \cdot \mathbf{e}^{\{i\}}$, while the normal one reads $g_3 = \mathbf{g} \cdot \mathbf{e}_3$. The gradient of a scalar function $f$ and the divergence of a generic vector $\mathbf{f} = f^{\{i\}}\mathbf{e}_i$ respectively read (Irgens 2019):

$$\nabla f = \partial_i f \, \mathbb{G}^{\{ij\}}\mathbf{e}_j = \partial^{\{i\}} f \, \mathbf{e}_j, \quad \nabla \cdot \mathbf{f} = w^{-1}\partial_i (w f^{\{i\}}). \tag{2.3}$$

The above-defined quantities and differential operators are enough to describe the coating problem on a generic substrate, introduced in the following section.

### 2.2. *Lubrication equation and drainage solution*

We consider a thin viscous film, flowing on a substrate $h^0$, of thickness $h$ measured along the direction perpendicular to the substrate itself. The fluid properties are the density $\rho$, viscosity $\mu$ and the surface tension coefficient $\gamma$. In the absence of inertia, the lubrication model for a generic curved substrate was first derived by (Roy *et al.* 2002) via central manifold theory. We non-dimensionalize the thickness with $h_i$ and the tangential directions with $R$, i.e. a characteristic film thickness (e.g. the initial one, if uniform) and a relevant length of the substrate (e.g. its equatorial radius), respectively. We introduce the drainage time scale $\tau = \mu R/(\rho g h_i^2)$. Upon non-dimensionalization, the equation in coordinate-free form reads

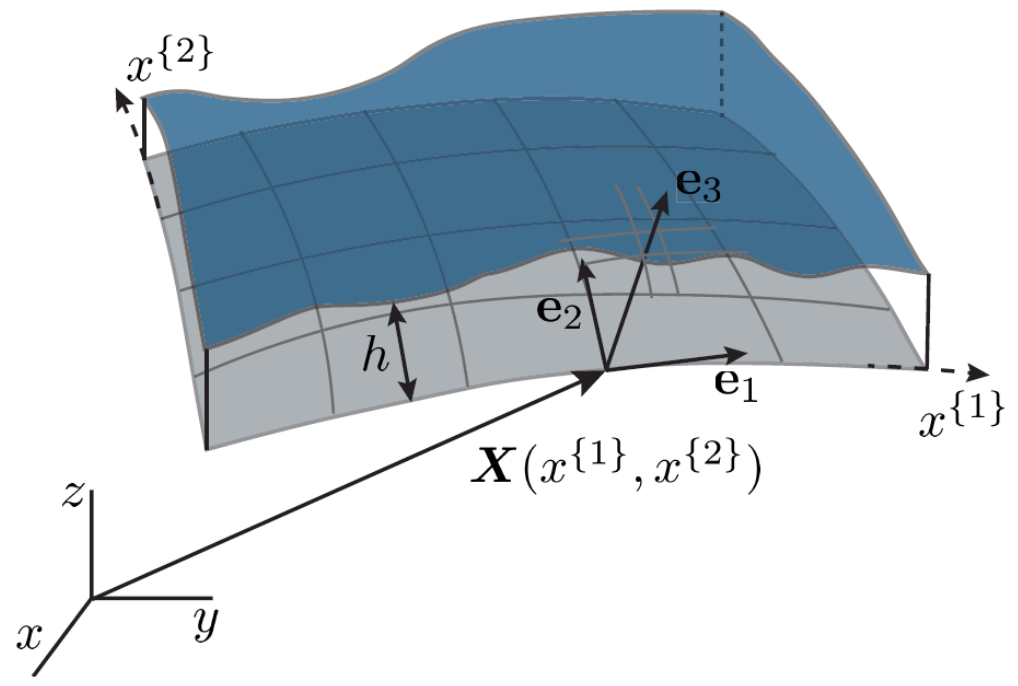

Figure 2: Sketch of the coordinates systems employed in this analysis. A global Cartesian reference frame $(x, y, z)$ is considered. At each point, the position of the substrate is identified by the vector $\boldsymbol{X}$, which depends on the chosen parameterization $(x^{\{1\}}, x^{\{2\}})$ of the substrate. The derivatives of the position vector identify the local reference frame on the substrate, on which the lubrication equation is solved.

(Roy *et al.* 2002; Howell 2003; Roberts & Li 2006; Thiffeault & Kamhawi 2006):

$$\begin{aligned}(1 - \delta\mathcal{K}h + \delta^2\mathcal{G}h^2)\frac{\partial h}{\partial t} + \frac{1}{3Bo}\nabla\cdot\left[h^3\left(\nabla\tilde{\kappa} - \frac{\delta}{2}h(2\mathcal{K}\mathbb{I} - \mathbb{K})\cdot\nabla\mathcal{K}\right)\right] \\ + \frac{1}{3}\nabla\cdot\left[h^3\left(\boldsymbol{g}_t - \delta h\left(\mathcal{K}\mathbb{I} + \frac{1}{2}\mathbb{K}\right)\cdot\boldsymbol{g}_t + \delta \mathrm{g}_3\nabla h\right)\right] = 0,\end{aligned} \tag{2.4}$$

where $\tilde{\kappa} = \mathcal{K} + \delta(\mathcal{K}^2 - 2\mathcal{G})h + \delta\nabla^2 h$ is the free-surface curvature, $Bo = (\rho g R^2)/\gamma$ is the Bond number, $\delta = h_i/R$ is the aspect ratio of the thin film, and $\boldsymbol{g}_t$ and $\mathrm{g}_3$ identify the gravity vector components tangent and normal to the substrate, respectively. The terms in the first and second bracket represent the flux induced by capillary and gravity effects, respectively. Capillary flow is induced, at leading order, by variations of the mean curvature of the substrate $\mathcal{K}$ and leads to film thinning and thickening in the neighborhood of local maximum and minimum values of the curvature (Roy *et al.* 2002). Corrections at order $O(\delta)$ introduce (i) free-surface curvature variations and (ii) higher order terms of the substrate curvature. Gravity-induced fluxes are instead related, at leading order, by the gravity components tangential to the substrate $\boldsymbol{g}_t$. Corrections at $O(\delta)$ introduce hydrostatic pressure gradients along the thin film. This equation can be written in compact form as follows:

$$(1 - \delta\mathcal{K}h + \delta^2\mathcal{G}h^2)\frac{\partial h}{\partial t} + \frac{1}{3}\nabla\cdot\mathbf{q} = 0, \tag{2.5}$$

so-called *conservation form*, where $\mathbf{q} = q^{\{1\}}\mathbf{e}_1 + q^{\{2\}}\mathbf{e}_2$ is the flux. The so-called drainage solution is identified in the limit $\delta \ll 1$ and when the Bond number is very large $Bo \to \infty$, and results from the following problem:

$$\frac{\partial h}{\partial t} + \frac{1}{3}\nabla\cdot\left[h^3\boldsymbol{g}_t\right] = 0. \tag{2.6}$$

In this case, the flux per unit length is defined as $\mathbf{q} = q^{\{1\}}\mathbf{e}_1 + q^{\{2\}}\mathbf{e}_2 = h^3 g_t^{\{1\}}\mathbf{e}_1 + h^3 g_t^{\{2\}}\mathbf{e}_2$. The solution of the drainage problem requires only the knowledge of $w$ and the tangential gravity vector components $g_t^{\{i\}}$. The drainage problem therefore relies on two assumptions. The Bond number is assumed to be very large, i.e. $R^2/\ell_c^2 \to \infty$, where $\ell_c = \sqrt{\gamma/(\rho g)}$ is the capillary length. After this first assumption, the problem accounts only for gravity effects resulting from drainage and hydrostatic pressure gradients. The latter terms are important when the aspect ratio $\delta = h_i/R$ is not negligible, i.e. for a thick film on a large substrate (compared to the capillary length) with small radius of curvature (compared to the film thickness). A limit case occurs when the substrate is locally flat. The leading order solution is given by the hydrostatic pressure terms, since drainage is absent. A case in which both capillary and hydrostatic effects cannot be neglected occurs when the radius of curvature of the substrate is comparable to the film thickness, i.e. regions of extremely large curvature such as the tip of a cone. In the following, we restrict ourselves to the situation in which the film is very thin and the substrate does not present regions of extremely large curvature, thus enabling the employment of the drainage equation (2.6).

The numerical implementation of the lubrication equation (2.4) is performed in the finite-element solver COMSOL Multiphysics, in which the lubrication equation is implemented in its conservation form (2.5). Quadratic lagrangian elements are exploited for the numerical discretization, while the time-marching is performed with the built-in BDF solver. In the case of equation (2.4), we solve for the variables $(h, \tilde{\kappa})$. We refer to the corresponding sections for more detail about the boundary conditions for the different substrates.

The validation procedure consists of a first mesh size validation. We thus verify the faithfulness of the employed parameterization $\boldsymbol{X}(x^{\{1\}}, x^{\{2\}})$ by a comparison with the parameterization $\boldsymbol{X} = (x, y, h^0(x, y))$, so-called Monge parameterization (Thiffeault & Kamhawi 2006; Mayo *et al.* 2015), reported in the Electronic Supplementary Material (ESM) together with the other parameterizations employed in this work. We also verify the non-dimensionalization by solving the dimensional equation (2.4) and comparing the solution for each substrate with the non-dimensional model. To illustrate and complement the theoretical results, we finally compare in Section 6 the drainage problem results to experiments performed following the procedure outlined in Lee *et al.* (2016) and Jones *et al.* (2021), for diverse substrates.

### 2.3. *Asymptotic theory - general expression for the thickness at a local maximum*

The employed substrate-free expression of the lubrication equation is suitable for analytical results. In this section, we develop a general expression for the thickness at a local maximum of the substrate. In the vicinity of the local maximum, the smooth substrate is described through a Monge parameterization of the substrate, i.e. $(x^{\{1\}} = x, x^{\{2\}} = y)$, see ESM for further detail. The generic substrate position vector in the vicinity of the maximum reads $\boldsymbol{X} = (x, y, h^0(x, y))$. The square root of the determinant of the metric tensor $w$ and the tangential gravity components respectively read:

$$w = \sqrt{1 + \left(\partial_x h^0\right)^2 + \left(\partial_y h^0\right)^2}, \quad g_t^{\{1\}} = -\frac{\partial_x h^0}{w^2}, \quad g_t^{\{2\}} = -\frac{\partial_y h^0}{w^2}. \tag{2.7}$$

We expand the drainage solution in the vicinity of the maximum identified by the point $\boldsymbol{x} = (x, y) = \boldsymbol{0}$ by employing an asymptotic expansion in the spatial variables, i.e.

$$h(x, y, t) = H_0(t) + H_{11}(t)x + H_{12}(t)y + \ldots \tag{2.8}$$

Upon substitution of the decomposition (2.8) in equation (2.6), the $\mathcal{O}(1)$ problem for $H_0(t)$ reads:

$$\frac{H_0'}{H_0^3} = \left.\frac{\left(\partial_{xx}h^0\left(1 + \left(\partial_y h^0\right)^2\right) - 2\partial_x h^0 \partial_{xy}h^0 \partial_y h^0 + \partial_{yy}h^0\left(\left(\partial_x h^0\right)^2 + 1\right)\right)}{3\left(\left(\partial_y h^0\right)^2 + \left(\partial_x h^0\right)^2 + 1\right)^2}\right|_{\boldsymbol{x}=\boldsymbol{0}} = -\left(\frac{1}{3}\left(\mathbf{g}\cdot\mathbf{e}_3\right)\mathcal{K}\right)|_{\boldsymbol{x}=\boldsymbol{0}}, \tag{2.9}$$

with the initial condition $H_0(0) = 1$ in the case of a unitary initial thickness. At the maximum location, $\partial_x h^0 = \partial_y h^0 = 0$, i.e. the normal vector and gravity are aligned. Therefore, the RHS of equation (2.9) simplifies to $\partial_{xx}h^0 + \partial_{yy}h^0 = -\mathcal{K}_p$, where $\mathcal{K}_p$ is the opposite of the mean curvature at the maximum. The resulting problem and associated solution read:

$$\frac{H_0'(t)}{H_0(t)^3} = -\frac{1}{3}\mathcal{K}_p \rightarrow H_0(t) = \frac{1}{\sqrt{\frac{2\mathcal{K}_p t}{3} + 1}}. \tag{2.10}$$

A general expression for the drainage in the vicinity of a local maximum is obtained. Note that the mean curvature at the local maximum is negative, and thus $\mathcal{K}_p > 0$. The thickness at a local maximum depends on the mean curvature. From a geometrical point of view, the mean curvature represents variations of the tangential vectors along the surface. Since the normal to the surface and the gravity vector are aligned, the mean curvature determines the evolution of the tangential components of the gravity field, in the vicinity of the local maximum. In particular, an increase of $\mathcal{K}_p$ implies larger values of gravity in the tangent plane of the substrate moving away from the pole and thus a faster drainage and a lower thickness, and vice versa. This solution allows one to identify the limits of the considered drainage model. If $\mathcal{K}_p = 0$, i.e. a locally flat substrate, there is no drainage and, therefore, the thickness is constant and equal to $H_0 = 1$. In this case, hydrostatic effects cannot be neglected since they are the leading order effect and lead to a time-dependent drainage (Huppert & Simpson 1980). Therefore, the drainage model is not suitable to describe locally flat substrates. A second limiting case occurs when $\mathcal{K}_p \rightarrow \infty$, i.e. the radius of curvature in the vicinity of the local maximum tends to zero. A classical example is the tip of a cone, parameterized with the radius $x^{\{1\}} = r$ and the azimuth $x^{\{2\}} = \varphi$. In this case, an exact solution $h \propto \sqrt{r}$ is obtained (see ESM), which presents a zero thickness at the pole, in accordance with solution (2.10), which tends to zero as $\mathcal{K}_p \rightarrow \infty$. In that case, hydrostatic and capillary effects are crucial to define the thickness distribution in the vicinity of the tip. Another important limitation comes from the considered geometry. When the fluid is located below the substrate, the solution is formally analogous to equation (2.10), and predicts a progressive thinning. However, it is well known, in these situations, that hydrostatic pressure gradients and capillary effects play a key role since the Rayleigh-Taylor instability can occur (Balestra *et al.* 2018*b*). In the case of converging geometries with a local minimum, the solution is formally analogous, but with $\mathcal{K}_p < 0$. In this case, the thickness progressively increases and tends to infinity for $t = -3/(2\mathcal{K}_p) > 0$, long after hydrostatic and capillary effects should have been considered. These combined effects contribute indeed to the Rayleigh Taylor instability when the fluid lies below the substrate and for a leveling and flattening of the interface when it lies above.

Turning back to the situation where $\mathcal{K}_p > 0$ and remains finite, one can take the limit for $t \rightarrow \infty$ of solution (2.10), leading to:

$$H_0(t) = \frac{1}{\sqrt{\frac{2\mathcal{K}_p t}{3}}}, \tag{2.11}$$

i.e. the solution is independent of the initial condition. This result is in agreement with the analysis in (Lee *et al.* 2016), where the authors theoretically and experimentally showed an insensitivity of the film thickness with respect to the initial conditions.

In the following, we investigate the spatial evolution of the thin film thickness when moving away from the pole. We initially consider the case of an axisymmetric substrate, the spheroid.

## 3. Drainage and spreading on axisymmetric substrates: coating of a spheroid

### 3.1. *Drainage problem*

In this section, we consider the drainage of a thin film flowing on an spheroidal substrate of equatorial radius $R$ (i.e. $a = b = 1$) and height $cR$. We non-dimensionalize the in-plane directions and substrate variables with the equatorial radius $R$. We parameterize the spheroidal surface via the zenith (or colatitude) $x^{\{1\}} = \vartheta$ and the azimuth $x^{\{2\}} = \varphi$:

$$\boldsymbol{X}(\vartheta, \varphi) = (\sin\vartheta\cos\varphi, \sin\vartheta\sin\varphi, c\cos\vartheta) \tag{3.1}$$

A complete description of the metric and curvature tensors is reported in the ESM. The gravity term $g_t^{\{1\}}$ and $w$ are:

$$g_t^{\{1\}}(\vartheta) = \frac{c\sin(\vartheta)}{c^2\sin^2(\vartheta)+\cos^2(\vartheta)}, \quad w(\vartheta) = \frac{1}{\sqrt{2}}\sin(\vartheta)\sqrt{(1-c^2)\cos(2\vartheta)+1+c^2}, \tag{3.2}$$

while $g_t^{\{2\}} = 0$. In the case $c = 1$, we recover the evolution equation for the spherical case, reported in the ESM. Following the previous section, we consider as initial condition a constant thickness on the substrate, i.e. $h(\vartheta, 0) = 1$. The problem is solved through an asymptotic expansion in the vicinity of the pole. We expand the solution at different orders in $\vartheta$:

$$h(\vartheta, t) = H_0(t) + \vartheta^2 H_2(t) + \vartheta^4 H_4(t) + \vartheta^6 H_6(t) + \ldots \tag{3.3}$$

We introduce this ansatz and expand in powers of $\vartheta$. At each order $O(\vartheta^n)$, one obtains an ordinary differential equation for $H_n$. The problem at order $O(1)$ and $O(\vartheta^2)$ together with their solution read:

$$\frac{2}{3}cH_0^3 + H_0' = 0, \quad H_0(0) = 1 \to H_0 = \left(\frac{4ct}{3}+1\right)^{-1/2}. \tag{3.4}$$

$$\begin{aligned}&(\frac{2c}{3} - c^3)H_0^3 + 4cH_0^2H_2 + H_2' = 0, \quad H_2(0) = 0,\\ &\quad \to H_2 = \frac{(3c^2-2)(64\sqrt{3}c^3t^3 + 144\sqrt{3}c^2t^2 + 108\sqrt{3}ct + 27(\sqrt{3}-\sqrt{4ct+3}))}{10(4ct+3)^{7/2}}.\end{aligned} \tag{3.5}$$

Applying the same procedure at orders $O(\vartheta^4)$ and $O(\vartheta^6)$, the solution up to $O(\vartheta^6)$ and for $t \to \infty$, reads (see Appendix A for further detail):

$$\begin{aligned}h(\vartheta, t) = \sqrt{\frac{3}{4\mathrm{t}}}\frac{1}{\sqrt{c}}\Bigg(1 + \frac{1}{10}\left(3c^2-2\right)\vartheta^2 - \frac{\left(336c^4 - 408c^2 + 31\right)\vartheta^4}{4800}\\ + \frac{\left(58464c^6 - 115368c^4 + 62667c^2 - 4576\right)\vartheta^6}{1584000}\Bigg) + O(\vartheta^8) + O\left(\frac{1}{t^{3/2}}\right)\end{aligned} \tag{3.6}$$

Note that the $O(1)$ large-time solution is formally analogous to equation (2.10) with $\mathcal{K}_p = 2c$, i.e. the opposite of the mean curvature at the pole.

We perform numerical simulations of equation (2.6), in the region $0 < \vartheta < \pi/2$. Owing to the hyperbolic nature of the equation, no boundary conditions are necessary at $\vartheta = 0$ and $\vartheta = \pi/2$, and thus we impose only the initial condition $h(\vartheta, 0) = 1$. Numerical convergence is achieved with the characteristic element size $\Delta\vartheta = 1°$. Figure 3 shows a comparison of the numerical solution of equation (2.6) at $t = 100$ with the analytical ones at order at order $O(\vartheta^2)$ (solid lines) and $O(\vartheta^6)$ (dashed lines), which shows an overall agreement. The solution at order $O(\vartheta^6)$ gives a better agreement with the numerics in a larger range of $\vartheta$. For $c > 1.2$, the numerical and analytical solutions start to deviate for $\vartheta > 60°$. The agreement with the solution at second order is good in most cases for $\vartheta < 60°$. The second order term in equation (3.6) vanishes when $c^* = \sqrt{2/3} \approx 0.81$. Under these conditions, the solution at $O(\vartheta^2)$ is constant along the zenith. The approximation at order $O(\vartheta^6)$ does not admit a constant solution. However, the minimum variation of its integral in the region $0 < \vartheta < \pi/2$, with respect to the constant value given by employing $H_0(t)$, is obtained for $c \approx 0.74$. Independently of the considered order of the solution, for very small (respectively very large) values of $c$ the thickness decreases (respectively increases) when moving away from the pole. Moreover, for $c < c^*$, the numerical solution and the analytical one at order $O(\vartheta^6)$ present a non-monotonous behavior, as shown in figure 3 for $c = 0.4, 0.6$, with an initial decrease followed by a slight increase for $\vartheta > 70°$. The solutions for $c > c^*$ monotonically increase.

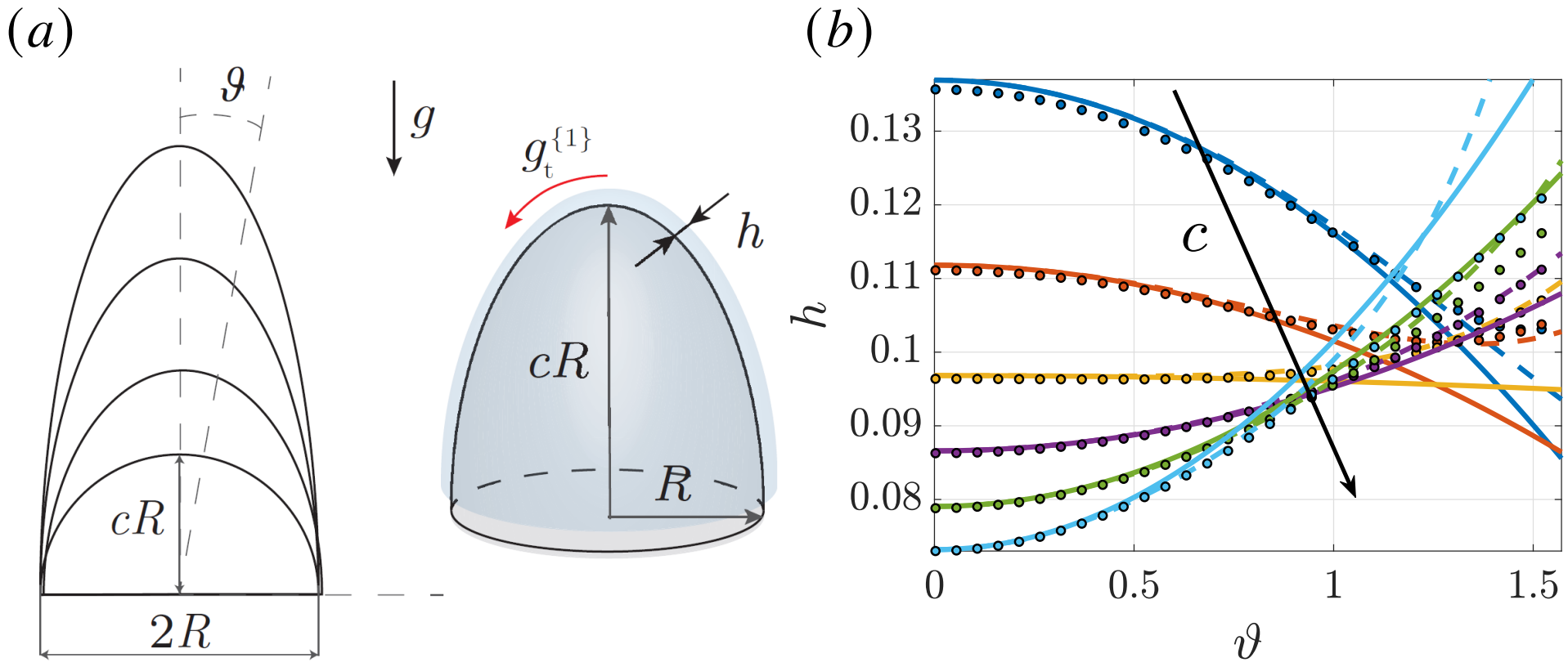

Figure 3: ($a$) Sketch of the spheroidal substrate with varying $c$. ($b$) Comparison of the numerical solution at $t = 100$ of equation (2.6) (colored dots) with the analytical ones at order $O(\vartheta^2)$ (solid lines) and $O(\vartheta^6)$ (dashed lines). Different colours identify different values of $c$: $c = 0.4$ (blue), $c = 0.6$ (orange), $c = 0.8$ (yellow), $c = 1$ (purple), $c = 1.2$ (green), $c = 1.4$ (cyan).

The leading order large time analytical solution presents a temporal decay $h \sim t^{-1/2}$. The spherical case is recovered by imposing $c = 1$ (Couder *et al.* 2005; Lee *et al.* 2016; Qin *et al.* 2021). The large-time solution is independent of the initial thickness $h_i$. It is interesting to note the good agreement between the analytical and numerical solutions for $c < 1.2$ and $\vartheta > 1$, which is out of the expected range of validity of the asymptotic expansion. The relative size of the terms in the asymptotic expansion decreases as higher orders are considered, thus suggesting that the power series expansion may converge to the exact solution in the considered range of $\vartheta$.

A decrease of $c$ implies a reduction of the gravity component parallel to the substrate and thus a reduction of the pole thickness, for a given time horizon. The film thinning or thickening moving downstream of the pole is also related to the considered geometry, as the consequence of two competing effects, already shown close to the pole (Section 2.3), where the thickness evolution depends on the normal gravity component multiplied by the local mean curvature $(\mathbf{g} \cdot \mathbf{e}_3)\mathcal{K}$. While at the pole gravity is aligned with the substrate normal, i.e. $\mathbf{g} \cdot \mathbf{e}_3 = 1$, as we move away the normal gravity component $\mathbf{g} \cdot \mathbf{e}_3$ decreases with the zenith owing to the slope increase of the substrate, leading to a first inhomogeneity mechanism. Moving away from the pole, this slope increase leads to a slower decrease of the thickness with time. This explains why, in the case of constant curvature, e.g. spheres or cylinders, the thickness increases moving toward the equator, in agreement with the results of Lee *et al.* (2016) and Balestra *et al.* (2018*a*). The second mechanism at hand is the evolution of the curvature $\mathcal{K}$ along the zenith direction. Curvature variations induce an accumulation of fluid in regions of lower curvature, characterized by a slower decrease of the thickness with time. Spheroids with small height are characterized by a mean curvature that increases away from the pole, and vice versa for spheroids of large height. The former are thus likely to present a decreasing thickness moving downstream, and vice versa, as observed in figure 3. As a result, the thickness distribution is a result of the competition between variations of slope and mean curvature, which may induce thinning or thickening of the fluid layer.

Therefore, the transition does not occur when the mean curvature is constant (i.e. $c = 1$), but when there is a balance between the thickness variations due to the change in mean curvature and those induced by slope variations. This value can be obtained by considering the quantity on the RHS of equation (2.9), i.e. $(\mathbf{g} \cdot \mathbf{e}_3)\,\mathcal{K}$, which, in the vicinity of the pole, reads:

$$(\mathbf{g} \cdot \mathbf{e}_3)\,\mathcal{K} \approx -\left(2c - c\left(3c^2 - 2\right)\vartheta^2 + O(\vartheta^4)\right), \tag{3.7}$$

which is constant for $c = c^*$, i.e. the value that causes the $O(\vartheta^2)$ contribute to vanish. Note that the same transition value can be obtained by evaluating how the quantity $\frac{1}{w}\partial_1(w g_t^{\{1\}})$ perturbs the $O(1)$ solution. The non-monotonous behaviors at large $\vartheta$ for $c < c^*$ are related to higher order terms.

### 3.2. *Spreading problem*

Typical coating applications involve the spreading of an initial volume of fluid located close to the top of the considered geometry (Takagi & Huppert 2010). Here, following previous works (Huppert 1982*a*), we recover some typical relevant quantities such as the position and thickness of the spreading front. An initial volume of fluid $V$ is released on the substrate. We impose the conservation of mass in general coordinates, under the assumption $\delta = 0$ (Roy *et al.* 2002):

$$\int\int_{\mathbb{S}} h(x^{\{1\}}, x^{\{2\}}, t)\, w\, \mathrm{d}x^{\{1\}}\mathrm{d}x^{\{2\}} = V, \tag{3.8}$$

where $V$ is the initial volume released on the substrate and $\mathbb{S}$ is the region of the substrate, parameterized with $(x^{\{1\}}, x^{\{2\}})$, which contains the fluid and varies with time because of the moving front. Far from the contact line, the drainage solution $h(x^{\{1\}}, x^{\{2\}}, t)$ can be employed, while capillary effects are relevant only in the close vicinity of the front (Huppert 1982*a*). For a fixed substrate geometry, $w$ is known, and thus relation (3.8) is an implicit equation with the front position as an unknown. A typical assumption to simplify the analysis is the employment of the large-time drainage solution.

We consider an initial volume of fluid of constant height $h_i = 1$ released at $t = 0$ in the region $0 < \varphi < 2\pi$, $0 < \vartheta < \vartheta_0$. Owing to the invariance along the azimuthal direction, the conservation of the initial fluid volume (equation (3.8)) reads:

$$\int_0^{\vartheta_F(t)} h(\vartheta, t) w(\vartheta)\mathrm{d}\vartheta = \int_0^{\vartheta_0} 1\, w(\vartheta)\mathrm{d}\vartheta, \tag{3.9}$$

where $\vartheta_F(t)$ is the front angle; the analytical expression (3.6) for $h(\vartheta, t)$ is employed. Equation (3.9) is numerically solved in Matlab via the built-in function "fsolve". Figure 4(*a*) shows the evolution of the front angle $\vartheta_F$ with time, for different values of $\vartheta_0$ and $c$. An increase in $\vartheta_0$ leads to larger values of $\vartheta_F$, for fixed time. At small times, an increase in $c$ leads to larger $\vartheta_F$; however, at large times, the opposite behavior is observed. In figure 4(*b*) we report the thickness at the front $h_F = h(\vartheta_F(t), t)$, which presents slight variations with $c$.

We approximate these results by considering an expansion for $\vartheta \ll 1$, by employing equation (3.4) for $t \to \infty$, and $w(\vartheta) = \vartheta + O(\vartheta^2)$. In this case, both the RHS and LHS of equation (3.9) can be analytically integrated and an explicit relation for $\vartheta_F$ is found, together

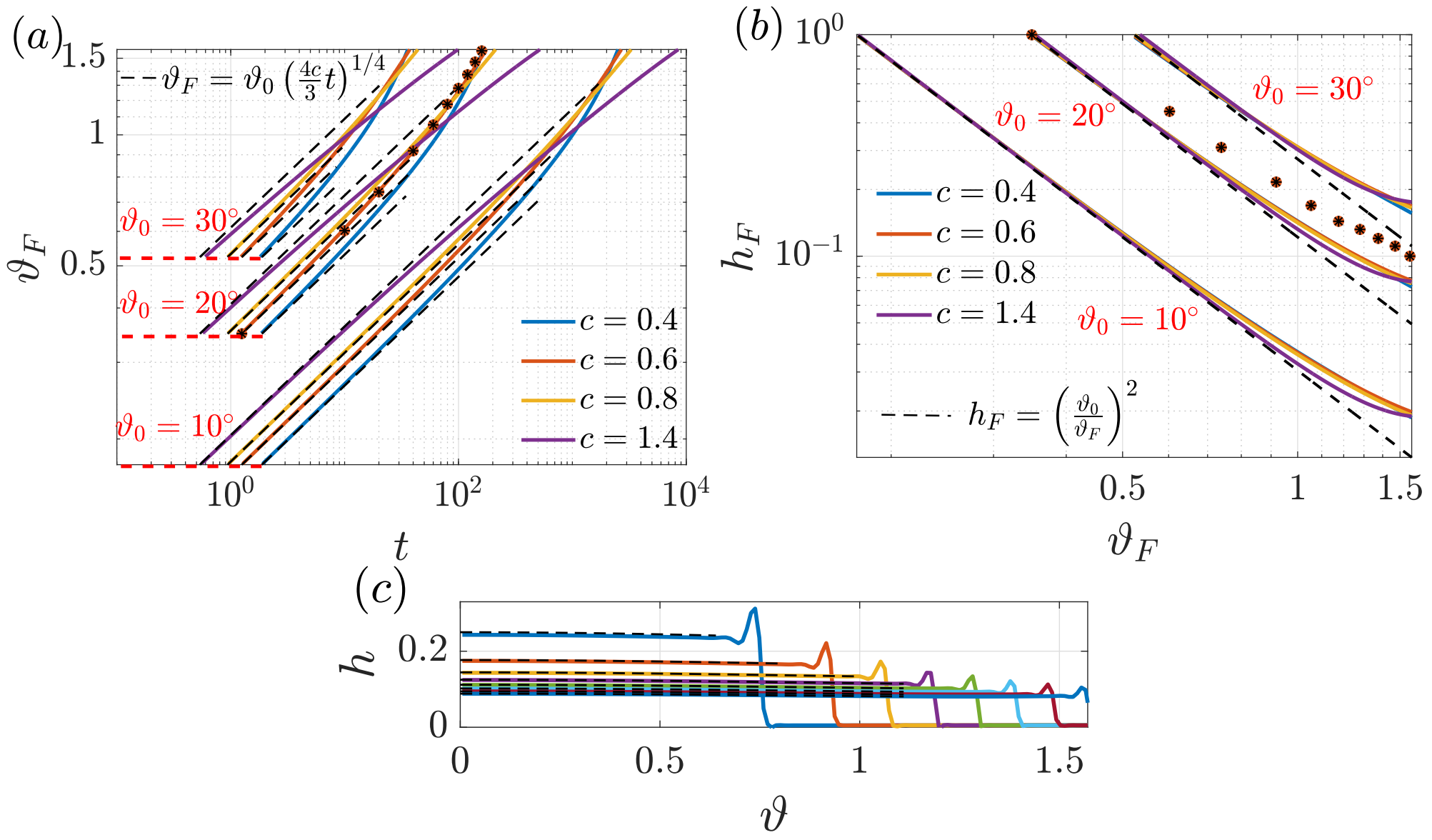

Figure 4: Spreading of an initial volume of fluid on an spheroid. ($a$) Variation of the front angle $\vartheta_F$ with time and ($b$) of the thickness at the front $h_F$ with $\vartheta_F$, for different values of $c$ (coloured lines) and $\vartheta_0$ (different clusters of curves). The black dashed lines correspond to the analytical approximation of the relation $\vartheta_F(t)$ and $h_F(\vartheta_F)$, while the stars are the values recovered by a numerical simulation of the complete model with $c = 0.6$, $Bo = 500$, $\delta = 10^{-3}$. ($c$) Numerical thickness distribution obtained from the complete model with $c = 0.6$, $Bo = 500$, $\delta = 10^{-3}$ as a function of $\vartheta$ at different times: $t = 20$ (blue), $t = 40$ (orange), $t = 60$ (yellow), $t = 80$ (purple), $t = 100$ (green), $t = 120$ (cyan), $t = 140$ (maroon), $t = 160$ (black). The black dashed lines denote the corresponding leading order large time drainage solution.

with an expression of the thickness at the front $h_F$:

$$\sqrt{\frac{3}{4\mathrm{t}}}\frac{1}{\sqrt{c}}\frac{\vartheta_F^2}{2}=\frac{\vartheta_0^2}{2}\longrightarrow\vartheta_F=\vartheta_0\left(\frac{4ct}{3}\right)^{1/4}, \quad h_F=\left(\frac{\vartheta_0}{\vartheta_F}\right)^2. \tag{3.10}$$

Note that this expression with $c = 1$ coincides with the solution on a sphere (Takagi & Huppert 2010). These results, reported in black dashed line in figure 4($a, b$), well agree with the implicit equation for small values of $\vartheta$. The velocity of the front $U_F = \mathrm{d}\vartheta_F/\mathrm{d}t = (c/192)^{1/4}\,t^{-3/4}$ decreases with time. Therefore, the front slows down as moving downstream toward the equator, for all values of $c$.

We verify the faithfulness of this approach by comparing it with the numerical results of the complete model (2.4) with parameters $c = 0.6$, $Bo = 500$ and $\delta = 10^{-3}$ (figure 4($c$)). To simulate the spreading on the substrate, we consider a precursor film of size $h_{pr} = 0.005$ (Troian *et al.* 1989*b*; Kondic & Diez 2002) with the following initial condition (Balestra *et al.* 2019):

$$h(\vartheta,0)=\frac{h_i-h_{pr}}{2}\left(1-\tanh\left(100\left(\vartheta-\vartheta_0\right)\right)\right)+h_{pr}. \tag{3.11}$$

Figure 4($c$) shows the evolution with time of the film thickness, with $\vartheta_0 = 20°$. In the vicinity of the front, a capillary ridge connects the film to the precursor one. Far from the front, the drainage solution well approximates the thin film evolution. In figure 4($a, b$), we report also the position and the values of the maximum thickness at the ridge, with a good agreement with the analytical approach.

The spreading velocity decreases with time and is proportional to $c^{1/4}$, in the vicinity of the pole. As $c$ increases, for fixed $\vartheta_F$, the tangential gravity component increases while the area invaded by the fluid does not vary, at leading order ($w \approx \vartheta$), close to the pole. The propagation velocity therefore increases since a faster drainage is observed with increasing $c$. Nevertheless, at large times, spheroids with smaller $c$ present larger values of $\vartheta_F$. Close to the equator, the tangential gravity component is almost vertical and thus the film velocity is not strongly affected by $c$. Nevertheless, for fixed equatorial radius, the distance covered for a small increment $\mathrm{d}\vartheta_F$ increases with $c$, at large $\vartheta_F$, therefore implying a reduction of the spreading velocity $\mathrm{d}\vartheta_F/\mathrm{d}t$.

In this section, we described the competition between the substrate's slope and curvature in defining the drainage and spreading patterns on an axisymmetric substrate, the spheroid. In the ESM, we report also the case of a paraboloid, which instead always shows a decreasing mean curvature and thus an increasing thickness moving away from the pole. The spheroid analysis was simplified thanks to the absence of odd terms in the asymptotic expansion in $\vartheta$. To better understand the role of the curvature in modifying the drainage, we now consider the torus, a substrate in which the symmetry with respect to $\vartheta$ is broken.

## 4. Non-symmetric drainage and spreading: coating of a torus

### 4.1. *Drainage problem*

In this section, we consider the drainage of a thin film flowing on a toroidal substrate of tube radius $R$ and distance $dR$ between the axis of revolution and the center of the tube (see figure 5($a$)). The torus is thus generated by the rotation along the azimuthal direction of a circular cross-section whose center is located at a distance $d$ from the axis of rotation. Non-dimensionalizing the in-plane directions and substrate variables with $R$, the following parameterization based on the zenith $\vartheta$ and the azimuth $\varphi$ is employed:

$$\boldsymbol{X}(\vartheta,\varphi) = ((d+\sin\vartheta)\cos\varphi, (d+\sin\vartheta)\sin\varphi, \cos\vartheta) \tag{4.1}$$

The position along the cylinder, at each azimuthal circular cross-section, is defined through the zenith $\vartheta$. Two limiting cases are identified; the first one occurs for $d \rightarrow \infty$, which leads to the cylindrical case, reported in the ESM. The second case occurs for $d = 1$, in which the points at $\vartheta = -90°$ are in contact, leading to the so-called horn torus. The gravity term $g_t^{\{1\}}$ and $w$ read

$$g_t^{\{1\}}(\vartheta) = \sin(\vartheta), \quad w(\vartheta) = d + \sin(\vartheta). \tag{4.2}$$

The same procedure employed for the drainage solution of the spheroidal case is adopted. However, in this case we cannot *a priori* neglect the odd terms in the asymptotic expansion, i.e. $h(\vartheta,t) = H_0(t) + \vartheta H_1(t) + \vartheta^2 H_2(t)\ldots$. The resulting problems, at different orders in $\vartheta$, are reported in Appendix B. For the sake of brevity, the large time solution at $O(\vartheta^4)$ reads:

$$h = \sqrt{\frac{3}{2t}}\left(\frac{19377\vartheta^4}{176000d^4} - \frac{1409\vartheta^3}{11000d^3} - \frac{7477\vartheta^4}{147840d^2} + \frac{31\vartheta^2}{200d^2} + \frac{91\vartheta^3}{2640d} - \frac{\vartheta}{5d} + \frac{43\vartheta^4}{10752} + \frac{\vartheta^2}{16} + 1\right) + O(\vartheta^5) + O\left(\frac{1}{t^{3/2}}\right) \tag{4.3}$$

The cylinder thickness distribution is recovered for $d \rightarrow \infty$ (Balestra *et al.* 2018$a$). The $O(1)$ solution is analogous to the cylinder case for any value of $d$. The drainage problem is numerically solved in the domain $-\pi/2 < \vartheta < \pi/2$. Numerical convergence is achieved with $\Delta\vartheta = 0.5°$. Figure 5 shows a comparison between the numerical and large-time analytical solutions of the drainage problem, for different values of $d$ in the range $-\pi/2 < \vartheta < \pi/2$. The distribution is not symmetric with respect to $\vartheta = 0$. In particular, the thickness is larger for negative values of $\vartheta$, i.e. on the inner side of the torus, while for $\vartheta > 0$ the thickness is almost constant. These differences are enhanced as $d$ decreases. The numerical solution well compares with the analytical one at $O(\vartheta^4)$ while, at $O(\vartheta^2)$, the agreement is good only in the vicinity of the top.

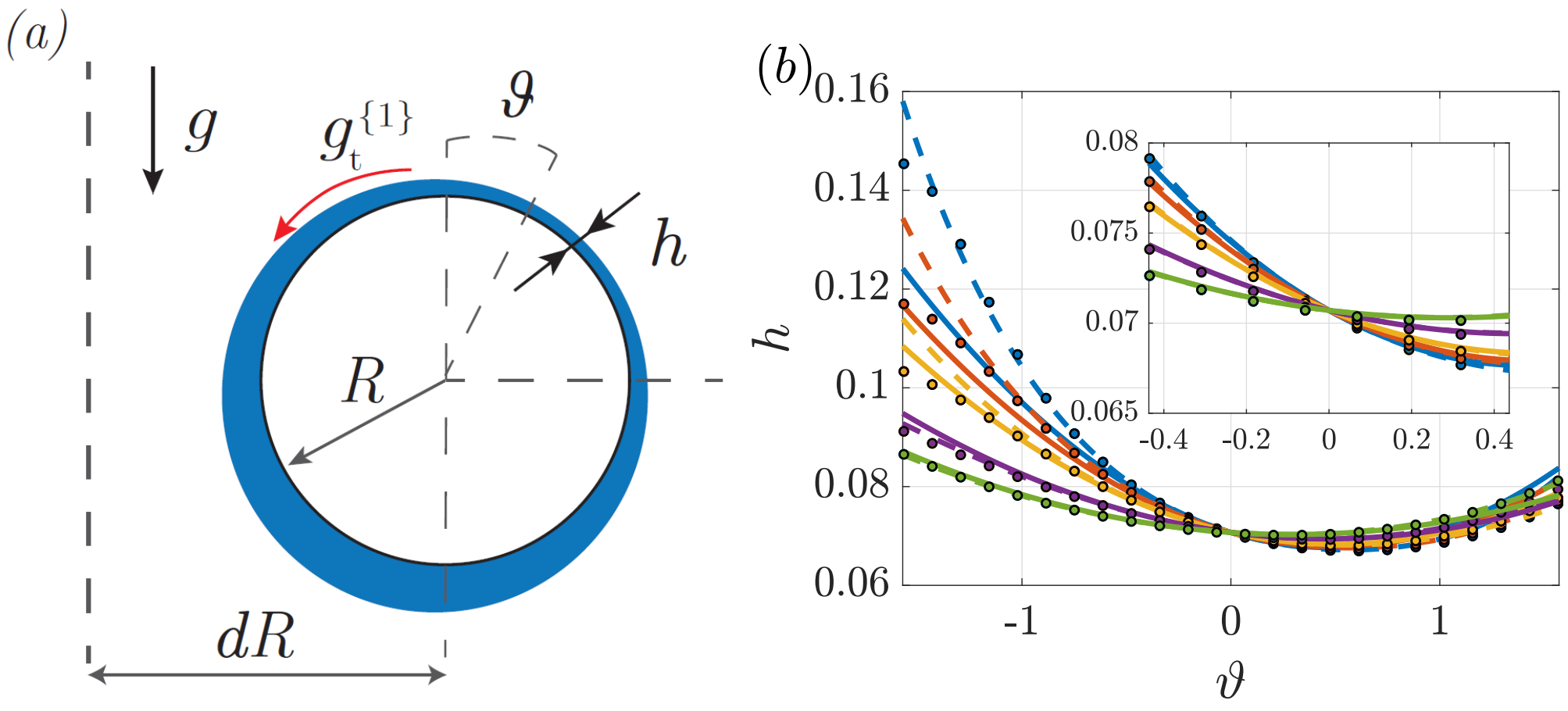

Figure 5: ($a$) Sketch of the axisymmetric flow configuration for the coating of a torus. ($b$) Drainage solution on a torus at $t = 300$, numerical (colored dots) and analytical solutions at order $O(\vartheta^2)$ (solid lines) and $O(\vartheta^4)$ (dashed lines), for $d = 1.1$ (blue), $d = 1.25$ (orange), $d = 1.5$ (yellow), $d = 2.5$ (purple), $d = 5$ (green).

At the top ($\vartheta = 0$), $\mathcal{K} = -1$ and therefore the film drains as in the cylinder case, locally. The different thickness distributions in the two sides of the circular cross-section of the torus result from the non-symmetric drainage with respect to $\vartheta$. While the slope is symmetric with respect to $\vartheta$, the mean curvature decreases on the inner part and decreases on the outer part. Following the discussion of Section 3.1, a decreasing (respectively increasing) curvature induces an increasing (respectively decreasing) thickness. Therefore, much larger thicknesses are attained on the inner part than on the outer one, where the thickness slightly decreases, in the vicinity of the top. The slight increase on the outer part observed at large $\vartheta$ is due to the saturation of the mean curvature value, which remains almost constant, while the substrate's slope increases. From a quantitative point of view, we consider the product between the normal component of gravity and the mean curvature:

$$(\mathbf{g} \cdot \mathbf{e}_3)\,\mathcal{K} \approx -\left(\left(-\frac{1}{a^2} - \frac{1}{2}\right)\vartheta^2 + \frac{\vartheta}{a} + 1\right). \tag{4.4}$$

which shows a decrease on the inner part, thus highlighting an accumulation of fluid downstream, and vice versa. The same result could be obtained by considering how the flux perturbs the $O(1)$ solution.

## 4.2. *Spreading problem*

We now present the results for the spreading of a volume of fluid on a torus. We consider an initial volume of fluid of thickness $h = 1$ in the region $-\vartheta_0 < \vartheta < \vartheta_0$. The breaking of symmetry with respect to $\vartheta = 0$ results in two different spreading fronts for $\vartheta < 0$ (inner side) and $\vartheta > 0$ (outer side). However, at $\vartheta = 0$, the drainage gravity component is exactly zero, i.e. $q^{\{1\}} = h^3 g_t^{\{1\}} h^3 = 0$. Therefore, the total volume on each side of the torus is conserved since there is no flux at $\vartheta = 0$. Note that, when hydrostatic or capillary effects are considered, the flux is not exactly zero at the top. A preliminary analysis showed that appreciable variations of the mass on the two sides (of the order of 2%) are observed for $Bo = 250$ and $\delta = 0.1$, when either pure capillary or pure hydrostatic effects are considered, in addition to drainage. For larger values of $Bo$ or smaller values of $\delta$, these differences rapidly decrease. In the limit $Bo \to \infty$ and $\delta = 0$ (i.e. the considered drainage problem), a zero flux at the top of the torus is numerically observed.

The conservation of mass for the two regions reads:

$$\int_0^{\vartheta_F^O(t)} h(\vartheta,t)w(\vartheta)\mathrm{d}\vartheta = \int_0^{\vartheta_0} w(\vartheta)\mathrm{d}\vartheta, \quad \int_{-\vartheta_F^I(t)}^{0} h(\vartheta,t)w(\vartheta)\mathrm{d}\vartheta = \int_{-\vartheta_0}^{0} w(\vartheta)\mathrm{d}\vartheta, \tag{4.5}$$

where $\vartheta_F^O(t)$ and $\vartheta_F^I(t)$ are the front angle on the outer and inner part, respectively, $w(\vartheta) = d + \sin(\vartheta)$, and $h(\vartheta,t)$ is given by equation (4.3). Note that the two integrals on the RHS do not assume the same value, since $w(\vartheta)$ is not symmetric with respect to $\vartheta = 0$. Equations (4.5) are implicit integrals that are solved in Matlab through the built-in function "fsolve". A first analytical approximation is found by taking the $O(1)$ approximation, leading to:

$$\frac{\vartheta_F^O}{\vartheta_0} = \frac{\vartheta_F^I}{\vartheta_0} = \sqrt{\frac{2t}{3}}, \tag{4.6}$$

i.e. the solution of $O(1)$ does not depend on $d$ and is analogous to the spreading on a cylinder (Balestra *et al.* 2018*a*). The thickness at the front thus reads $h_F = \vartheta_0/\vartheta_F$. A better approximation that includes the curvature of the torus can be obtained by considering the $O(\vartheta)$ approximation of the integrand:

$$\int_0^{\vartheta_F^O(t)} \sqrt{\frac{3}{2t}}(\frac{4}{5}\vartheta+d)\mathrm{d}\vartheta = \int_0^{\vartheta_0}(d+\vartheta)\mathrm{d}\vartheta \to \vartheta_F^{O2}+\frac{5d}{2}\vartheta_F^O-\left(\frac{5}{2}\left(d\vartheta_0+\vartheta_0^2/2\right)\right)\sqrt{\frac{2t}{3}} = 0,$$

$$\to \vartheta_F^O(t) = \frac{1}{2}\left(-\frac{5d}{2}+\sqrt{\frac{25d^2}{4}+4(d\vartheta_0+\vartheta_0^2/2)\frac{5}{2}\sqrt{\frac{2t}{3}}}\right), \tag{4.7}$$

$$\int_{-\vartheta_F^I(t)}^{0} \sqrt{\frac{3}{2t}}(\frac{4}{5}\vartheta+d)\mathrm{d}\vartheta = \int_{-\vartheta_0}^{0}(d+\vartheta)\mathrm{d}\vartheta \to \vartheta_F^{I2}-\frac{5d}{2}\vartheta_F^I+\left(\frac{5}{2}\left(d\vartheta_0-\vartheta_0^2/2\right)\right)\sqrt{\frac{2t}{3}} = 0,$$

$$\to \vartheta_F^I(t) = \frac{1}{2}\left(\frac{5d}{2}-\sqrt{\frac{25d^2}{4}-4(d\vartheta_0-\vartheta_0^2/2)\frac{5}{2}\sqrt{\frac{2t}{3}}}\right). \tag{4.8}$$

Figure 6($a,b$) shows the behaviors of $\vartheta_F^O$, $\vartheta_F^I$ and the front thicknesses $h_F^O$ and $h_F^I$ on the inner and outer sides of the torus, respectively, for different values of $d$ and $\vartheta_0$. As concerns panel ($a$), for a fixed time, the front angle on the inner side is always larger than the one on the outer side. An increase of $d$ leads to a decrease (respectively increase) of $\vartheta_F$ on the inner (respectively outer) side. The front thickness does not strongly depend on $d$, even if some differences can be appreciated on the inner side, for large values of $\vartheta_F^O$. The $O(1)$ approximation gives a reasonable agreement in the prediction of the front angle and thickness. In particular, it appears to be the lower (respectively upper) limit for the inner (respectively outer) sides, as $d$ increases. The order $O(\vartheta)$ approximations well follow the implicit relations (4.5). We compare these analytical results with a numerical simulation of the complete model (2.4) with parameters $d = 1.25$, $Bo = 500$, $\delta = 10^{-3}$, $h_{pr} = 0.005$, initial condition

$$h(\vartheta,0) = \frac{h_i-h_{pr}}{2}\left(1-\tanh(100(\vartheta-\vartheta_0)\right)+h_{pr}, \quad \text{for } \vartheta > 0, \tag{4.9}$$

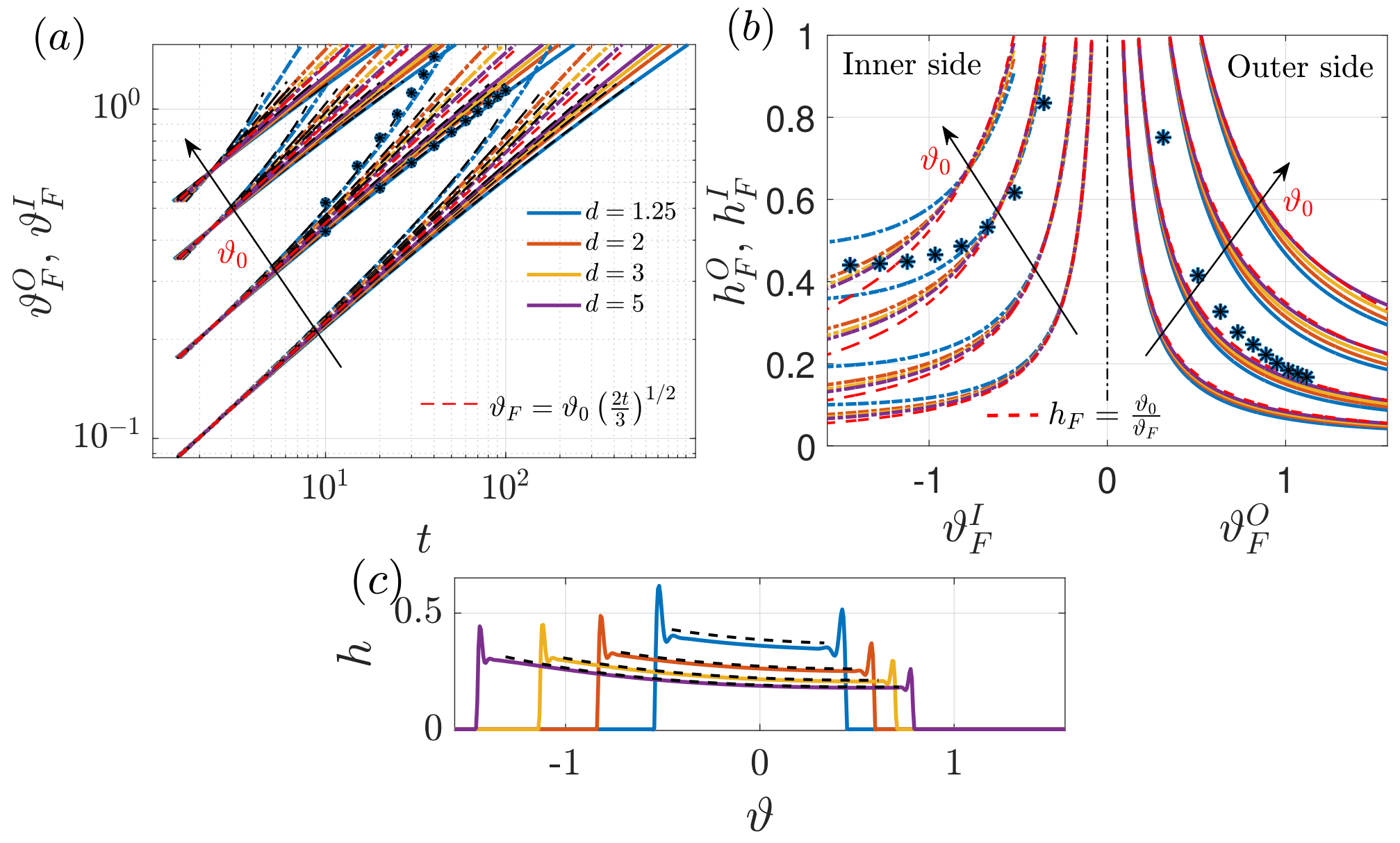

Figure 6: Spreading of an initial volume of fluid on a torus. $(a)$ Variation of the front angle $\vartheta_F$ with time and $(b)$ of the thickness at the front $h_F$ with $\vartheta_F$, for different values of the initial angle $\vartheta_0$ and $d$. The solid and dot-dashed lines denote the values of $\vartheta_F$ and $h_F$ on the outer and inner sides, respectively. The black and red dashed lines correspond to the $O(1)$ and $O(\vartheta)$ analytical approximations of the relation $\vartheta_F(t)$ and $h_F(\vartheta_F)$, respectively, while the stars are the values recovered by a numerical simulation of the complete model with $d = 1.25$, $Bo = 500$, $\delta = 10^{-3}$, precursor film thickness $h_{pr} = 0.005$. $(c)$ Numerical thickness distribution obtained from the complete model with $d = 1.25$, $Bo = 500$, $\delta = 10^{-3}$ as a function of $\vartheta$ at different times: $t = 10$ (blue), $t = 20$ (orange), $t = 30$ (yellow), $t = 40$ (purple). The black dashed lines denote the corresponding large-time drainage solutions.

$$h(\vartheta, 0) = \frac{h_i - h_{pr}}{2}\,(1 - \tanh(100(-\vartheta - \vartheta_0)) + h_{pr}, \quad \text{for } \vartheta < 0, \tag{4.10}$$

and $\vartheta_0 = 10°$ (see figure 6$(c)$). The agreement between the numerical front angle, given by the maximum thickness location, and the theoretical one is very good, and also the maximum thickness well follows the front thickness predicted by the theory.

In analogy with the drainage solution, the faster spreading attained on the inner region is related to the substrate geometry. For a fixed angular distance from the top, the area covered by the spreading fluid is larger on the outer region than on the inner one. Therefore, for a fixed time, the fluid spreads faster on the inner region, reaching larger values of $\vartheta_F$ than on the outer region. Interestingly, the solution at $O(1)$ does not capture the symmetry breaking, since, at the top, a torus locally coincides with a cylinder. Nevertheless, the $O(\vartheta)$ approximation already captures the asymmetry of the substrate.

The torus case shows non-symmetric drainage and spreading along the zenith direction. In the following, we present how these analyses can be extended to non-axisymmetric substrates which are characterized by a three-dimensional, non-uniform along the azimuthal direction, drainage. We chose as a testing ground an ellipsoid with three different axes.

## 5. Three-dimensional drainage and spreading: coating of an ellipsoid

### 5.1. *Numerical drainage solution*

In this section, we study the coating of an ellipsoidal substrate of horizontal semiaxes $aR$, $bR$ and vertical semiaxis $R$ (see figure 1); gravity is pointing downward. In non-dimensional

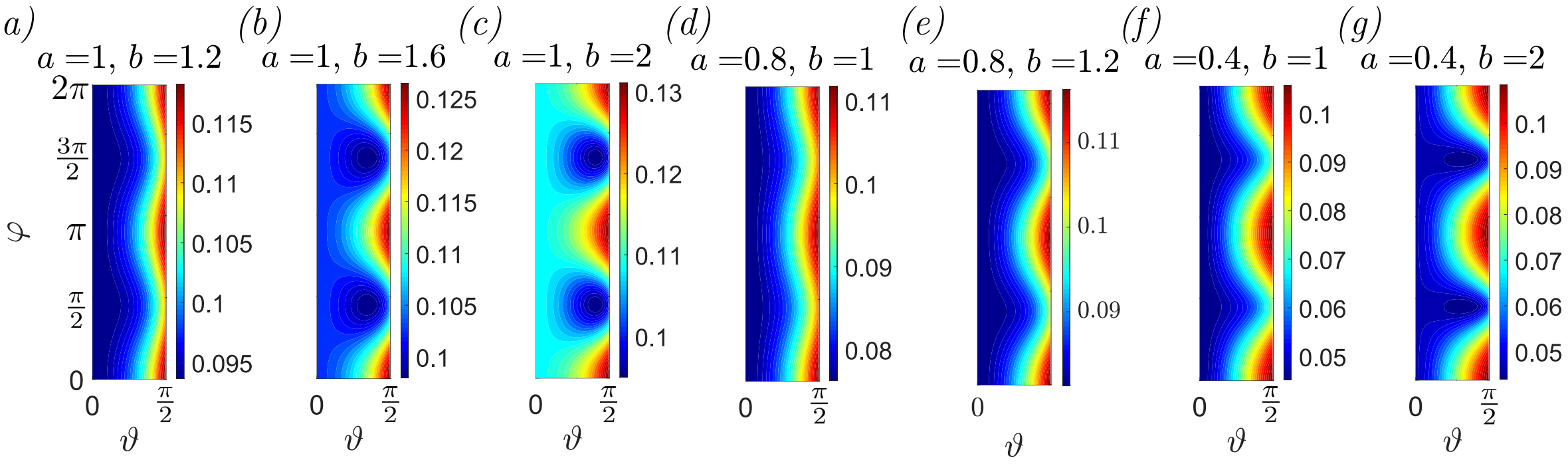

Figure 7: Numerical solution of equation (2.6) at $t = 100$ as a function of $(\vartheta, \varphi)$, for different values of the semiaxes $a$ and $b$, with $a \leqslant 1$ and $b \geqslant 1$.

form, the following parameterization holds:

$$\boldsymbol{X}(\vartheta, \varphi) = (a \sin\vartheta \cos\varphi, b \sin\vartheta \sin\varphi, \cos\vartheta) \tag{5.1}$$

We identify different limiting cases, depending on the values of $a$ and $b$. If $a = b = 1$, we recover the spherical case; if $a = b \neq 1$ the resulting substrate is an axisymmetric ellipsoid of unitary height and equatorial radius $a = b$, whose results can be recovered from those of Section 3.1. Note that the time scale is different since the in-plane directions and substrate variables are non-dimensionalized with the height and not with the equatorial radius, in the present section. When $a \neq b$ the axisymmetry is broken since the two axes at the equator are different. In the following, we assume that $b \geqslant a$ and consider the range $0.4 < a, b < 2$. Note that the solutions for $a > b$ can be recovered by simply translating of $\varphi = 90°$ the solution for $b \geqslant a$ (obtained by swapping the desired values of $a$ and $b$).

The metric components vary along the $\varphi$ direction; in particular, the metric tensor is not diagonal (see ESM). The local coordinates system defined by the parameterization is thus non-orthogonal and a second gravity component $g_t^{\{2\}}(\vartheta, \varphi)$, appears. The square root of the determinant of the metric and the gravity terms now read:

$$w(\vartheta, \varphi) = \sqrt{\sin^4(\vartheta)\left(a^2 \sin^2(\varphi) + b^2 \cos^2(\varphi)\right) + a^2 b^2 \sin^2(\vartheta)\cos^2(\vartheta)}, \tag{5.2}$$

$$g_t^{\{1\}}(\vartheta, \varphi) = \frac{\sin(\vartheta)\left(a^2 \sin^2(\varphi) + b^2 \cos^2(\varphi)\right)}{\sin^2(\vartheta)\left(a^2 \sin^2(\varphi) + b^2 \cos^2(\varphi)\right) + a^2 b^2 \cos^2(\vartheta)}, \tag{5.3}$$

$$g_t^{\{2\}}(\vartheta, \varphi) = \frac{\sin(\varphi)\cos(\varphi)\left(a^2 \cos(\vartheta) - b^2 \cos(\vartheta)\right)}{a^2 b^2 \cos^2(\vartheta)\cos^2(\varphi) + a^2 b^2 \cos^2(\vartheta)\sin^2(\varphi) + a^2 \sin^2(\vartheta)\sin^2(\varphi) + b^2 \sin^2(\vartheta)\cos^2(\varphi)}. \tag{5.4}$$

We solve equation (2.6) by imposing periodic boundary conditions in $0 < \varphi < 2\pi$ and the initial condition $h(\vartheta, \varphi, 0) = 1$. Numerical convergence is achieved with a characteristic mesh size of 0.9°. Figure 7 and figure 8 respectively show the resulting film distributions and a section at $\vartheta = \pi/4$ for different values of $a$ and $b$, at $t = 100$. We first increase the value of $b$, with $a = 1$. For $b = 1.2$ (panel $(a)$), the thickness presents modulations along the azimuthal direction, with a maximum thickness localized at $\varphi = k\pi$ ($k = 0, 1, 2$), i.e. along the direction of the smaller axis $a$. These modulations are enhanced as $b$ increases (panel $(b)$), with larger values of the attained thickness. Two regions of low thickness are localized at $\varphi = \pi/2 + k\pi$, along the larger axis $b$. The same trends are observed further increasing $b$ (panel $(c)$). When $b = 1$ and $a$ decreases (panel $(d)$), the thickness also presents modulations along the azimuthal direction, but the thickness always increases moving downstream. The thickness decreases as $a$ decreases. Similar patterns are also obtained when small values of $a$ and large values of $b$ are considered. The numerical solution of equation (2.6) shows the presence of modulations of the thickness along the azimuthal direction. According to Section 3.1, spheroids with small (respectively large) height were characterized by a decrease (respectively increase) of the thickness. We can extend these considerations to an ellipsoid by considering the drainage along the principal directions defined by $(x, y)$, see figure 1($c$). Since the drainage component along the azimuthal direction is identically zero along the two principal semiaxes, the flow locally behaves like the spheroidal case of Section 3.1. Therefore, we expect to follow these trends along the two semiaxes, depending on $a$ and $b$. In the axisymmetric case, the thickness increases downstream for height-radius ratios larger than 0.74, which corresponds to $a, b \lessapprox 1.35$. Therefore, when $a, b \lessapprox 1.35$ we always observe an increase of the thickness with $\vartheta$, as observed in figure 9($a$-$c$) (see also figure 3 for the cases with $c > c^*$). However, the thickness presents clear modulations owing to the non-uniform drainage when $a \neq b$. Similarly, when $a, b \gtrapprox 1.35$ one expects a decrease of the thickness followed by a slight increase at large $\vartheta$, with modulations if $a \neq b$, as shown in figure 9($d$-$g$). The intermediate situation occurs when $a \lessapprox 1.35$ and $b \gtrapprox 1.35$, characterized by an increase of the thickness along the $x$ direction and a decrease along the $y$ direction, as observed in figure 7($b, c, g$).

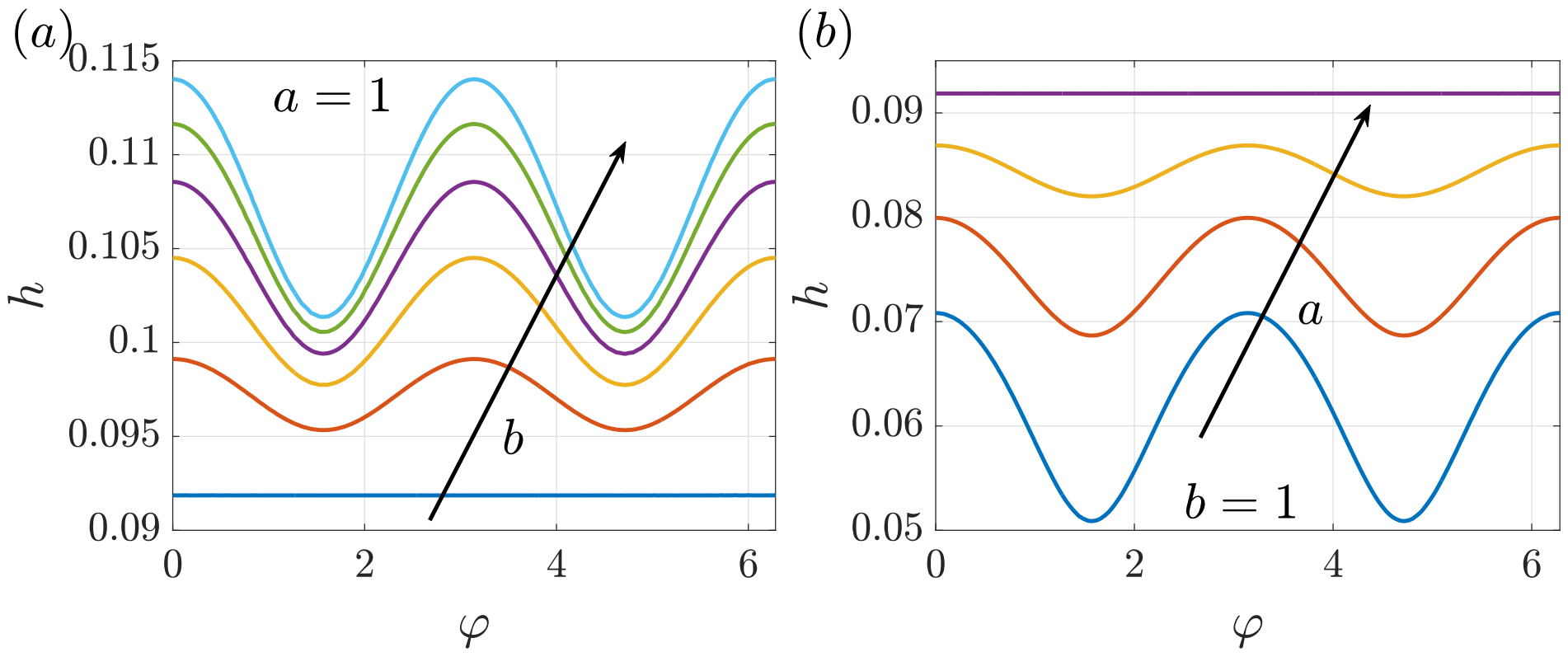

Figure 8: Numerical solution of equation (2.6) at $t = 100$ and $\vartheta = \pi/4$ as a function of ($\varphi$): (*a*) $a = 1$ and $b = 1$ (blue), $b = 1.2$ (orange), $b = 1.4$ (yellow), $b = 1.6$ (purple), $b = 1.8$ (green), $b = 2$ (cyan); (*b*) $b = 1$ and $a = 0.4$ (blue), $a = 0.6$ (orange), $a = 0.8$ (yellow), $a = 1$ (purple).

The modulations of the thickness distribution are related to the variation of drainage with the azimuth. In the vicinity of the minor semiaxis, the tangential gravity component along the zenith is larger than close to the major semiaxis. Higher velocities are thus attained along the minor semiaxis, displacing more fluid downstream than along the major semiaxis. This process induces transport of fluid from progressively farther and farther regions and thus a secondary flow from the major semiaxis (associated with low velocities) to the minor semiaxis (associated with large velocities). In the light of this discussion, one may wonder if these patterns persist with time or merely represent a snapshot of a more intricate evolution. In the following, we also aim at clarifying this aspect by deriving an analytical solution for the drainage problem.

### 5.2. *Analytical drainage solution*

In this section, we derive an analytical drainage solution and compare it with the numerical results of the previous section. In analogy with Section 3.1, we perform an asymptotic expansion in powers of $\vartheta$, with $\vartheta \ll 1$. The solution at order $O(1)$ does not depend on $\varphi$ since the solution at the pole has to be unique. We thus consider the following expansion, in

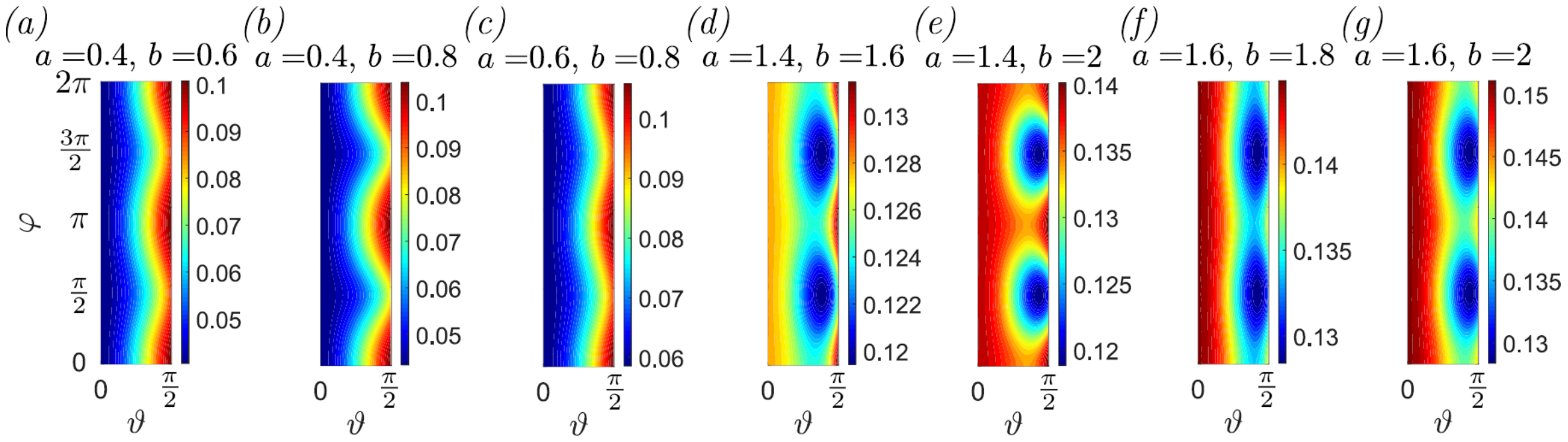

Figure 9: Numerical solution of equation (2.6) at $t = 100$ as a function of $(\vartheta, \varphi)$, for different values of the semiaxes $a$ and $b$, with $a, b < 1$ and $a, b > 1$.

which the odd terms have been removed because of symmetry:

$$h(\vartheta, \varphi, t) = H_0(t) + \vartheta^2 H_2(\varphi, t) + \vartheta^4 H_4(\varphi, t) + \vartheta^6 H_6(\varphi, t) + \ldots \tag{5.5}$$

We expand equation (2.6) at various orders in $\vartheta$. At order $O(1)$, one obtains the following ODE:

$$\frac{1}{3}\left(\frac{1}{a^2} + \frac{1}{b^2}\right) H_0(t)^3 + H_0'(t) = 0 \longrightarrow H_0(t) = \frac{1}{\sqrt{\frac{2}{3}t\left(\frac{1}{a^2} + \frac{1}{b^2}\right) + 1}} = \frac{1}{\sqrt{\alpha t + 1}}, \tag{5.6}$$

where $\alpha = \frac{2}{3}\left(1/a^2 + 1/b^2\right)$. Also in this case, the $O(1)$ solution reduces to $\left(\frac{3}{2\mathcal{K}_p t}\right)^{1/2}$ at late time, with $\mathcal{K}_p = \left(1/a^2 + 1/b^2\right)$. The equation at order $O(\vartheta^2)$ reads:

$$\frac{\partial H_2(\varphi, t)}{\partial t} = \frac{H_0(t)^2\left(\left(b^2 - a^2\right)\sin(2\varphi)\frac{\partial H_2}{\partial \varphi} + 2H_2\left(\left(a^2 - b^2\right)\cos(2y) - 2\left(a^2 + b^2\right)\right)\right)}{2a^2b^2}$$
$$+ \frac{H_0(t)^3\left(\left(a^4\left(b^2 - 2\right) - a^2b^4 + 2b^4\right)\cos(2\varphi) + 2\left(a^4\left(-\left(b^2 - 1\right)\right) + a^2\left(b^2 - b^4\right) + b^4\right)\right)}{6a^4b^4}, \tag{5.7}$$

which is a parabolic PDE in $H_2(\varphi, t)$. We numerically solve equation (5.7) with initial condition $H_2(\varphi, 0) = 0$. The periodic boundary conditions at $\varphi = [0, 2\pi]$ are automatically imposed thanks to a Fourier spectral collocation method implemented in Matlab. The time-stepping is performed by employing the built-in function "ode23t", with a tolerance of $10^{-6}$. Numerical convergence is achieved with 100 collocation points.

Figure 10($a$) shows the spatiotemporal evolution of the second order solution $H_2(\varphi, t)$, for $a = 0.5$ and $b = 1.5$. An initial growth in absolute value until $t \approx 0.3$ is followed by a slow decay at large times. In figure 10($b$) we report the $H_2$ profiles rescaled with $H_0$, at different times in the slow-decay regime. The second-order solution $H_2$ is $\pi$-periodic and the maximum is attained at $\varphi = k\pi$, i.e. along the smaller axis of the ellipsoid. At $\varphi = k\pi/2$, i.e. along the larger axis of the ellipsoid, the correction reaches much smaller values. As time increases, the profiles collapse on a single curve, suggesting that a large-time solution characterized by a separation of variables is possible, i.e. $H_2(\varphi, t) = H_0(t) H_2^*(\varphi)$. We introduce this decomposition in equation (5.7). Exploiting equation (5.6), the temporal

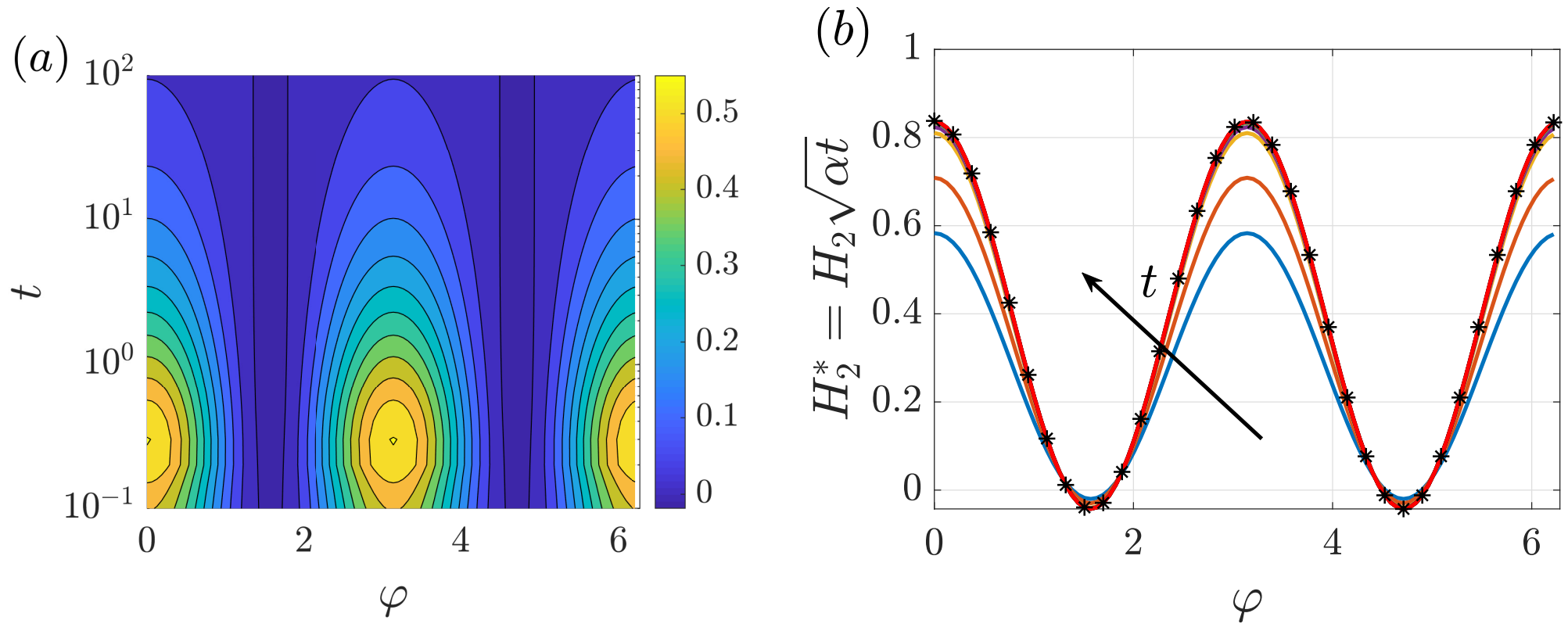

Figure 10: Drainage along an ellipsoid with $a = 0.5$ and $b = 1.5$. ($a$) Spatiotemporal evolution of $H_2$: iso-contours of $H_2$ in the $(\varphi, t)$ plane. ($b$) Second order correction $H_2^* = H_2/H_0 \approx H_2\sqrt{\alpha t}$ as a function of $\varphi$ at different times: $t = 0.4$ (blue), $t = 1$ (orange), $t = 5$ (yellow), $t = 10$ (purple) $t = 30$ (green), $t = 50$ (cyan), $t = 70$ (maroon), $t = 90$ (black), $t = 100$ (red). The black stars denote the late-time analytical solution for $H_2^*$ from equation (5.9).

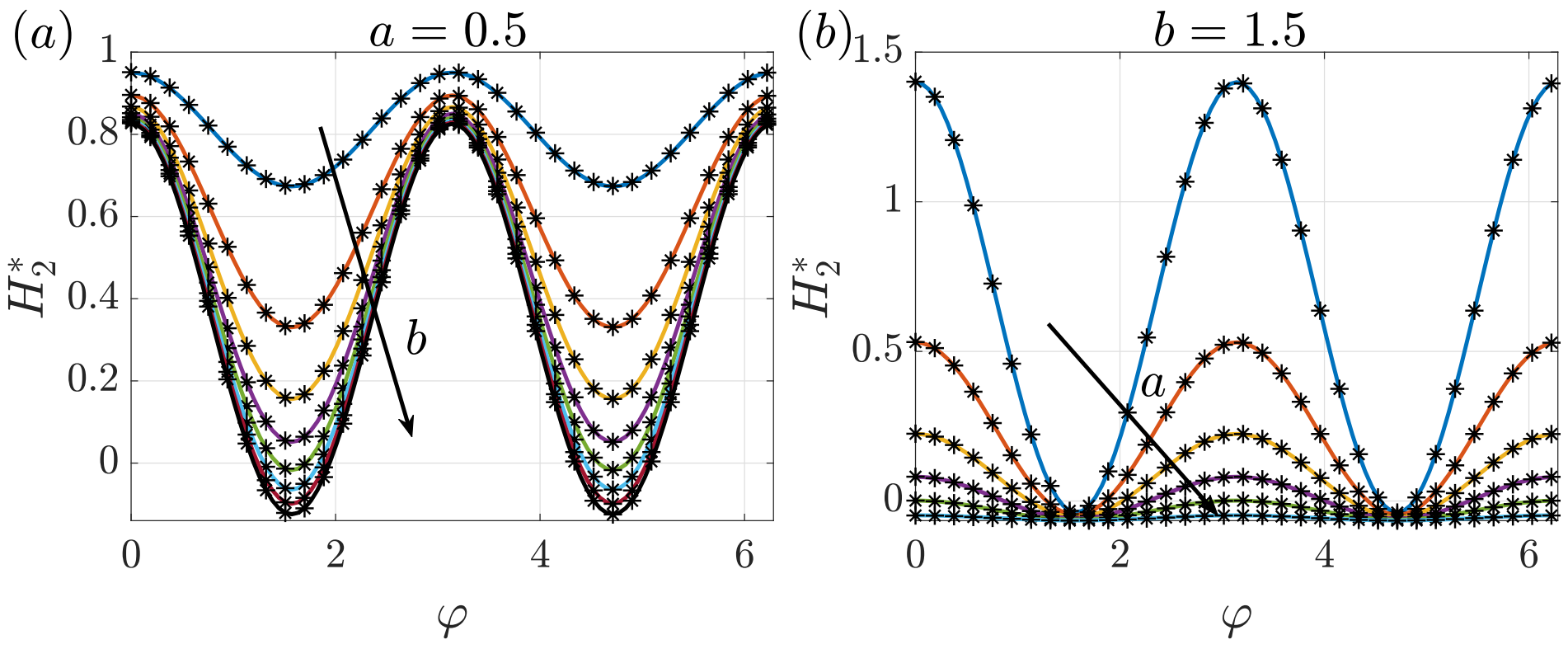

Figure 11: ($a$) Second order correction $H_2^* = H_2/H_0 \approx H_2\sqrt{\alpha t}$ as a function of $\varphi$ at $t = 100$, for $a = 0.5$ and increasing $b$: $b = 0.6$ (blue), $b = 0.8$ (orange), $b = 1$ (yellow), $b = 1.2$ (purple), $b = 1.4$ (green), $b = 1.6$ (cyan), $b = 1.8$ (maroon), $b = 2$ (black). ($b$) $H_2^*$ as a function of $\varphi$ at $t = 100$, for $b = 1.5$ and increasing $a$: $a = 0.4$ (blue), $a = 0.6$ (orange), $a = 0.8$ (yellow), $a = 1$ (purple), $a = 1.2$ (green), $a = 1.4$ (cyan). The black stars denote the late-time analytical solution for $H_2^*$ from equation (5.9).

dependence disappears and the following ODE for $H_2^*(\varphi)$ is obtained:

$$-a^4b^2\cos(2\varphi) + 2a^4b^2 + 2a^4\cos(2\varphi) - 2a^4 + a^2b^4\cos(2\varphi) + 2a^2b^4 + 3a^2b^2\left(a^2 - b^2\right)\sin(2\varphi)H_2^{*'}(\varphi)$$
$$- 2a^2b^2H_2^*(\varphi)\left(3\left(a^2 - b^2\right)\cos(2\varphi) - 5\left(a^2 + b^2\right)\right) - 2a^2b^2 - 2b^4\cos(2\varphi) - 2b^4 = 0, \quad (5.8)$$

whose solution reads:

$$H_2^*(\varphi) = C_1\sin(2\varphi)\sin^{-\frac{5\left(a^2+b^2\right)}{3\left(a^2-b^2\right)}}(\varphi)\cos^{\frac{5\left(a^2+b^2\right)}{3\left(a^2-b^2\right)}}(\varphi) - \frac{1}{8a^2b^2\left(4a^2+b^2\right)\left(a^2+4b^2\right)}\left(a^6\left(7b^2-4\right)\right.$$
$$\left. + 26a^4\left(b^4 - b^2\right) + a^2b^4\left(7b^2 - 26\right) + (a^2 - b^2)\left(a^4\left(b^2+4\right) + a^2b^2\left(b^2+14\right) + 4b^4\right)\cos(2\varphi) - 4b^6\right), \quad (5.9)$$

where $C_1$ is a constant to be determined. However, it is observed that $C_1 \neq 0$ implies an unbounded behavior. Therefore, we impose $C_1 = 0$ to prevent non-physical solutions. The analytical result for $H_2^*$ is reported in figure 10($b$), with an excellent agreement with the numerical solution. We then investigate the effect of $a$ and $b$ by considering different cases at $t = 100$, reported in figure 11. An increase of $b$ for fixed $a = 0.5$ (panel $(a)$) leads to a decrease of $H_2^*$ in the region $\varphi = k\pi/2$, while an increase in $a$ for fixed $b = 1.5$ (panel $(b)$) shows an overall decrease of $H_2^*$. Also for these cases, an excellent agreement with the analytical solution is observed. At $O(\vartheta^2)$, the large time analytical solution can be written in compact form as:

$$h \approx H_0(t)\left(1 + (f(a,b) + g(a,b)\cos(2\varphi))\vartheta^2\right). \tag{5.10}$$

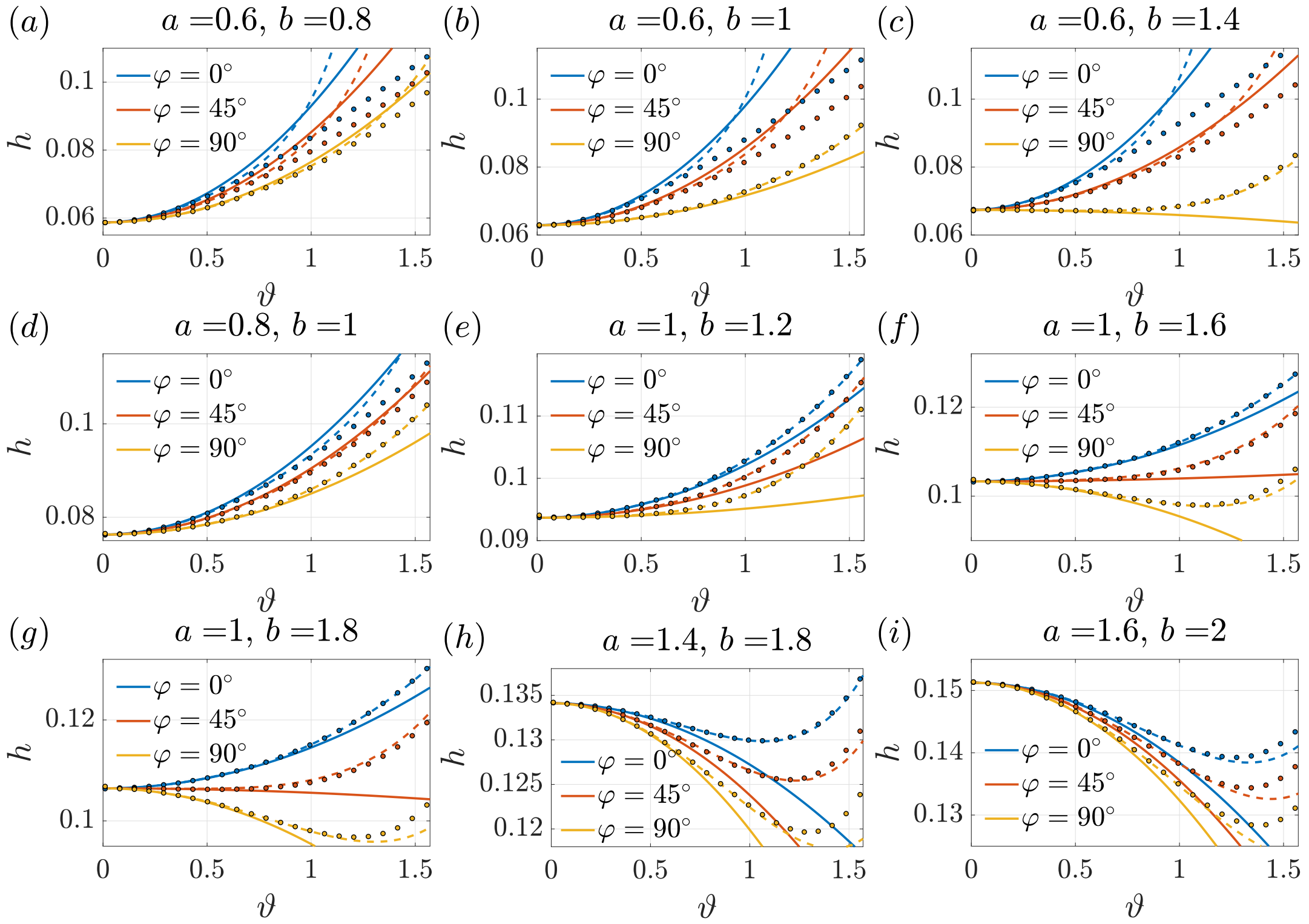

Figure 12: Comparison at three different azimuthal sections between the numerical (colored dots) and the quasi-analytical solution for an ellipsoid at $O(\vartheta^2)$ (solid lines) and $O(\vartheta^6)$ (dashed lines) at $t = 100$, for different values of $a$ and $b$.

The modulations observed in the numerical simulations of the previous section are captured by the $O(\vartheta^2)$ term, which is a $\pi-$periodic function of the azimuth. These modulations are present as long as $g(a,b) = (a^2 - b^2)\left(a^4\left(b^2+4\right) + a^2b^2\left(b^2+14\right) + 4b^4\right) \neq 0$. The only case in which modulations are absent occurs when $g(a,b) = 0$ and thus $a = b$, i.e. the spheroidal case. In the latter case, the solution reads $H_2^* = \frac{1}{10}(\frac{3}{b^2} - 2)$ and is formally analogous to the second order solution of the spheroid with $c = 1/b$ (see Section 3.1).

The faithfulness of the analytical solution is verified against the numerical simulations of Section 5.1 in figure 12. For the comparison, we consider the solution at orders $O(\vartheta^2)$ and $O(\vartheta^6)$. The higher order problems, together with their solutions $H_4$ and $H_6$, are reported in Appendix C. The same large-time behavior is observed. In general, the analytical solutions at $O(\vartheta^6)$ compare well with the numerical ones, while those at $O(\vartheta^2)$ are accurate only in the vicinity of the pole. The analytical solution at $O(\vartheta^6)$ deviates from the numerical one for $a < 0.8$. The agreement for $a > 0.8$ is satisfactory for any value of $b$.

In this section, we derived an analytical approximation for the drainage problem. The problem was solved by employing an asymptotic expansion in a first stage, followed by a separation of variables at each order of the expansion. The final structure of the analytical approximation (5.10) is characterized by a time-dependence separated by the spatial one, similarly to the previous cases. Nevertheless, the power-series expansion presents terms that depend on the azimuth. The simple form of (5.10) well captures the $\pi$-periodicity of the drainage solution, induced by the differences in drainage along the minor and major axes. In the spreading problem, these modulations may play a crucial role.

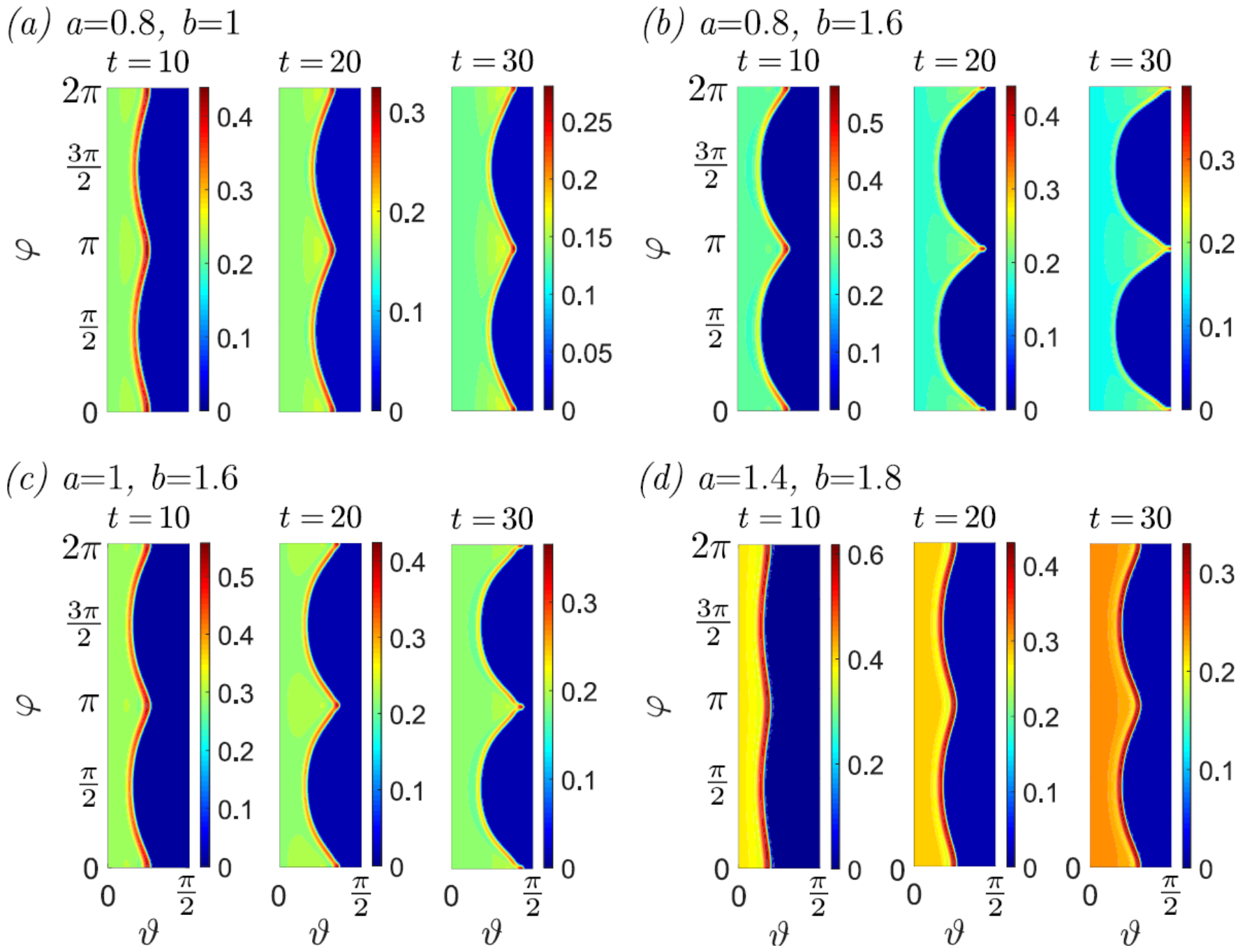

Figure 13: Iso-contours of the numerical solution for the spreading of equation (2.4), at different times, for an ellipsoid and $Bo = 100$, $\delta = 10^{-2}$, precursor film $h_{pr} = 0.02$ and $\vartheta_0 = 20°$. (*a*) $a = 0.8,\ b = 1$, (*b*) $a = 0.8,\ b = 1.6$, (*c*) $a = 1,\ b = 1.6$, (*d*) $a = 1.4,\ b = 1.8$.

### 5.3. *Spreading problem*

In this section, we consider the spreading of an initial volume of fluid of height $h_i = 1$ contained in the region $0 < \vartheta < \vartheta_0$, $0 < \varphi < 2\pi$. Figure 13 shows the evolution of the film thickness with time, for different values of $a$ and $b$, obtained employing the complete model (2.4) with initial condition formally analogous to equation (3.11), i.e. invariant along the azimuthal direction. For $t = 10$, the maximum thickness position, at which the front is located, is modulated along the azimuthal direction. This modulation accentuates with time and a region of large thickness forms at $\varphi = k\pi$ (along the shorter axis) while the thickness is much lower at $\varphi = k\pi/2$. Therefore, the front presents two peaks of large thickness aligned along the shorter axis. This effect is enhanced when larger (respectively lower) values of $b$ (respectively $a$) are considered.

In figure 14, we report a zoom in the region $0 < \varphi < \pi$ for one simulation of the complete model (2.4) with $Bo = 1000$ and $\delta = 10^{-2}$, together with a three-dimensional rendering of the spreading on the ellipsoid, viewed from the top. The black lines denote the streamlines of the flux $\mathbf{q} = q^{\{1\}}\mathbf{e}_1 + q^{\{2\}}\mathbf{e}_2$. The flux streamlines are almost parallel to the azimuthal direction at low values of $\vartheta$, then bend and align along the zenith direction as $\vartheta$ increases. An exception to this behavior is observed at $\varphi = 0, \pi/2$, in which the flow streamlines are always parallel to the zenith direction. The three-dimensional rendering highlights the formation of two, finger-like, front peaks of large thickness along the shorter axis, while the fluid slowly spreads along the larger axis.

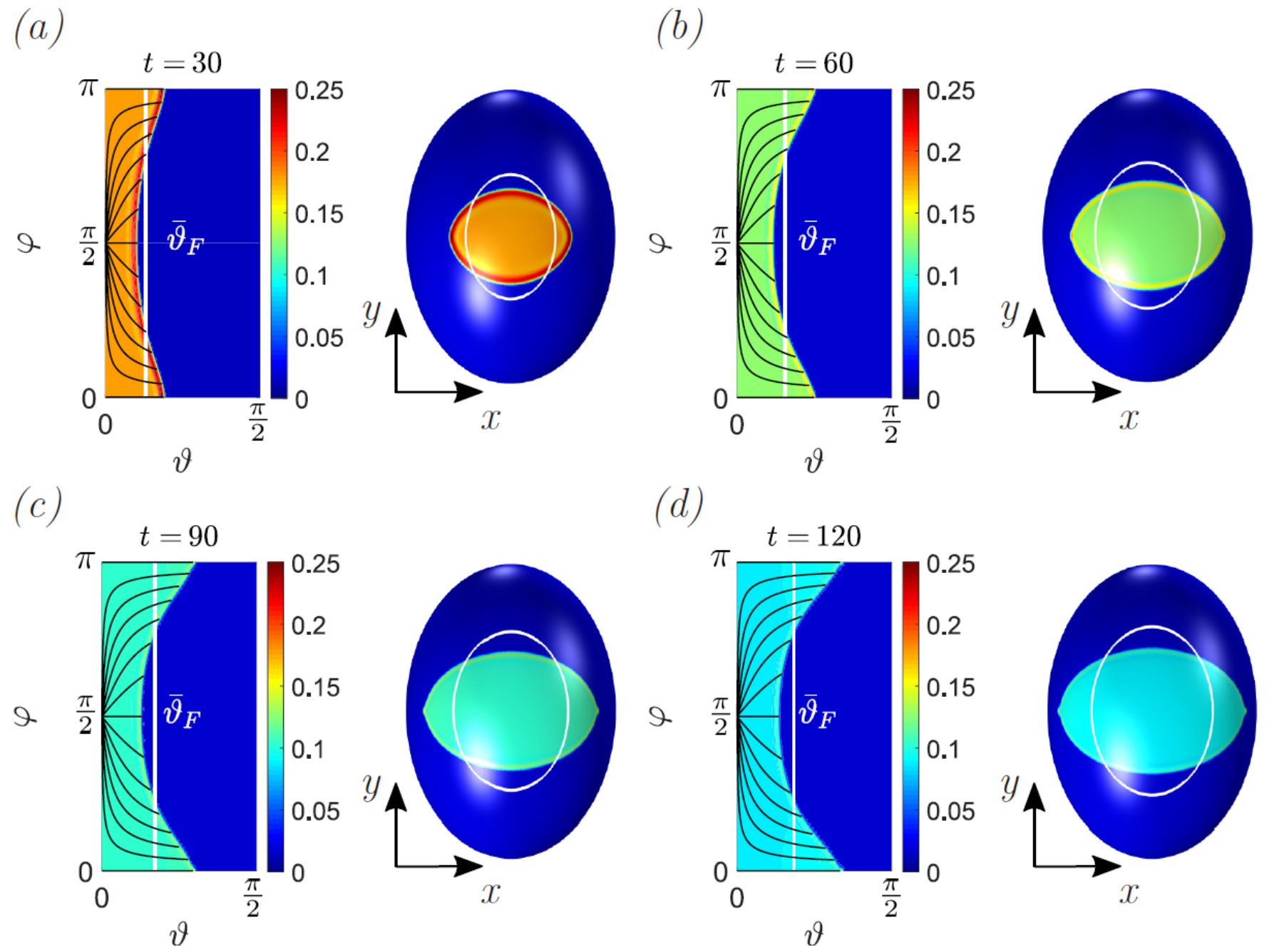

Figure 14: Iso-contours of the numerical spreading solution at different times of equation (2.4) for an ellipsoid with $a = 1$, $b = 1.4$, $Bo = 1000$, $\delta = 10^{-2}$, precursor film $h_{pr} = 0.02$ and $\vartheta_0 = 10°$. The black solid lines denote the streamlines of the volume flux per unit length $\mathbf{q}$ and the white line the average front $\bar{\vartheta}_F$.

A scaling law for the spreading front and thickness is obtained by neglecting the modulations of the front, and assuming a constant average value along the azimuth, i.e. $\bar{\vartheta}_F = 1/(2\pi)\int_0^{2\pi}\vartheta_F(\varphi,t)\mathrm{d}\varphi$. The conservation of volume reads:

$$\int_0^{2\pi}\int_0^{\bar{\vartheta}_F(t)} h(\vartheta,\varphi,t)w(\vartheta)\mathrm{d}\vartheta\mathrm{d}\varphi = \int_0^{2\pi}\int_0^{\vartheta_0} w(\vartheta)\mathrm{d}\vartheta\mathrm{d}\varphi, \tag{5.11}$$

where $h = H_0(t) + \vartheta^2 H_2(\varphi,t) + \vartheta^4 H_4(\varphi,t) + \vartheta^6 H_6(\varphi,t)$ is the asymptotic solution obtained in the previous section, and $w$ is given by equation (5.2). Also in this case, an analytical approximation is found by employing the large-time $O(\vartheta)$ approximation (5.6), with $w = ab\vartheta + O(\vartheta^2)$, leading to the following expressions for the average front position and thickness:

$$\bar{\vartheta}_F = \bar{\vartheta}_0\left(\frac{2}{3}\left(\frac{1}{a^2}+\frac{1}{b^2}\right)t\right)^{1/4}, \quad \bar{h}_F = \left(\frac{\bar{\vartheta}_0}{\bar{\vartheta}_F}\right)^2. \tag{5.12}$$

The azimuth-averaged numerical solution of equation (5.11) and the analytical approximation (5.12) are reported in figure 15, displaying a good agreement for low values of $\vartheta_0$ and large values of $a$, while the results start to diverge for large $\vartheta_0$ and small $a$.

Figure 16 shows the evolution of the front position and thickness with time, obtained from a numerical simulation of the complete model with with $a = 1$, $b = 1.4$, $Bo = 1000$, $\delta = 10^{-2}$, precursor film $h_{pr} = 0.02$ and $\vartheta_0 = 10°$. The values of front position and thickness are averaged and compared with the analytical prediction. The analytical and numerical simulation results show similar trends. However, at large times, the modulations of the front are very large and the front travels much faster along the shorter axis than along the longer one.

The spreading problem on an ellipsoid is characterized by a different front speed along the azimuthal direction, which leads to an accumulation of fluid and a faster spreading along the smaller axis. Peaks of large thickness form together with a modulation of the front, prior to any fingering instability. These modulations are similar to those observed in the previous Section for the drainage solution. As already explained, larger velocities induce transport of fluid from regions of lower velocity to regions of larger velocity. As a result, fluid accumulates and forms the observed peaks of large thickness. These velocity differences lead to a progressively more pronounced bending of the front. Therefore, a fingering instability analysis necessarily needs to consider the non-uniform spreading of the fluid along the ellipsoid, which may lead to the preferential formation of fingers. While this analysis focused on the spreading in the absence of surface tension, further studies may involve the formation of fingers resulting from the driven contact line instability.

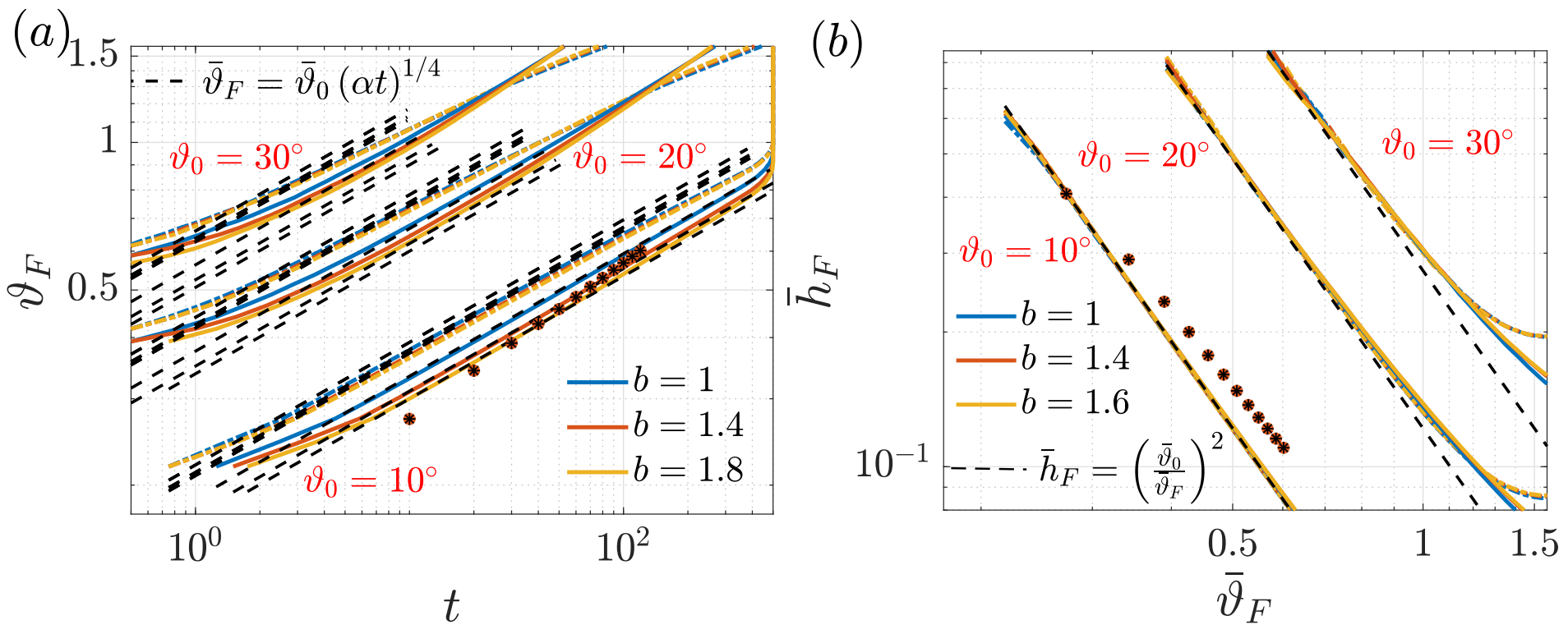

Figure 15: Spreading of an initial volume of fluid on an ellipsoid. $(a)$ Variation of the average front angle $\bar{\vartheta}_F$ with time and $(b)$ of the average thickness at the front $\bar{h}_F$ with $\bar{\vartheta}_F$, for $a = 0.6$ (dash-dotted lines), $a = 1$ (solid lines) and different values of the initial angle $\vartheta_0$ and $b$. The black dashed lines correspond to the analytical approximation of the relation $\bar{\vartheta}_F(t)$ and $\bar{h}_F(\vartheta_F)$, while the stars are the values recovered by a numerical simulation of the complete model with $a = 1$, $b = 1.4$, $Bo = 1000$, $\delta = 10^{-2}$, precursor film $h_{pr} = 0.02$ and $\vartheta_0 = 10°$.

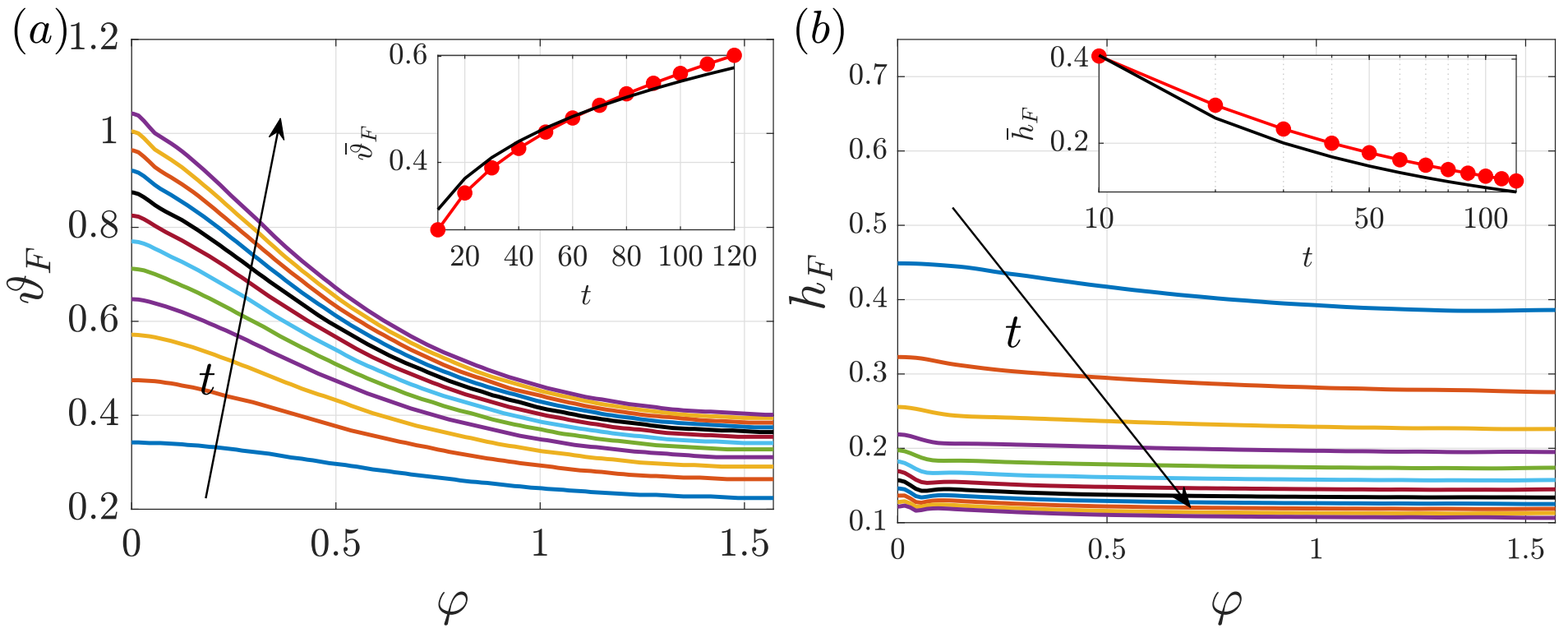

Figure 16: Maximum thickness $(a)$ position and $(b)$ value recovered from the numerical spreading simulation with $a = 1$, $b = 1.4$, $Bo = 1000$, $\delta = 10^{-2}$, precursor film $h_{pr} = 0.02$ and $\vartheta_0 = 10°$. Different colours denote different times $10 \leqslant t \leqslant 120$, with step size $\Delta t = 10$. In the insets, we report a comparison between theoretical (black dashed line) and numerical (red dots) $(a)$ average front position and $(b)$ average front thickness.

In this section, we described the drainage and spreading solution for the coating on an ellipsoid. We obtained an analytical solution that well compares with the numerical one. We showed the potential of general coordinates and asymptotic expansions to obtain a two-dimensional analytical solution suitable for a physical interpretation of the drainage and spreading process, in complement to the previous results for axisymmetric geometries.

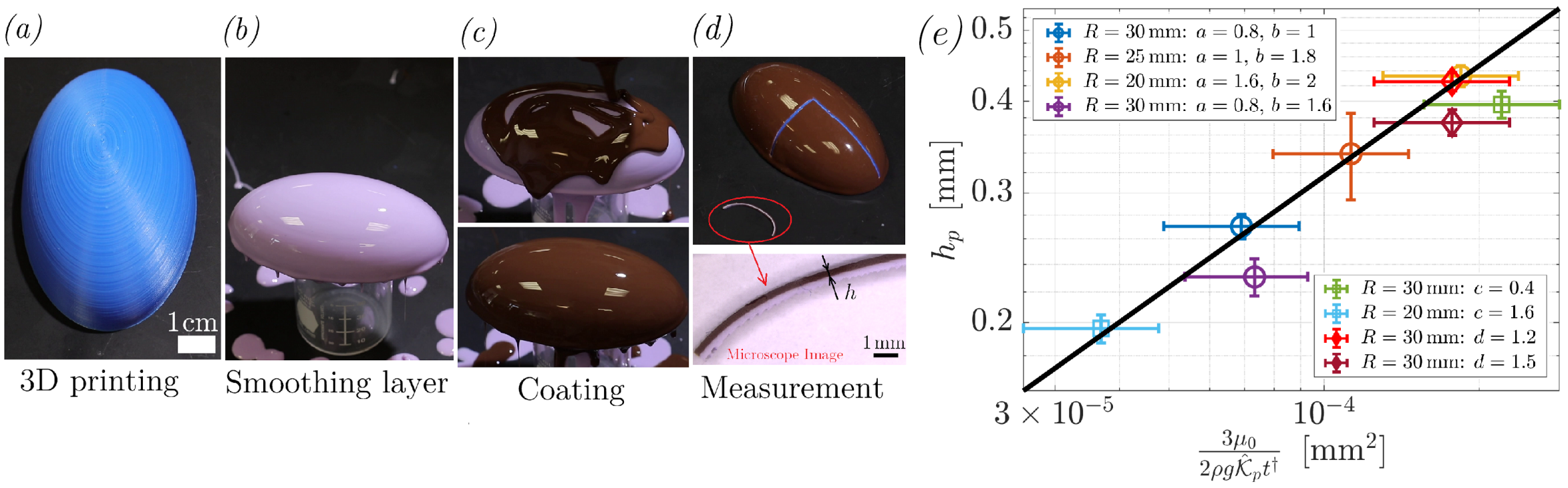

Figure 17: Different steps of the experimental procedure. ($a$) 3D printing of the molds. ($b$) Smoothing of the mold via a first layer of polymer. ($c$) Coating and curing of the second layer. ($d$) Peeling of a thin stripe, whose thickness is measured through a microscope. The thickness of the second layer is compared with the analytical and numerical solutions. ($e$) Pole thickness, measurements for ellipsoids (circles), spheroids (squares) and tori (diamonds). The black line denotes the theoretical prediction.

A large-time solution characterized by the separation of temporal and the two spatial dependencies was obtained. Modulations of the drainage solution and spreading front were explained in terms of the different slopes along the principal semiaxes, which induce an accumulation of fluid along the minor axis.

## 6. An experimental comparison

Our work focuses on developing analytical and numerical treatments of gravity-driven coatings on curved substrates. In this section, we compare our predictions to experiments. Rather than using Newtonian fluids, we use curable elastomers which drain until they solidify (Lee *et al.* 2016; Jones *et al.* 2021). As shown in Lee *et al.* (2016), this allows to easily measure the final film thickness distribution by peeling off the solidified layer. Moreover, because of the large amount of fluid poured on the surface (20 g, leading to an initial thickness of $\approx$ 2 mm) and since the time required for the elastomer to solidify ($\sim$ 10 min) is much longer than the characteristic drainage time ($\tau \sim 10$ s), the solidified film thickness becomes insensitive to initial condition and does not depend on the pouring condition (see Lee *et al.* (2016)). The experimental film thickness is compared to the late time drainage solution, here modified to account for the change of viscosity of the elastomer melt over time (Lee *et al.* 2016; Jones *et al.* 2021). The experimental procedure is shown in figure 17 (and Movie 1). We start by 3D printing a mold with the desired geometry (Anycubic i3 Mega). The resulting surface is rough, with vertical steps of the order of the printer layer height: 0.1mm (figure 17($a$)). We smooth the surface by applying a first coating using a rapidly curing elastomer (Zhermack VPS-16, see Jones *et al.* 2021 for more details on the elastomer mixing procedure). This first layer is sufficiently thin compared to the substrate characteristic size ($\hat{h}/R \sim 10^{-3}$) so that we assume that the substrate curvature remains unchanged after coating. After solidification of the first layer (figure 17($b$)), we proceed to the experiment and coat the sample with a second layer of elastomer (Zhermack VPS-32, figure 17($c$)). After solidification of the second layer, thin strips of the solid shell (containing both layers) are cut, peeled from the substrate and imaged with a microscope (figure 17($d$)). Dyes are mixed to both elastomers to enhance contrast thereby allowing us to automatically extract the second layer thickness as a function of the arc-length $\hat{h}(\hat{s})$. The errors introduced through the cutting procedure and subsequent image analysis are smoothed by binning the thickness over 50 pixels in the horizontal direction. The standard deviation within each bin defines the experimental uncertainty. Finally, we map the dimensionless arc-length $s$ back to the zenith angle $\vartheta$ with the relation:

$$s(\vartheta) = \int_0^{\vartheta} \sqrt{a^2 \sin^2(\vartheta') \cos^2 \varphi + b^2 \sin^2(\vartheta') \sin^2 \varphi + c^2 \cos^2 \vartheta'} \mathrm{d}\vartheta'. \tag{6.1}$$

| Polymer | $\mu_0$( Pa.s ) | $\alpha$ | $\beta\left(\times 10^{-3}\right)$ | $\tau_c$(s) |
|---|---|---|---|---|
| VPS-32 | $7.1 \pm 0.2$ | $5.3 \pm 0.7$ | $2.06 \pm 0.09$ | $574 \pm 11$ |

Table 1: Properties of VPS-32, extracted from Lee *et al.* (2016).

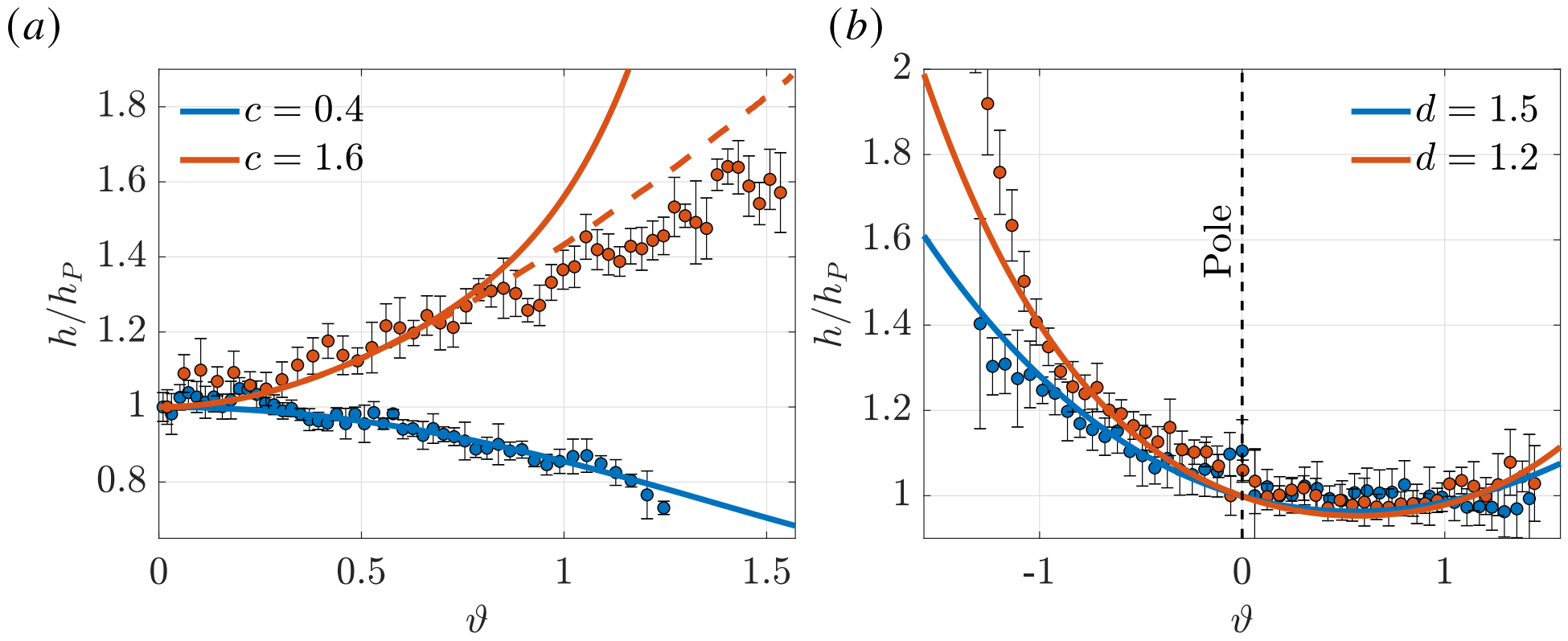

Figure 18: Comparison between experimental measurements of $h/h_p$ (colored dots) and the theoretical prediction from $(a)$ Section 3.1 for two spheroids with $c = 0.4$ and $R = 30$ mm, $c = 1.6$ and $R = 20$ mm, and $(b)$ Section 4 for two tori with $R = 30$ mm, and $d = 1.2, 1.5$. The colored solid thick lines denote the analytical solutions, while the dashed ones the numerical solutions.

In all cases considered, the Bond number is in between $177 < Bo = R^2/\ell_c^2 < 400$, where $\ell_c \approx 1.5$ mm is the capillary length of the polymer, while the final thickness is of order $10^{-1}$ mm, leading to $\delta \sim 10^{-2} - 10^{-3}$. These values of $Bo$ and $\delta$ ensure the accuracy of the drainage solution everywhere except close to the edge of the mold where capillary effects play a central role by creating a rim or bead. We exclude from our results this rim, intrinsically induced by capillarity. Following the results of the asymptotic expansion for $\vartheta \ll 1$, the dimensional large-time thickness can be written as $\hat{h} = h_p f(\text{geometry})$, where $f$ embeds the spatial distribution and depends only on the geometry, and $h_p$ is the thickness at the pole which depends on the rheology of the polymer melt during the drainage. For a Newtonian fluid the pole thickness is given by (2.11), or in dimensional units $h_p = \sqrt{3\mu/2\rho g \hat{\mathcal{K}}_p t}$ with $\hat{\mathcal{K}}_p$ the dimensional pole curvature. For a solidifying elastomer, we must account for the change of viscosity of the melt during curing $\mu(t)$ and the pole thickness is given by

$$h_p = \sqrt{\frac{3}{2\rho g \hat{\mathcal{K}}_p \int_{\tau_w}^{\infty} \frac{1}{\mu(t)} \mathrm{d}t}}, \tag{6.2}$$

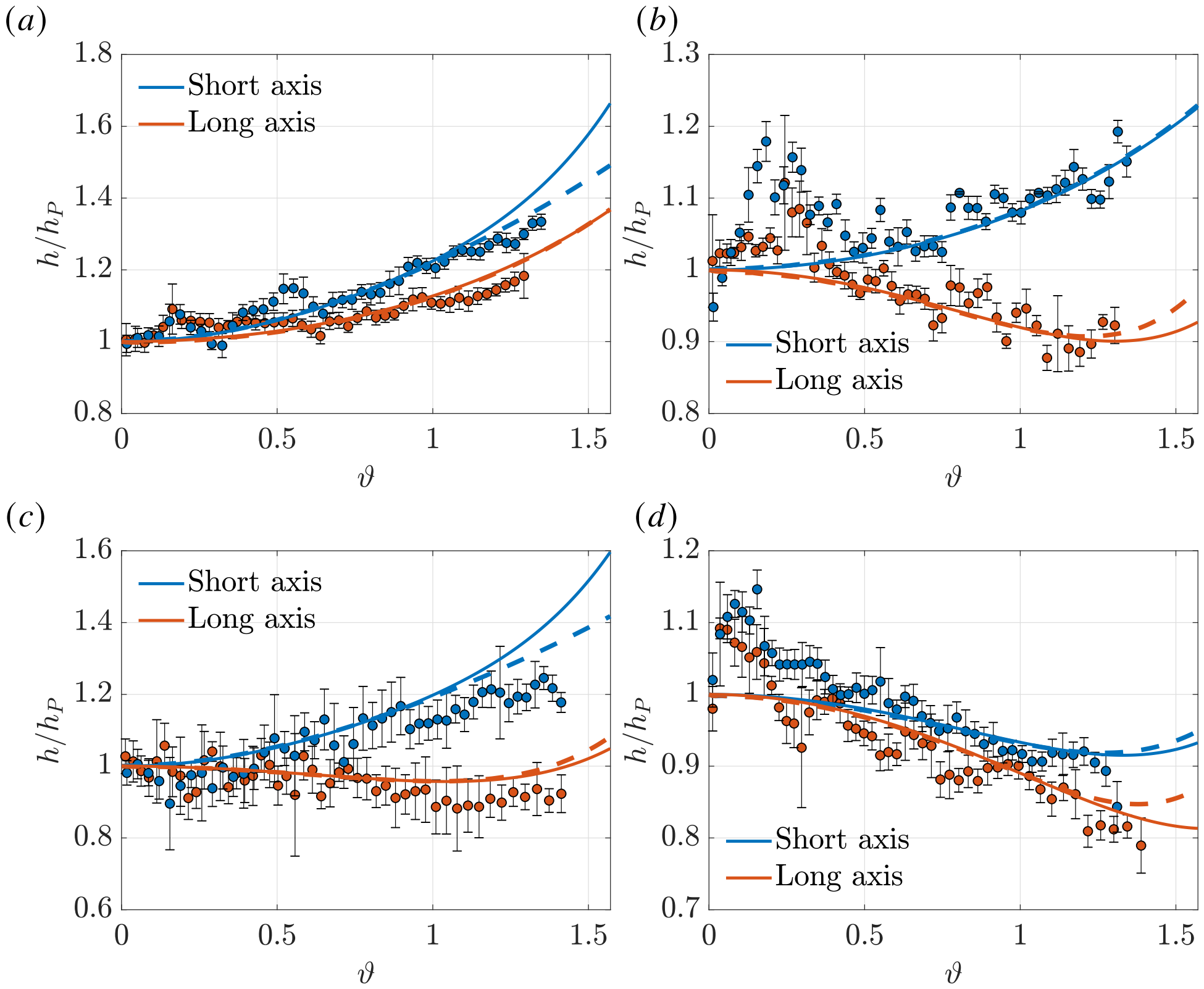

Figure 19: Comparison between analytical (solid lines), numerical (dashed lines) solutions and experimental measurements, in analogy with figure 19, for ellipsoids with ($a$) $a = 0.8$, $b = 1$ and $R = 30$ mm, ($b$) $a = 1$, $b = 1.8$ and $R = 25$ mm, ($c$) $a = 0.8$, $b = 1.6$ and $R = 30$ mm, ($d$) $a = 1.6$, $b = 2$ and $R = 20$ mm.

where $\tau_w$ is the time after mixing at which we start the drainage, $\tau_w \approx 6$ min in our experiments.The rheology of VPS-32 is reported in Lee *et al.* (2016):

$$\mu(t) = \begin{cases} \mu_0 \exp(\beta t), & \text{if } t \leqslant \tau_c, \\ \mu_1 t^{\alpha}, & \text{if } t > \tau_c, \end{cases} \tag{6.3}$$

with $\mu_1 = \mu_0 \exp\left(\beta \tau_{\rm c}\right) \tau_{\rm c}^{-\alpha}$. Upon integration, the pole thickness reads: $h_p = \sqrt{3\mu_0/(2\rho g \hat{\mathcal{K}}_p t^\dagger)}$, where $t^\dagger = \left\{\left(e^{-\beta\tau_w} - e^{-\beta\tau_c}\right)/\beta\right\} + \left\{\tau_c e^{-\beta\tau_c}/(\alpha-1)\right\}$. The values of the parameters, together with the uncertainties, are reported in table 1. In Figure 17($e$), the theoretical prediction is compared with the experimental measurements for different substrates, showing an overall good agreement, valid for all substrates as highlighted in Section 2.3.

In the following, we rescale the measured thickness with the pole thickness so as to compare the spatial distributions, independently of the fluid rheology $\hat{h}/h_p = f(\text{geometry})$. Figure 18 shows experimental measurements (dots) for two spheroids ($a$) and two tori ($b$) compared to the numerical (dashed line) and analytical (solid line) solutions at order $O(\vartheta^6)$ for the spheroids, and $O(\vartheta^4)$ for the tori. In all cases, the trend of analytical, numerical and experimental results are similar. For the spheroids, the thickness decreases when moving from the apex to the equator for $c = 0.4$. Instead, for $c = 1.6$ the thickness is found to increase. For the latter case, a favorable agreement for large $\vartheta$ is obtained with the numerical solution (dashed line), in agreement with previous discussions. Similar favorable agreements are obtained for tori. In particular, the analytical solution captures the increase in thicknesses observed for $\vartheta < 0$, *i.e.* in the inner part of the torus. In figure 19, we show experimental measurements (dots) for the thickness along the long ($\varphi = \pi/2$) and short ($\varphi = 0$) axis of ellipsoids with various aspect ratios and compare it to the numerical (dashed lines) and analytical (solid lines) solutions. We recover the three thickness distributions predicted, i.e. ($a$) thickness increasing both on the short and long axis for $a, b \lessapprox 1.35$, ($b, c$) thickness increasing along the short axis and decreasing on the long axis for $a \lessapprox 1.35 \lessapprox b$, ($d$) thickness decreasing along both axes $a, b \gtrapprox 1.35$. In all cases, the experimental measurements agree reasonably well with the numerical solutions.

## 7. Conclusion

This work studied the coating problem on a generic substrate with a focus on three-dimensional drainage and spreading. We analyzed different substrate geometries and derived analytical solutions for the drainage and spreading of an initial volume of fluid, under the assumption of very large Bond number and very thin film compared to the substrate characteristic length. We derived a general solution for the thickness evolution on a local maximum of the substrate. The thickness was found to be inversely proportional to the square root of the mean curvature at the pole, i.e. $h = \sqrt{3/(2\mathcal{K}_p t)}$. The latter represents how the components of the gravitational field, tangential to the substrate, vary across the surface in the vicinity of the pole. Therefore, a larger drainage and a faster decrease of the thickness are obtained with increasing mean curvature. We then investigated the role of the substrate geometry in modifying the thickness distribution away from the local maximum. We considered as a test-case the coating on a spheroid of dimensionless height $c$, whose solution was derived through an asymptotic expansion in the vicinity of the pole. For $c > c^*$, the thickness decreases as we move away from the pole while it increases for $c > c^*$. These thickness variations result from a competition between the slope and curvature which balance each other for $c = c^* = \sqrt{2/3}$. In particular, the fluid tends to accumulate in regions of lower curvature while a slope increase induces film thickening. The drainage solution was then employed to study the spreading of an initial volume of fluid contained in a region close to the pole. The spreading velocity was found to increase with the spheroid height. We related this behavior to the increase of the drainage gravity component with $c$ in the vicinity of the pole. We then studied the coating of a substrate in which the symmetry of the spreading is broken, i.e. the torus. The coating solution presented much larger values of the thickness on the inner part than on the outer part. The inner part presents a decreasing mean curvature, thus leading to larger values of the thickness than those observed on the outer part, where the mean curvature slightly increases since gravity is symmetric. The spreading of an initial volume of fluid occurred much faster on the inner region than on the outer region since the area to be invaded is smaller on the inner region, giving rise to two different spreading fronts. We concluded the analysis by applying the method to the three-dimensional spreading problem on a non-axisymmetric ellipsoidal substrate, i.e. with three different axes. We first derived a large-time analytical drainage solution which well agrees with the numerical simulations. Depending on the ellipsoid geometry, the thickness can increase or decrease away from the pole, with a behavior similar to the spheroid one along the principal axes. The solution was characterized by $\pi$-periodic modulations along the azimuthal direction, related to the different drainage along the two principal axes of the ellipsoid, which tend to move fluid from the major axis to the minor one. These modulations reflect in a spreading which does not occur uniformly along the azimuthal direction, but shows an accumulation of fluid and a faster spreading along the shorter axis. These modulations in the front position occur prior to any fingering instability. We obtained a scaling for the average front which fairly agrees with numerical results. We finally compared the spreading results with experimental measurements and found a good agreement in terms of spatial distributions.

The scope of the present work is to give a coherent and formal framework for the study of the drainage and coating on generic substrates based on the generalization and targeted application of previous analytical developments. These analyses show a crucial effect of the substrate curvature in defining the leading order thickness distribution and the spreading front of a gravity-driven coating. The natural extension of this work is the focus on the destabilization of these spreading fronts. While previous works focused on the fingering instability of two-dimensional fronts (Troian *et al.* 1989*a*; Bertozzi & Brenner 1997; Balestra *et al.* 2019), similar studies in which the primary front can bend and evolve together with fingering instabilities still need to be pursued. These analyses are not necessarily constrained by the considered configuration, but can be also extended to converging flows and more complex substrates. Besides, the performed analyses are valid in the absence of capillary and hydrostatic pressure effects. This assumption is respected when the film is very thin and the substrate does not present regions with infinite or zero curvature. If one of these hypotheses is violated, then other effects may play a crucial role. Hydrostatic pressure gradients become non negligible if the film becomes thicker or the substrate presents flat regions, e.g. a saddle point. Following the work of Lister (1992) for a flat substrate, the role of hydrostatic pressure gradients in these situations still needs to be investigated. These findings may find several applications both in environmental studies and thin film technologies. The interweaving between differential geometry and asymptotic theory showed great potential in the evaluation of analytical and numerical solutions for the coating on complex geometries, which may find further developments not only in the study of contact line instabilities, but in several coating flow phenomena such as Marangoni, inertia-driven and Rayleigh-Taylor instabilities.

**Acknowledgements** P.G.L. is grateful to A. Bongarzone for the precious suggestions regarding the ellipsoid analytical solution.

**Funding**. This work was supported by the Swiss National Science Foundation (grant no. 200021_178971 to P.G.L.).

**Declaration of Interests.** The authors report no conflict of interest.

## Appendix A. Spheroid: higher order drainage problems

In this section, the higher order drainage problems are described. We report only the ODE to be solved since their expressions are cumbersome. The ODE at $O(\vartheta^4)$ reads:

$$H_0^2((\frac{11c}{3} - 5c^3)H_2 + 6cH_4) + \frac{1}{36}c(48c^4 - 66c^2 + 19)H_0^3 + 6cH_0H_2^2 + H_4' = 0, \quad H_4(0) = 0. \tag{A 1}$$

At $O(\vartheta^6)$, the problem reads:

$$-\frac{1}{3}cH_0H_2((21c^2-16)H_2-48H_4)+H_0^2(-7c^3H_4+(6c^5-\frac{17c^3}{2}+\frac{13c}{5})H_2+\frac{16}{3}cH_4+8cH_6) \tag{A 2}$$

$$+(-\frac{5c^7}{3}+\frac{31c^5}{9}-\frac{257c^3}{120}+\frac{49c}{135})H_0^3+\frac{8}{3}cH_2^3+H_6'=0, \quad H_6(0)=0. \tag{A 3}$$

## Appendix B. Torus: drainage solution

Also in this section, we report only the problems when their relative solution is cumbersome. The problems for increasing orders read:

$$H_0{}'(t)=-\frac{1}{3}H_0(t)^3, \quad H_0(0)=1 \rightarrow H_0(t)=\frac{1}{\sqrt{\frac{2t}{3}+1}}, \tag{B 1}$$

$$H_1'(t)=-\left(\frac{H_0(t)^3}{3d}+2H_0(t)^2H_1(t)\right), \quad H_1(0)=0,$$
$$\rightarrow H_1(t)=\frac{-8\sqrt{3}t^3-36\sqrt{3}t^2-54\sqrt{3}t+27\sqrt{2t+3}-27\sqrt{3}}{5d(2t+3)^{7/2}}, \tag{B 2}$$

$$H_2'(t)=-\left(-\frac{\left(d^2+2\right)H_0(t)^3}{6d^2}+\frac{H_0(t)^2(3dH_2(t)+H_1(t))}{d}+3H_0(t)H_1(t)^2\right), \quad H_2(0)=0,$$
$$\rightarrow H_2(t)=\frac{1}{50d^2(2t+3)^{11/2}}\{4\sqrt{3}\left(25d^2+62\right)t^5+30\sqrt{3}\left(25d^2+62\right)t^4+90\sqrt{3}\left(25d^2+62\right)t^3$$
$$+27t^2\left(125\sqrt{3}d^2+8\sqrt{2t+3}+310\sqrt{3}\right)+81t\left(25\sqrt{3}d^2+8\left(\sqrt{2t+3}+5\sqrt{3}\right)\right)$$
$$-486\left(\sqrt{3}-\sqrt{2t+3}\right)\}, \tag{B 3}$$

$$72d^3H_0(t)H_1H_2-6d^3H_0(t)^2H_1+36d^3H_0(t)^2H_3(t)+12d^3H_1^3+9d^3H_3'(t)+9d^2$$
$$H_0(t)H_1^2+9d^2H_0(t)^2H_2-2d^2H_0(t)^3-9dH_0(t)^2H_1+3H_0(t)^3=0, \quad H_3(0)=0, \tag{B 4}$$

$$720d^4H_0(t)H_1(t)H_3(t)-60d^4H_0(t)H_1(t)^2+360d^4H_0(t)H_2(t)^2-60d^4H_0(t)^2H_2(t)$$
$$+360d^4H_0(t)^2H_4(t)+d^4H_0(t)^3+360d^4H_1(t)^2H_2(t)+72d^4H_4'(t)+144d^3H_0(t)H_1(t)H_2(t)$$
$$-48d^3H_0(t)^2H_1(t)+72d^3H_0(t)^2H_3(t)+24d^3H_1(t)^3-72d^2H_0(t)H_1(t)^2-72d^2H_0(t)^2H_2(t)$$
$$+20d^2H_0(t)^3+72dH_0(t)^2H_1(t)-24H_0(t)^3=0, \quad H_4(0)=0, \tag{B 5}$$

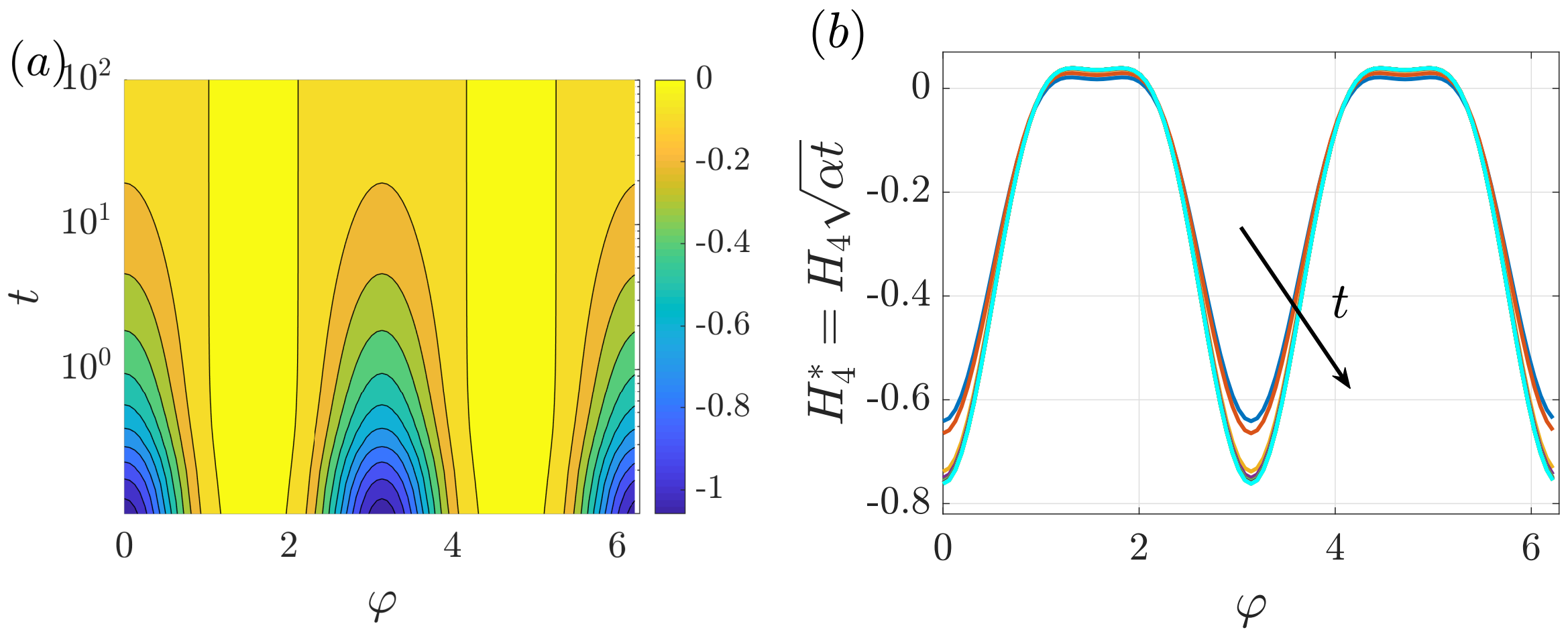

[!b]

Figure 20: Drainage along an ellipsoid with $a = 0.5$ and $b = 1.5$. ($a$) Spatiotemporal evolution of $H_4$: iso-contours of $H_2$ in the $(\varphi, t)$ plane. ($b$) Second order correction $H_4^* = H_4/H_0 \approx H_4\sqrt{\alpha t}$ as a function of $\varphi$ at different times: $t = 0.4$ (blue), $t = 1$ (orange), $t = 5$ (yellow), $t = 10$ (purple) $t = 30$ (green), $t = 50$ (cyan), $t = 70$ (maroon), $t = 90$ (black), $t = 100$ (red).

## Appendix C. Ellipsoid: higher order drainage solutions

The PDE for $H_4(\varphi, t)$ reads:

$$
\begin{aligned}
&\frac{7\cos(2\varphi)H_0^3}{36a^2} + \frac{\cos(2\varphi)H_0^3}{2a^6} + \frac{\cos(2\varphi)H_0^3}{6a^4b^2} + \frac{13\cos(2\varphi)H_0^3}{18b^4} + \frac{\cos(4\varphi)H_0^3}{8a^6} + \frac{\cos(4\varphi)H_0^3}{6a^2b^2} \\
&+ \frac{\cos(4\varphi)H_0^3}{8b^6} + \frac{19H_0^3}{72a^2} - \frac{\cos(4\varphi)H_0^3}{12a^4} - \frac{13\cos(2\varphi)H_0^3}{18a^4} - \frac{23H_0^3}{36a^4} + \frac{3H_0^3}{8a^6} - \frac{7\cos(2\varphi)H_0^3}{36b^2} + \frac{19H_0^3}{72b^2} \\
&- \frac{5H_0^3}{9a^2b^2} - \frac{\cos(4\varphi)H_0^3}{8a^4b^2} + \frac{7H_0^3}{24a^4b^2} - \frac{\cos(4\varphi)H_0^3}{12b^4} - \frac{23H_0^3}{36b^4} - \frac{\cos(2\varphi)H_0^3}{6a^2b^4} - \frac{\cos(4\varphi)H_0^3}{8a^2b^4} \\
&+ \frac{7H_0^3}{24a^2b^4} - \frac{\cos(2\varphi)H_0^3}{2b^6} + \frac{3H_0^3}{8b^6} + \frac{4\cos(2\varphi)H_2H_0^2}{3a^2} + \frac{2\cos(2\varphi)H_2H_0^2}{b^4} + \frac{\cos(4\varphi)H_2H_0^2}{2a^2b^2} + \frac{11H_2H_0^2}{6a^2} \\
&+ \frac{11H_2H_0^2}{6b^2} + \frac{2\cos(2\varphi)H_4H_0^2}{a^2} + \frac{3H_4H_0^2}{a^2} + \frac{3H_4H_0^2}{b^2} + \frac{\cos(2\varphi)\sin(2\varphi)\frac{\partial H_2}{\partial\varphi}H_0^2}{4a^4} \\
&+ \frac{\cos(2\varphi)\sin(2\varphi)\frac{\partial H_2}{\partial\varphi}H_0^2}{4b^4} + \frac{\sin(2\varphi)\frac{\partial H_2}{\partial\varphi}H_0^2}{4a^4} + \frac{\sin(2\varphi)\frac{\partial H_2}{\partial\varphi}H_0^2}{4b^2} + \frac{\sin(2\varphi)\frac{\partial H_4}{\partial\varphi}H_0^2}{2b^2} - \frac{\sin(2\varphi)\frac{\partial H_4}{\partial\varphi}H_0^2}{2a^2} \\
&- \frac{\sin(2\varphi)\frac{\partial H_2}{\partial\varphi}H_0^2}{4a^2} - \frac{2\cos(2\varphi)H_2H_0^2}{a^4} - \frac{\cos(4\varphi)H_2H_0^2}{4a^4} - \frac{7H_2H_0^2}{4a^4} - \frac{2\cos(2\varphi)H_4H_0^2}{b^2} - \frac{4\cos(2\varphi)H_2H_0^2}{3b^2} \\
&- \frac{3H_2H_0^2}{2a^2b^2} - \frac{\cos(2\varphi)\sin(2\varphi)\frac{\partial H_2}{\partial\varphi}H_0^2}{2a^2b^2} - \frac{\cos(4\varphi)H_2H_0^2}{4b^4} - \frac{7H_2H_0^2}{4b^4} - \frac{\sin(2\varphi)\frac{\partial H_2}{\partial\varphi}H_0^2}{4b^4} + \frac{2\cos(2\varphi)H_2^2H_0}{a^2} \\
&+ \frac{3H_2^2H_0}{a^2} + \frac{3H_2^2H_0}{b^2} + \frac{H_2\sin(2\varphi)\frac{\partial H_2}{\partial\varphi}H_0}{b^2} - \frac{H_2\sin(2\varphi)\frac{\partial H_2}{\partial\varphi}H_0}{a^2} - \frac{2\cos(2\varphi)H_2^2H_0}{b^2} + \frac{\partial H_4}{\partial t} = 0 \qquad \text{(C 1)}
\end{aligned}
$$

whose numerical solution is reported in figures 20 and 21. Figure 20($b$) shows that the values of the rescaled fourth order solution $H_4^* = H_4\sqrt{\alpha t}$ collapse to the same curve as time increase, thus suggesting that also in this case a large-time separation of variables is possible.

Because of its size, we do not write the PDE for $H_6$ here; however, the solutions are shown in figures 22 and 23. In analogy with the solutions at order $O(\vartheta^2)$ and $O(\vartheta^4)$, the behavior of the large-time solution suggests that a solution $H_n^*(\varphi) = H_n(\varphi, t)/H_0(t)$ satisfies the problem.

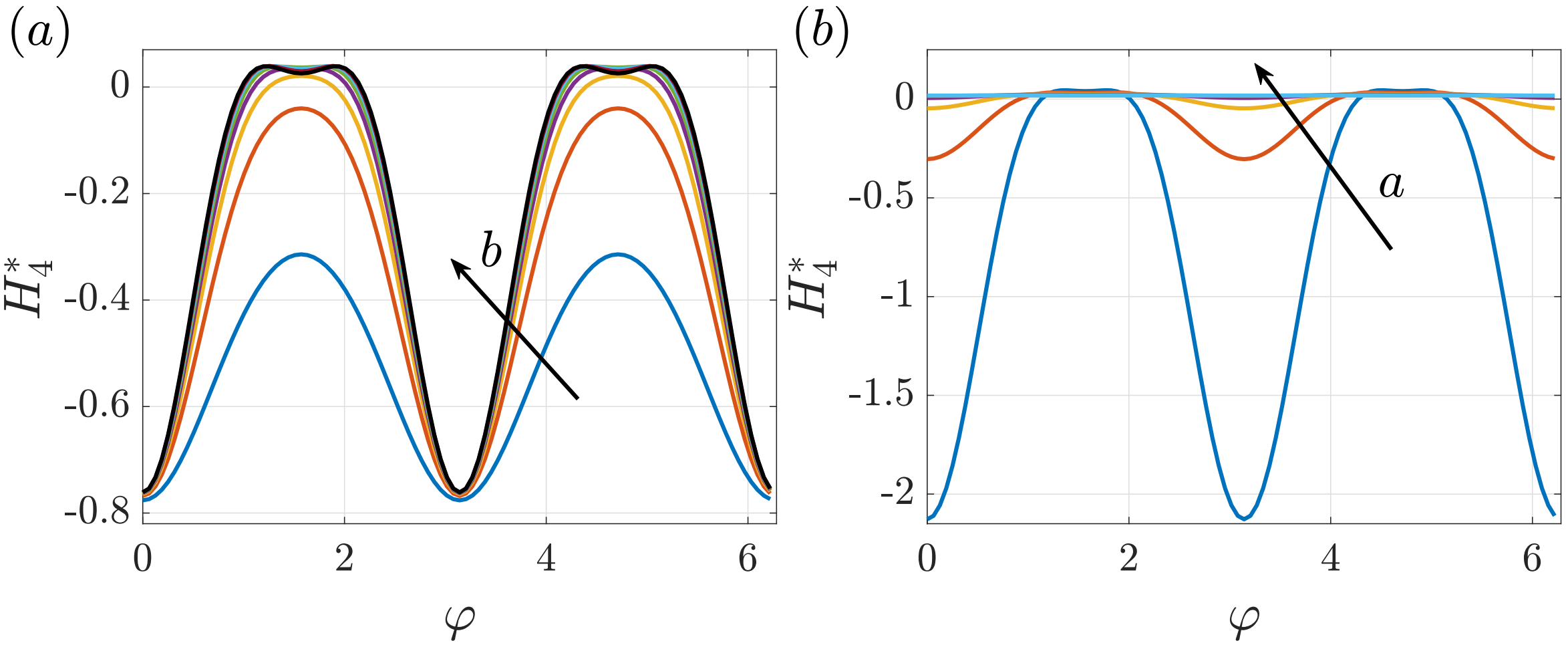

Figure 21: ($a$) Fourth order correction $H_4^* = H_4/H_0 \approx H_4\sqrt{\alpha t}$ as a function of $\varphi$ at $t = 100$, for $a = 0.5$ and increasing $b$: $b = 0.6$ (blue), $b = 0.8$ (orange), $b = 1$ (yellow), $b = 1.2$ (purple), $b = 1.4$ (green), $b = 1.6$ (cyan), $b = 1.8$ (maroon), $b = 2$ (black). ($b$) Second order correction $H_4^*$ as a function of $\varphi$ at $t = 100$, for $b = 1.5$ and increasing $a$: $a = 0.4$ (blue), $a = 0.6$ (orange), $a = 0.8$ (yellow), $a = 1$ (purple), $a = 1.2$ (green), $a = 1.4$ (cyan).

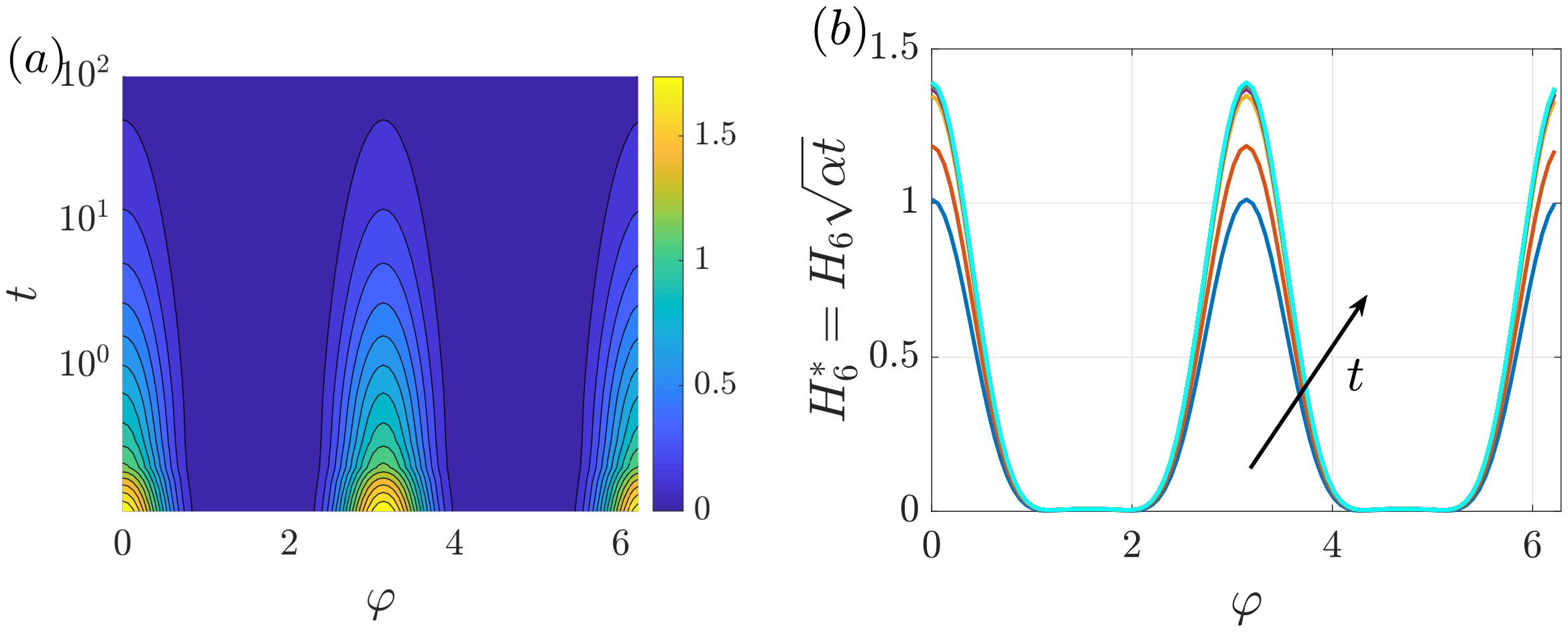

Figure 22: Drainage along an ellipsoid with $a = 0.5$ and $b = 1.5$. ($a$) Spatiotemporal evolution of $H_6$: iso-contours of $H_6$ in the $(\varphi, t)$ plane. ($b$) Second order correction $H_6^* = H_6/H_0 \approx H_6\sqrt{\alpha t}$ as a function of $\varphi$ at different times: $t = 0.4$ (blue), $t = 1$ (orange), $t = 5$ (yellow), $t = 10$ (purple) $t = 30$ (green), $t = 50$ (cyan), $t = 70$ (maroon), $t = 90$ (black), $t = 100$ (red).

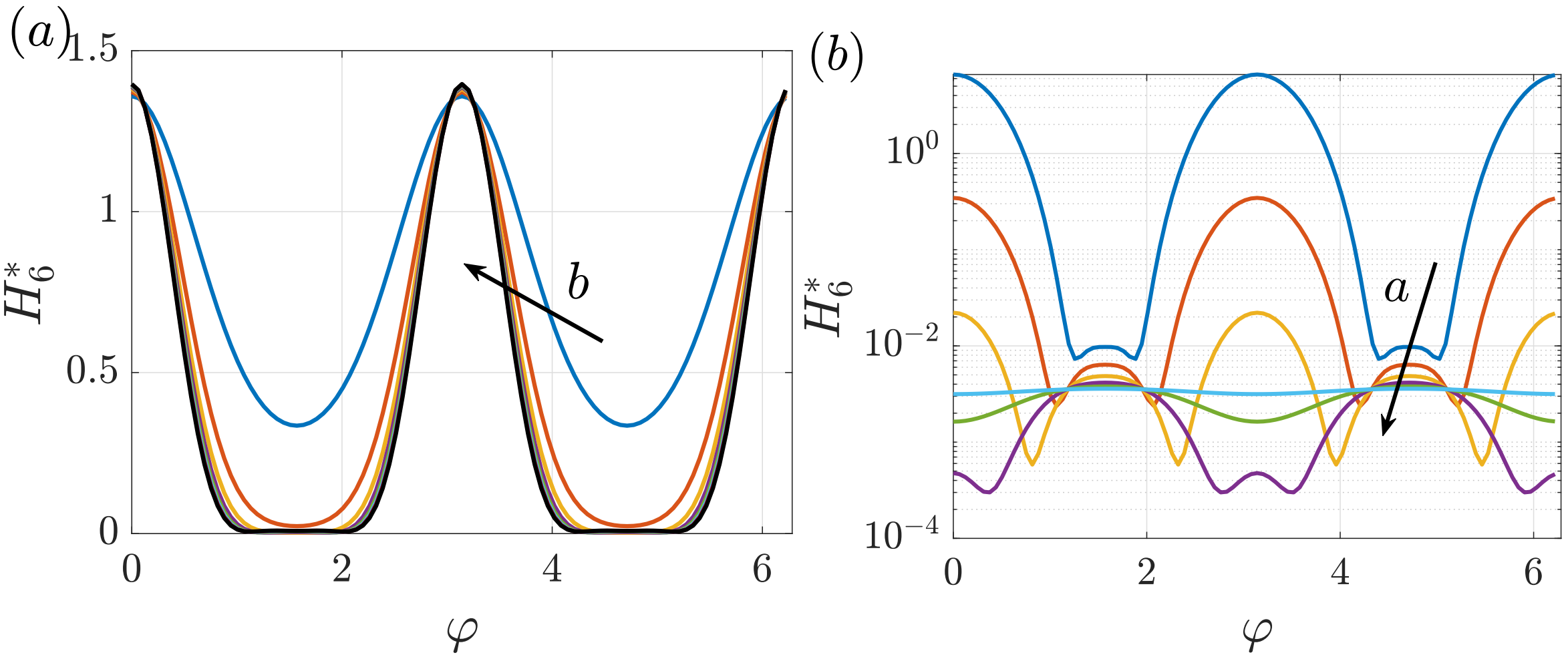

Figure 23: ($a$) Sixth order correction $H_6^* = H_6/H_0 \approx H_6\sqrt{\alpha t}$ as a function of $\varphi$ at $t = 100$, for $a = 0.5$ and increasing $b$: $b = 0.6$ (blue), $b = 0.8$ (orange), $b = 1$ (yellow), $b = 1.2$ (purple), $b = 1.4$ (green), $b = 1.6$ (cyan), $b = 1.8$ (maroon), $b = 2$ (black). ($b$) Second order correction $H_6^*$ as a function of $\varphi$ at $t = 100$, for $b = 1.5$ and increasing $a$: $a = 0.4$ (blue), $a = 0.6$ (orange), $a = 0.8$ (yellow), $a = 1$ (purple), $a = 1.2$ (green), $a = 1.4$ (cyan).

# Supporting information

## Appendix A. Parametrization, metric and curvature tensor for different substrates

### A.1. *Monge parameterization of a generic substrate*

In the *Monge parameterization*, the substrate height is described through the function

$$\boldsymbol{X} = (x, y, h^0(x, y)), \tag{S1}$$

i.e. the parameters that describe the surface are the $x$ and $y$ directions themselves, $(x^{\{1\}} = x, x^{\{2\}} = y)$. This form is convenient when the substrate thickness is known as a function of the global reference frame, and fails when the substrate cannot be represented as a function of $(x, y)$, e.g. in the case of vertical tangent.

The derivatives of the height function $h^0(x, y)$ read:

$$\begin{gathered} h^0_{(1,0)} = \partial_x h^0, \quad h^0_{(0,1)} = \partial_y h^0 \\ h^0_{(2,0)} = \partial_x \partial_x h^0, \quad h^0_{(1,1)} = \partial_x \partial_y h^0, \quad h^0_{(0,2)} = \partial_y \partial_y h^0 \end{gathered} \tag{S2}$$

The tangent and normal vectors to the substrate are:

$$\begin{gathered} \mathbf{e}_1 = \partial_1 \boldsymbol{X} = \left[1, 0, h^0_{(1,0)}\right], \quad \mathbf{e}_2 = \partial_2 \boldsymbol{X} = \left[0, 1, h^0_{(0,1)}\right], \\ \mathbf{e}_3 = \frac{\mathbf{e}_1 \times \mathbf{e}_2}{|\mathbf{e}_1 \times \mathbf{e}_2|} = \frac{\mathbf{e}_1 \times \mathbf{e}_2}{n} = \frac{1}{n}[-h^0_{(1,0)}, -h^0_{(0,1)}, 1] \end{gathered} \tag{S3}$$

The covariant vectors are found solving the linear system for the i-th covariant vector $\mathbf{e}^{\{i\}} \cdot \mathbf{e}_j = \delta_{ij},\ i, j = 1, 2, 3$. Note that $\mathbf{e}^{\{3\}} = \mathbf{e}_3$, since the normal-to-the-substrate vector is normalized by construction. The metric tensor and the inverse metric tensor therefore read:

$$\begin{gathered} \mathbb{G}_{ij} = \mathbf{e}_i \cdot \mathbf{e}_j = \begin{pmatrix} 1 + \left(h^0_{(1,0)}\right)^2 & h^0_{(1,0)} h^0_{(0,1)} \\ h^0_{(1,0)} h^0_{(0,1)} & 1 + \left(h^0_{(0,1)}\right)^2 \end{pmatrix}, \\ \mathbb{G}^{\{ij\}} = \mathbf{e}^{\{i\}} \cdot \mathbf{e}^{\{j\}} = \frac{1}{n^2} \begin{pmatrix} 1 + \left(h^0_{(0,1)}\right)^2 & -h^0_{(1,0)} h^0_{(0,1)} \\ -h^0_{(1,0)} h^0_{(0,1)} & 1 + \left(h^0_{(1,0)}\right)^2 \end{pmatrix}. \end{gathered} \tag{S4}$$

The curvature tensor reads:

$$\mathbb{K}_i^{\{j\}} = \frac{1}{n^3} \begin{pmatrix} \left(1 + \left(h^0_{(0,1)}\right)^2\right) h^0_{(2,0)} - h^0_{(1,0)} h^0_{(0,1)} h^0_{(1,1)} & \left(1 + \left(h^0_{(1,0)}\right)^2\right) h^0_{(1,1)} - h^0_{(1,0)} h^0_{(0,1)} h^0_{(2,0)} \\ \left(1 + \left(h^0_{(0,1)}\right)^2\right) h^0_{(1,1)} - h^0_{(1,0)} h^0_{(0,1)} h^0_{(0,2)} & \left(1 + \left(h^0_{(1,0)}\right)^2\right) h^0_{(0,2)} - h^0_{(1,0)} h^0_{(0,1)} h^0_{(1,1)} \end{pmatrix}, \tag{S5}$$

from which the mean and Gaussian curvature are obtained:

$$\begin{aligned} \mathcal{K} &= \frac{1}{n^3} \left( \left(1 + \left(h^0_{(0,1)}\right)^2\right) h^0_{(2,0)} - 2 h^0_{(1,0)} h^0_{(0,1)} h^0_{(1,1)} + \left(1 + \left(h^0_{(1,0)}\right)^2\right) h^0_{(0,2)} \right) \\ \mathcal{G} &= \frac{1}{n^4} \left( h^0_{(2,0)} h^0_{(0,2)} - \left(h^0_{(1,1)}\right)^2 \right) \end{aligned} \tag{S6}$$

The gravity vector, which reads $\boldsymbol{g} = [0, 0, -1]$ in the $(x, y, z)$ coordinate system, is projected on the contravariant base:

$$\begin{aligned} g_t^{\{1\}} &= \boldsymbol{g} \cdot \mathbf{e}^{\{1\}} = -h^0_{(1,0)}/n^2, \\ g_t^{\{2\}} &= \boldsymbol{g} \cdot \mathbf{e}^{\{2\}} = -h^0_{(0,1)}/n^2, \\ g_3 &= \boldsymbol{g} \cdot \mathbf{e}_3 = -1/n. \end{aligned} \tag{S7}$$

These equations close the lubrication model.

### A.2. *Monge parameterization of a surface of revolution*

A cylindrical reference frame $(r, \varphi, z)$ is introduced. In this case, the Monge parameterization assumes the form:

$$\boldsymbol{X} = (r\cos\varphi, r\sin\varphi, h^0(r)). \tag{S8}$$

The tangent and normal to the substrate vectors are

$$\begin{gathered} \mathbf{e}_1 = \mathbf{e}_r = \partial_r \boldsymbol{X} = [\cos\varphi, \sin\varphi, h^{0\prime}], \quad \mathbf{e}_2 = \mathbf{e}_\varphi = \partial_\varphi \boldsymbol{X} = [-r\sin\varphi, r\cos\varphi, 0], \\ \mathbf{e}_3 = [-h^{0\prime}\cos\varphi, -h^{0\prime}\sin\varphi, 1]. \end{gathered} \tag{S9}$$

The covariant vectors are found solving the linear system for the $i$th covariant vector $\mathbf{e}^{\{i\}} \cdot \mathbf{e}_j = \delta_{ij},\ i, j = 1, 2, 3$. As before, the metric tensor and the inverse metric tensor are evaluated:

$$\mathbb{G}_{ij} = \mathbf{e}_i \cdot \mathbf{e}_j = \begin{pmatrix} 1 + \left(h^{0\prime}\right)^2 & 0 \\ 0 & r^2 \end{pmatrix}, \quad \mathbb{G}^{\{ij\}} = \mathbf{e}^{\{i\}} \cdot \mathbf{e}^{\{j\}} = \begin{pmatrix} \frac{1}{1+\left(h^{0\prime}\right)^2} & 0 \\ 0 & 1/r^2 \end{pmatrix}. \tag{S10}$$

The curvature tensor thus reads:

$$\mathbb{K}_i^{\{j\}} = \frac{1}{\sqrt{1 + \left(h^{0\prime}\right)^2}} \begin{pmatrix} \frac{-h^{0\prime\prime}}{1+\left(h^{0\prime}\right)^2} & 0 \\ 0 & -h^{0\prime}/r \end{pmatrix}, \tag{S11}$$

from which the mean and Gaussian curvature are obtained:

$$\mathcal{K} = \mathrm{Tr}(\mathbb{K}), \quad \mathcal{G} = \mathrm{Det}(\mathbb{K}). \tag{S12}$$

Considering the gravity direction $\boldsymbol{g} = [0, 0, -1]$, the components along the coordinate vectors read:

$$\begin{aligned} g_t^{\{1\}} &= \boldsymbol{g} \cdot \mathbf{e}^{\{1\}} = -\frac{h^{0\prime}}{1 + \left(h^{0\prime}\right)^2}, \\ g_t^{\{2\}} &= \boldsymbol{g} \cdot \mathbf{e}^{\{2\}} = 0, \\ g_3 &= \boldsymbol{g} \cdot \mathbf{e}_3 = -\frac{1}{\sqrt{1 + \left(h^{0\prime}\right)^2}}. \end{aligned} \tag{S13}$$

A.3. *Parameterization of spheroids*

We parameterize the ellipsoidal surface via the zenith $x^{\{1\}} = \vartheta$ and the azimuth $x^{\{2\}} = \varphi$:

$$\boldsymbol{X}(\vartheta, \varphi) = (\sin\vartheta\cos\varphi, \sin\vartheta\sin\varphi, c\cos\vartheta) \tag{S14}$$

The tangent and normal vectors to the substrate are:

$$\mathbf{e}_1 = [\cos(\vartheta)\cos(\varphi), \cos(\vartheta)\sin(\varphi), -c\sin(\vartheta)], \quad \mathbf{e}_2 = [-\sin(\vartheta)\sin(\varphi), \sin(\vartheta)\cos(\varphi), 0],$$

$$\mathbf{e}_3 = \left[\frac{\sqrt{2}c\sin^2(\vartheta)\cos(\varphi)}{\sqrt{\sin^2(\vartheta)\left(-\left(\left(c^2-1\right)\cos(2\vartheta)-c^2-1\right)\right)}}, \frac{\sqrt{2}c\sin^2(\vartheta)\sin(\varphi)}{\sqrt{\sin^2(\vartheta)\left(-\left(\left(c^2-1\right)\cos(2\vartheta)-c^2-1\right)\right)}}, \frac{\sqrt{2}\sin(\vartheta)\cos(\vartheta)}{\sqrt{\sin^2(\vartheta)\left(-\left(\left(c^2-1\right)\cos(2\vartheta)-c^2-1\right)\right)}}\right] \tag{S15}$$

The metric tensor and the inverse metric tensor therefore read:

$$\begin{aligned} \mathbb{G}_{ij} = \mathbf{e}_i \cdot \mathbf{e}_j &= \begin{pmatrix} \frac{1}{2}\left(-\left(c^2-1\right)\cos(2\vartheta)+c^2+1\right) & 0 \\ 0 & \sin^2(\vartheta) \end{pmatrix}, \\ \mathbb{G}^{\{ij\}} = \mathbf{e}^{\{i\}} \cdot \mathbf{e}^{\{j\}} &= \begin{pmatrix} \frac{2}{-\left(c^2-1\right)\cos(2\vartheta)+c^2+1} & 0 \\ 0 & \csc^2(\vartheta) \end{pmatrix}. \end{aligned} \tag{S16}$$

The curvature tensor is:

$$\mathbb{K}_i^{\{j\}} = \begin{pmatrix} -\frac{2\sqrt{2}c\sin^3(\vartheta)}{\left(\sin^2(\vartheta)\left(-\left(c^2-1\right)\cos(2\vartheta)+c^2+1\right)\right)^{3/2}} & 0 \\ 0 & -\frac{\sqrt{2}c\sin(\vartheta)}{\sqrt{\sin^2(\vartheta)\left(-\left(\left(c^2-1\right)\cos(2\vartheta)-c^2-1\right)\right)}} \end{pmatrix}, \tag{S17}$$

from which the mean and Gaussian curvature are obtained:

$$\begin{aligned} \mathcal{K} &= \frac{\sqrt{2}c\sin^3(\vartheta)\left(\left(c^2-1\right)\cos(2\vartheta)-c^2-3\right)}{\left(\sin^2(\vartheta)\left(-\left(c^2-1\right)\cos(2\vartheta)+c^2+1\right)\right)^{3/2}}, \\ \mathcal{G} &= \frac{4c^2}{\left(-\left(c^2-1\right)\cos(2\vartheta)+c^2+1\right)^2}. \end{aligned} \tag{S18}$$

The gravity components in the contravariant base read:

$$\begin{aligned} g_t^{\{1\}} &= \boldsymbol{g} \cdot \mathbf{e}^{\{1\}} = \frac{c\sin(\vartheta)}{c^2\sin^2(\vartheta)+\cos^2(\vartheta)}, \\ g_t^{\{2\}} &= \boldsymbol{g} \cdot \mathbf{e}^{\{2\}} = 0, \\ g_3 &= \boldsymbol{g} \cdot \mathbf{e}_3 = -\frac{\sqrt{2}\sin(\vartheta)\cos(\vartheta)}{\sqrt{\sin^2(\vartheta)\left(-\left(\left(c^2-1\right)\cos(2\vartheta)-c^2-1\right)\right)}}. \end{aligned} \tag{S19}$$

Note that imposing $c = 1$ we recover the case of a sphere of unitary radius.

A.4. *Parameterization of tori*

We consider a torus of unitary tube radius and distance $d$ between the center of the tube and the axis of revolution. The parameterization is performed using the colatitude $x^{\{1\}} = \vartheta$ and the azimuth $x^{\{2\}} = \varphi$:

$$\boldsymbol{X}(\vartheta, \varphi) = ((d + \sin\vartheta)\cos\varphi, (d + \sin\vartheta)\sin\varphi, \cos\vartheta) \tag{S20}$$

The tangent and normal vectors to the substrate are:

$$\begin{aligned}
\mathbf{e}_1 &= [\cos(\vartheta)\cos(\varphi), \cos(\vartheta)\sin(\varphi), -\sin(\vartheta)],\\
\mathbf{e}_2 &= [-(d + \sin(\vartheta))\sin(\varphi), (d + \sin(\vartheta))\cos(\varphi), 0],\\
\mathbf{e}_3 &= \left[\frac{\sin(\vartheta)\cos(\varphi)(d + \sin(\vartheta))}{d + \sin(\vartheta)}, \frac{\sin(\vartheta)\sin(\varphi)(d + \sin(\vartheta))}{d + \sin(\vartheta)}, \frac{\cos(\vartheta)(d + \sin(\vartheta))}{d + \sin(\vartheta)}\right].
\end{aligned} \tag{S21}$$

The metric tensor and the inverse metric tensor read:

$$\begin{aligned}
\mathbb{G}_{ij} = \mathbf{e}_i \cdot \mathbf{e}_j &= \begin{pmatrix} 1 & 0 \\ 0 & (d + \sin(\vartheta))^2 \end{pmatrix},\\
\mathbb{G}^{\{ij\}} = \mathbf{e}^{\{i\}} \cdot \mathbf{e}^{\{j\}} &= \begin{pmatrix} 1 & 0 \\ 0 & \frac{1}{(d+\sin(\vartheta))^2} \end{pmatrix}.
\end{aligned} \tag{S22}$$

The curvature tensor is:

$$\mathbb{K}_i^{\{j\}} = \begin{pmatrix} -1 & 0 \\ 0 & -\frac{\sin(\vartheta)}{d+\sin(\vartheta))} \end{pmatrix}, \tag{S23}$$

from which the mean and Gaussian curvature are obtained:

$$\begin{aligned}
\mathcal{K} &= -\frac{d + 2\sin(\vartheta)}{d + \sin(\vartheta)},\\
\mathcal{G} &= \frac{\sin(\vartheta)}{d + \sin(\vartheta)}.
\end{aligned} \tag{S24}$$

The gravity components read:

$$\begin{aligned}
g_t^{\{1\}} &= \boldsymbol{g} \cdot \mathbf{e}^1 = \sin(\vartheta),\\
g_t^{\{2\}} &= \boldsymbol{g} \cdot \mathbf{e}^{\{2\}} = 0,\\
g_3 = \boldsymbol{g} \cdot \mathbf{e}_3 &= -\frac{\cos(\vartheta)(d + \sin(\vartheta))}{d + \sin(\vartheta)}.
\end{aligned} \tag{S25}$$

A.5. *Parameterization of ellipsoids*

We consider an ellipsoid of axes $aR$, $bR$ and $R$, with the last one aligned along the vertical direction. The non-dimensionalization is performed by assuming as characteristic length the vertical axis $R$. The substrate is identified by:

$$\mathbf{X} = [a\sin(\vartheta)\cos(\varphi), b\sin(\vartheta)\sin(\varphi), \cos(\vartheta)] \tag{S26}$$

The metric tensor reads:

$$\mathbb{G}_{ij} = \begin{pmatrix} \cos^2(\vartheta)\left(a^2\cos^2(\varphi) + b^2\sin^2(\varphi)\right) + \sin^2(\vartheta) & \left(b^2 - a^2\right)\sin(\vartheta)\cos(\vartheta)\sin(\varphi)\cos(\varphi) \\ \left(b^2 - a^2\right)\sin(\vartheta)\cos(\vartheta)\sin(\varphi)\cos(\varphi) & \sin^2(\vartheta)\left(a^2\sin^2(\varphi) + b^2\cos^2(\varphi)\right) \end{pmatrix}, \tag{S27}$$

while the components of the inverse metric tensor are:

$$\mathbb{G}^{\{11\}} = \frac{a^2 \sin^2(\varphi) + b^2 \cos^2(\varphi)}{\sin^2(\vartheta)\left(a^2 \sin^2(\varphi) + b^2 \cos^2(\varphi)\right) + a^2 b^2 \cos^2(\vartheta)},$$

$$\mathbb{G}^{\{12\}} = \mathbb{G}^{\{21\}} = \frac{\left(a^2 - b^2\right) \cot(\vartheta) \sin(\varphi) \cos(\varphi)}{\sin^2(\vartheta)\left(a^2 \sin^2(\varphi) + b^2 \cos^2(\varphi)\right) + a^2 b^2 \cos^2(\vartheta)},$$

$$\mathbb{G}^{\{22\}} = \left(\csc^2(\vartheta)\left(a^2 \cot^2(\vartheta)\cos^2(\varphi) + b^2 \cot^2(\vartheta)\sin^2(\varphi) + 1\right)\right) / \left(a^2 b^2 \cot^2(\vartheta)\cos^4(\varphi) + a^2 \sin^2(\varphi)\left(b^2 \cot^2(\vartheta)\sin^2(\varphi) + 1\right) + b^2 \cos^2(\varphi)\left(2a^2 \cot^2(\vartheta)\sin^2(\varphi) + 1\right)\right). \quad \text{(S28)}$$

The mean and gaussian curvatures read:

$$\mathcal{K} = -\frac{ab\left[3\left(a^2 + b^2\right) + 2 + \left(a^2 + b^2 - 2\right)\cos(2\vartheta) - 2\left(a^2 - b^2\right)\cos(2\varphi)\sin^2\vartheta\right]}{4\left[a^2 b^2 \cos^2\vartheta + \left(b^2 \cos^2\varphi + a^2 \sin^2\varphi\right)\sin^2\vartheta\right]^{3/2}} \quad \text{(S29)}$$

$$\mathcal{G} = \frac{a^2 b^2}{\left[a^2 b^2 \cos^2\vartheta + \left(b^2 \cos^2\varphi + a^2 \sin^2\varphi\right)\sin^2\vartheta\right]^2} \quad \text{(S30)}$$

The gravity components in the contravariant base read:

$$g_t^{\{1\}} = \frac{\sin(\vartheta)\left(a^2 \sin^2(\varphi) + b^2 \cos^2(\varphi)\right)}{\sin^2(\vartheta)\left(a^2 \sin^2(\varphi) + b^2 \cos^2(\varphi)\right) + a^2 b^2 \cos^2(\vartheta)}, \quad \text{(S31)}$$

$$g_t^{\{2\}} = \frac{\sin(\varphi)\cos(\varphi)\left(a^2 \cos(\vartheta) - b^2 \cos(\vartheta)\right)}{a^2 b^2 \cos^2(\vartheta)\cos^2(\varphi) + a^2 b^2 \cos^2(\vartheta)\sin^2(\varphi) + a^2 \sin^2(\vartheta)\sin^2(\varphi) + b^2 \sin^2(\vartheta)\cos^2(\varphi)}, \quad \text{(S32)}$$

$$g_3 = -\frac{ab \sin(\vartheta)\cos(\vartheta)}{\sqrt{\sin^4(\vartheta)\left(a^2 \sin^2(\varphi) + b^2 \cos^2(\varphi)\right) + a^2 b^2 \sin^2(\vartheta)\cos^2(\vartheta)}}. \quad \text{(S33)}$$

## Appendix B. Classical drainage and spreading solutions: diverging flow on a sphere, a cylinder and a cone

We consider the coating of an initially uniform layer of fluid on a sphere of radius $R$, recently analyzed by Takagi & Huppert (2010); Lee *et al.* (2016) and Qin *et al.* (2021). In non-dimensional form, the parameterization of the spherical surface through the zenith (or colatitude) $x^{\{1\}} = \vartheta$ and azimuth $x^{\{2\}} = \varphi$ reads:

$$\boldsymbol{X}(\vartheta, \varphi) = (\sin\vartheta\cos\varphi, \sin\vartheta\sin\varphi, \cos\vartheta) \quad \text{(S1)}$$

The gravity components tangent to the substrate and $w$ respectively read:

$$g_t^{\{1\}}(\vartheta) = \sin(\vartheta), \quad g_t^{\{2\}} = 0, \quad w(\vartheta) = \sin(\vartheta). \quad \text{(S2)}$$

The drainage problem can be written as follows:

$$\frac{\partial h(\vartheta, t)}{\partial t} + \frac{1}{3w(\vartheta)}\frac{\partial}{\partial\vartheta}\left(g_t^{\{1\}}(\vartheta)w(\vartheta)h(\vartheta, t)^3\right) = 0 \rightarrow \frac{\partial h}{\partial t} + \frac{1}{3\sin(\vartheta)}\frac{\partial}{\partial\vartheta}\left(\sin^2(\vartheta)h^3\right) = 0. \quad \text{(S3)}$$

The problem admits a large-time solution by separation of variables, independent of the initial condition (Couder *et al.* 2005; Qin *et al.* 2021):

$$h(z,t) = \eta(z)t^{-1/2}, \quad z = \cos(\vartheta), \tag{S4}$$

where

$$\eta(z) = \frac{[f(1) - f(z)]^{1/2}}{(1-z^2)^{1/3}}, \quad f(z) = z\mathrm{Hy}\left(\frac{1}{3}, \frac{1}{2}; \frac{3}{2}; z^2\right), \tag{S5}$$

where Hy is the hypergeometric function. Takagi & Huppert (2010), with a scaling analysis around the pole, found that $h \approx \sqrt{\frac{3}{4}}t^{-1/2}$. A more formal approach which gives the solution at different orders in $\vartheta$ for a generic initial condition, was proposed by Lee *et al.* (2016) through an asymptotic expansion of equation (S3) around the pole, i.e. $h(\vartheta,t) = H_0(t) + \vartheta H_1(t) + \vartheta^2 H_2(t) + ....$, leading to the following solution, truncated at $O(\vartheta^2)$:

$$h\left(\vartheta,t\right) \approx \frac{1}{\sqrt{1+\frac{4}{3}t}}\left[1 + \frac{\vartheta^2}{10}\left(1 + C\left(1+\frac{4}{3}t\right)^{-5/2}\right)\right] + O(\vartheta^4), \tag{S6}$$

where the constant $C$ depends on the initial condition. Note that the odd terms in the asymptotic expansion are zero because of the symmetry with respect to $\vartheta = 0$. When $t \longrightarrow \infty$, the term that multiplies $C$ vanishes, at the leading order. The leading-order large-time solution at order $O(\vartheta^6)$ reads:

$$h(\vartheta,t) = \sqrt{\frac{3}{4t}}\left(1 + \frac{\vartheta^2}{10} + \frac{41\vartheta^4}{4800} + \frac{1187\vartheta^6}{1584000}\right) + O(\vartheta^8) + O\left(\frac{1}{t^{3/2}}\right). \tag{S7}$$

The large-time solution is thus independent of the initial condition and has a time dependence analogous to the one obtained by Takagi & Huppert (2010) and Qin *et al.* (2021). The resulting zenith dependence is also an approximation of the hypergeometric function. In figure S1($a$), the various orders large-time analytical solutions show a good agreement with the numerical solution of equation (S3).

We now consider the case of the spreading on a sphere (of radius $R$) of an initial mass of fluid of thickness $h_i = 1$, contained in the region $\vartheta < \vartheta_0$, $0 < \varphi < 2\pi$. Substituting the $O(\vartheta)$ approximation of equation (S6) and $w \approx \vartheta$ into the conservation of mass (equation (3.13) of the manuscript), one finds the following expressions for the front angle $\vartheta_F(t)$ and the thickness at the front $h_F(\vartheta(t))$:

$$\frac{\vartheta_F}{\vartheta_0} = \left(\frac{4}{3}t\right)^{1/4}, \quad h_F = \left(\frac{\vartheta_0}{\vartheta_F}\right)^{1/2}, \tag{S8}$$

which is the non-dimensional version of the results reported in Takagi & Huppert (2010).

In the case of the coating on a cylinder of radius $R$, the non-dimensional parameterization reads:

$$\boldsymbol{X} = (\sin\vartheta, y, \cos\vartheta), \tag{S9}$$

The tangential gravity component and $w$ read:

$$g_t^{\{1\}} = \sin(\vartheta), \quad g_t^{\{2\}} = 0, \quad w = 1. \tag{S10}$$

The drainage problem reads:

$$\frac{\partial h}{\partial t} + \frac{1}{3}\frac{\partial}{\partial\vartheta}\left(\sin(\vartheta)h^3\right) = 0. \tag{S11}$$

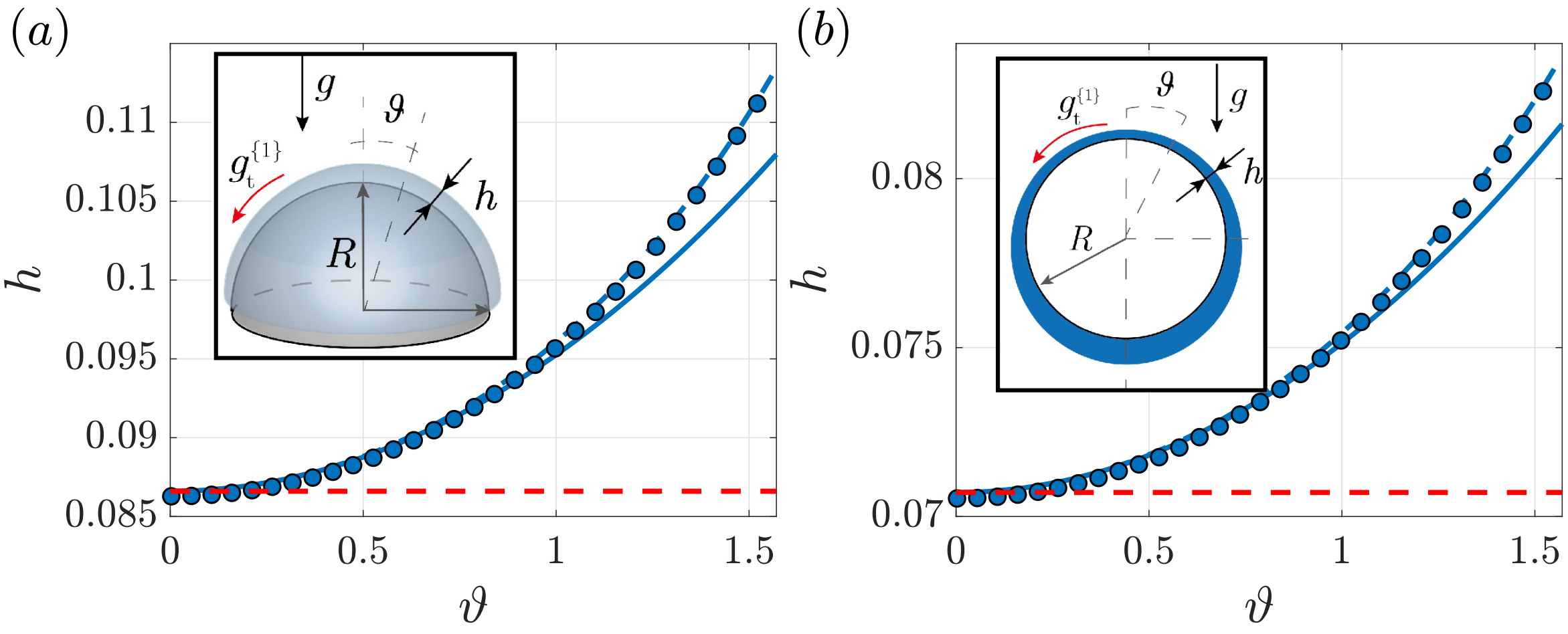

Figure S1: ($a$) Coating on a sphere: film thickness as a function of the zenith $\vartheta$ at $t = 100$, numerical simulation (dots), large-time analytical solution at order $O(1)$ (red dashed line), $O(\vartheta^2)$ (solid line) and $O(\vartheta^6)$ (blue dashed line). ($b$) Coating on a cylinder: film thickness as a function of the zenith $\vartheta$ at $t = 300$, numerical simulation (dots), large-time analytical solution at order $O(1)$ (red dashed line), $O(\vartheta^2)$ (solid line) and $O(\vartheta^6)$ (blue dashed line).

Performing an asymptotic expansion around the pole in powers of $\vartheta$, the large-time solution reads (Balestra *et al.* 2019):

$$h(\vartheta, t) = \sqrt{\frac{3}{2t}} \left(1 + \frac{\vartheta^2}{16} + \frac{43\vartheta^4}{10752} + \frac{109\vartheta^6}{368640}\right) + O(\vartheta^8) + O\left(\frac{1}{t^{3/2}}\right), \qquad \text{(S12)}$$

whose $O(1)$ approximation correspond to the result of Takagi & Huppert (2010). Note that the large-time solution is independent of the initial condition, and is in good agreement with the numerical solution of equation (S11) (see figure S1($b$)). Takagi & Huppert (2010) also studied the spreading on a cylinder. Following the same steps adopted for the case of the sphere, one obtains the following expressions for the front angle and thickness:

$$\frac{\vartheta_F}{\vartheta_0} = \left(\frac{2t}{3}\right)^{1/2}, \quad h_F = \frac{\vartheta_0}{\vartheta_F}. \qquad \text{(S13)}$$

We conclude by considering the classical problem of the coating on a conical substrate, of maximum radius $R$ and height $aR$. This problem is typically solved by employing a self-similar approach. A large-time solution can be obtained, similar to Qin *et al.* (2021) for the sphere coating. We consider a parameterization in cylindrical coordinates (see ESM for further details), which reads, in non-dimensional form:

$$\boldsymbol{X} = (r\cos\varphi, r\sin\varphi, -ar), \qquad \text{(S14)}$$

where $r$ is the radius and $\varphi$ is the azimuth. In this parameterization, the gravity terms and $w$ read:

$$g_t^{\{1\}} = \frac{a}{a^2+1}, \quad g_t^{\{2\}} = 0, \quad w(r) = \sqrt{(a^2+1)}r \qquad \text{(S15)}$$

The lubrication equation reads:

$$\frac{\partial h(r,t)}{\partial t} + \frac{1}{3\sqrt{(a^2+1)}r}\frac{\partial}{\partial r}\left(\sqrt{(a^2+1)}r\, g_t^{\{1\}}(r) h(r,t)^3\right) = 0 \qquad \text{(S16)}$$

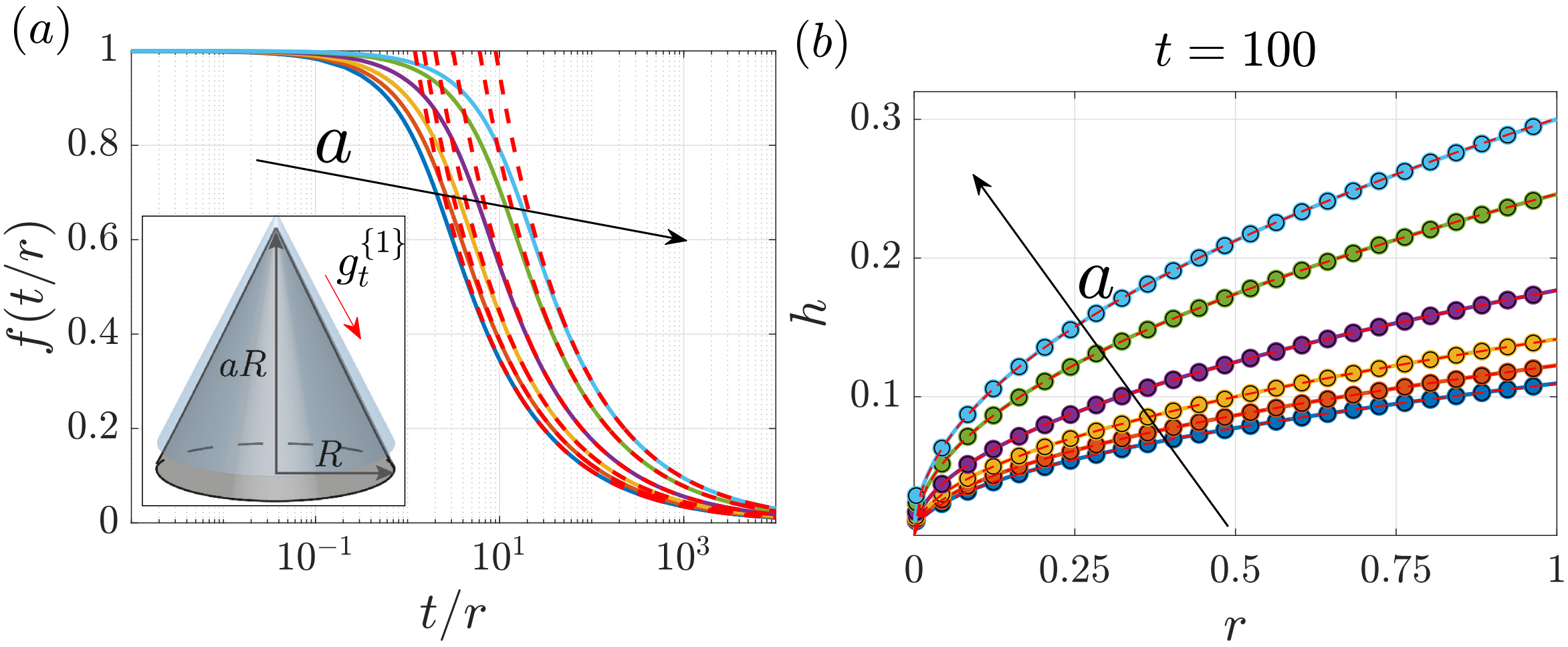

Figure S2: ($a$) Self-similar solution for the coating of a cone $h(r,t) = f(\eta)$ as a function of $\eta = t/r$, for different values of $a$. The red dashed lines denote the corresponding large-time approximation. ($b$) Comparison at $t = 100$ between the numerical (colored dots) self-similar (solid lines) and large-time (red dashed lines) solutions. The different colors correspond to $a = 1$ (blue), $a = 2$ (orange), $a = 3$ (yellow), $a = 5$ (purple), $a = 10$ (green), $a = 15$ (cyan).

Developing the derivatives, one obtains:

$$\frac{\partial h}{\partial t} + \frac{ah^2}{a^2+1}\frac{\partial h}{\partial r} + \frac{ah^3}{3a^2 r + 3r} = 0, \tag{S17}$$

which is completed with the initial condition $h(r,0) = 1$. This problem admits a self-similar solution of the form $h(r,t) = f(\eta)$, where $\eta = t/r$ is the self-similar variable. By substituting the self-similar ansatz in equation (S17), one obtains the following ordinary differential equation:

$$f'\left(1 - \frac{a}{1+a^2}\eta f^2\right) + \frac{1}{3}\frac{a}{1+a^2}f^3 = 0, \quad f(0) = 1, \tag{S18}$$

which can be numerically solved so as to find the solution of the initial value problem. The results are reported in figure S2. Moreover,the following expression satisfies equation (S18):

$$f(\eta) = f_0\eta^{-1/2}, \quad f_0 = \sqrt{\frac{3(a^2+1)}{5a}} \rightarrow h(r,t) = \sqrt{\frac{3(a^2+1)}{5a}}\sqrt{\frac{r}{t}}. \tag{S19}$$

The latter expression is well known in literature (Acheson 1990) and well agrees with the numerical self-similar solution as $\eta \rightarrow \infty$, i.e. $t \rightarrow \infty$ or $r \rightarrow 0$ (see figure S2). The large-time solution is thus independent of the initial condition.

In analogy to the previous cases, the spreading of an initial mass of fluid of height $h_i = 1$ contained in a region $r < r_0$ can be solved by employing the large-time solution: (S19):

$$h(r,t) = \sqrt{\frac{3(a^2+1)}{5a}}\sqrt{\frac{r}{t}}, \quad w(r) = \sqrt{(a^2+1)}r. \tag{S20}$$

Upon integration in $0 < \varphi < 2\pi$ and $0 < r < r_N$, one obtains the radial position of the front $r_N$:

$$r_N = \left(\frac{125V^2}{48\pi^2}\frac{a}{(a^2+1)^2}\right)^{1/5} t^{1/5}, \tag{S21}$$

where $V = \pi r_0^2 \sqrt{a^2+1}$ is the initial volume. Noting that $\frac{a}{(a^2+1)^2} = \sin^3(\beta)\cos(\beta)$, where $\beta$ is the inclination angle of the cone and the spreading distance from the tip of the cone is $s_N = r_N/\sin(\beta)$, one obtains the classical result for the spreading on a cone (Acheson 1990):

$$s_N = \left(\frac{125V^2}{48\pi^2}\frac{\cos\beta}{\sin^2\beta}\right)^{1/5} t^{1/5}, \tag{S22}$$

## Appendix C. Spreading and drainage on a paraboloid

### C.1. *Drainage problem*

In this section, we consider the coating of a paraboloid of characteristic radial extent $R$. We parameterize the paraboloidal surface as follows:

$$\boldsymbol{X} = (r\cos\varphi, r\sin\varphi, -er^2) \tag{S1}$$

The gravity term $g_t^{\{1\}}$ and $w$ now read:

$$g_t^{\{1\}}(r) = \frac{2er}{4e^2r^2+1}, \quad w(r) = r\sqrt{4e^2r^2+1} \tag{S2}$$

The lubrication equation reads:

$$\frac{\partial h}{\partial t} + \frac{2erh^2}{4e^2r^2+1}\frac{\partial h}{\partial r} + \frac{4\left(2e^3r^2+e\right)h^3}{3\left(4e^2r^2+1\right)^2} = 0. \tag{S3}$$

In analogy with the sphere case, we expand the solution in series of $r$:

$$h(r,t) = H_0(t) + r^2H_2(t) + r^4H_4(t) + r^6H_6(t) + \ldots \tag{S4}$$

Because of symmetry, all the odd terms in $r$ are zero. In Appendix **??** we report the analytical developments. The solution at order $O(r^6)$, for $t \to \infty$, reads:

$$h(r,t) = \frac{\sqrt{\frac{3}{2}}\sqrt{\frac{1}{t}}\left(3248e^6r^6 - 1540e^4r^4 + 1650e^2r^2 + 1375\right)}{2750\sqrt{e}} + O(r^8) + O\left(\frac{1}{t^{3/2}}\right). \tag{S5}$$

Also in this case, $H_0 = \left(\frac{3}{2\mathcal{K}_p t}\right)^{1/2}$, where $\mathcal{K}_p = 4e$ is the absolute value of the mean curvature at the top.

Equation (S3) is solved in the domain $0 < r < 1$ with initial condition $h(0,t) = 1$. Numerical convergence for all values of $e$ is achieved with a characteristic element size $\Delta r = 0.01$.

The comparison between the analytical and numerical drainage solution is reported in figure S3($a$). The solution is characterized by mild variations of the thickness with the radius. The comparison shows a good agreement for $e < 1$, while already at $e = 1$ the accuracy of the analytical solution rapidly degrades as $r > 0.5$.

An analytical approximation for larger values of $e$ can be obtained by assuming $er \gg 1$ in equation (S3):

$$\frac{\partial h}{\partial t} + \frac{h^2}{2er}\frac{\partial h}{\partial r} + \frac{h^3}{6er^2} = 0, \quad h(r,0) = 1. \tag{S6}$$

The problem admits a self-similar solution of the form $h(r,t) = f(\xi)$, where $\xi = t/r^2$. Following the same procedure of the cone problem, one obtains the following ordinary

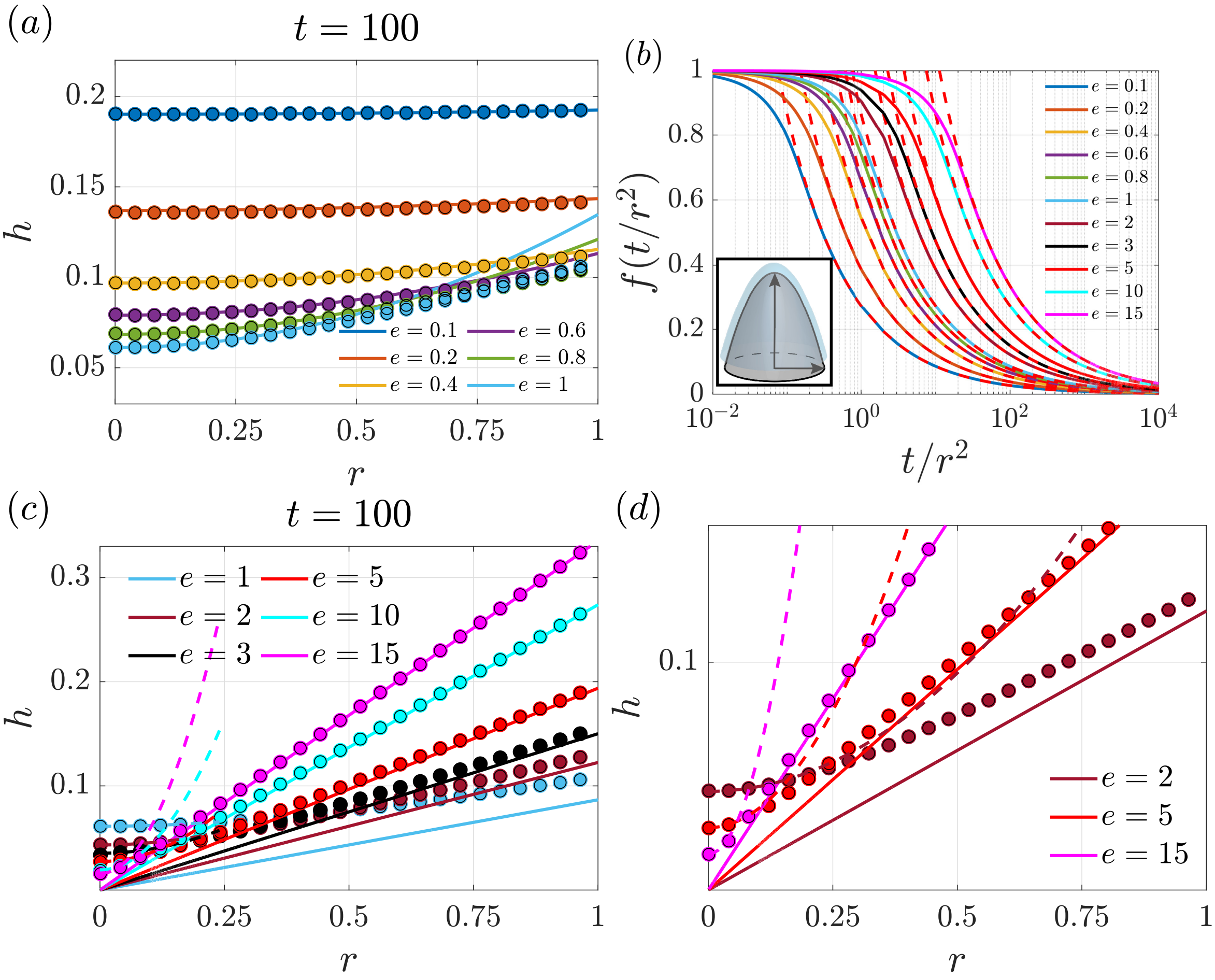

Figure S3: $(a)$ Comparison of the analytical solution (equation (S5), solid lines) against the numerical one (colored dots) for the drainage on a paraboloid. $(b)$ Self-similar solution for the coating on a paraboloid, valid for $er \gg 1$ (colored lines). The red dashed lines are the corresponding large-time approximations (equation (S8). $(c)$ Comparison of the numerical solution (colored dots) with the self-similar one for $er \gg 1$ (solid lines). The dashed lines denote the $O(r^6)$ approximation of equation (S5). $(d)$ Zoom in $(c)$ to highlight the range of validity of the two analytical solutions.

differential equation:

$$f'\left(1 - \frac{1}{e}\xi f^2\right) + \frac{1}{6e}f^3 = 0, \qquad f(0) = 1. \tag{S7}$$

The resulting equation is formally analogous to the self-similar problem of the cone (S18), with different coefficients. The numerical solution is reported in figure S3$(b)$. We find an approximate solution of the form $f(\xi) = f_0\xi^{-1/2}$, where $f_0 = \sqrt{3e}/2$:

$$h(r,t) = \frac{\sqrt{3e}}{2}\frac{r}{\sqrt{t}}, \tag{S8}$$

that well agrees with the solution of the self-similar initial value problem for $\xi \to \infty$, i.e. $t \to \infty$ (see red dashed lines in S3$(b)$). The agreement with the numerical solution for $e > 2$, shown in figure S3$(c)$ is very good, except in the close vicinity of $r = 0$, in which the expansion of equation (S5) can be employed.

Following the reasoning of Section 3.1 of the manuscript, the mean curvature decreases moving downstream, therefore inducing film thickening. Also the tangential gravity compo-nent increases, thus leading to an overall film thickening effect. Interestingly, the drainage problem was approached by employing the asymptotic expansion, the self-similar solution and the large-time scaling arguments.

### C.2. *Spreading problem*

We now focus on the spreading problem of a mass of fluid of initial thickness $h_i = 1$ contained in the region $r < r_0$. The conservation of mass reads:

$$\int_0^{2\pi}\int_0^{r_F(t)} h(r,t)w(r)\mathrm{d}r\mathrm{d}\varphi = \int_0^{2\pi}\int_0^{r_0} w(r)\mathrm{d}r\mathrm{d}\varphi \\ \rightarrow \int_0^{r_F(t)} h(r,t)w(r)\mathrm{d}r = \int_0^{r_0} w(r)\mathrm{d}r, \tag{S9}$$

where $w(r)$ is given by equations (S2). The thickness $h(r,t)$ is given by the two large-time analytical solutions (S5), valid for small values of $e$, and (S8), valid instead when large values of $e$ are considered, far from the pole. The resulting problems for the front radius $r_F$ and thickness $h_F$ are numerically solved via the built-in function "fsolve" in Matlab, and the results are reported in figure S4, for the two different analytical solutions. These results are compared with following two analytical approximations in the vicinity of ($r \ll 1$) and far from the pole ($er \gg 1$). When small values of $r$ and large times are considered, the thickness and $w$ can be approximated as follows:

$$h = \sqrt{\frac{3}{8et}} + O(r^2) + O\left(\frac{1}{t^{3/2}}\right), \quad w = r + O(r^2). \tag{S10}$$

Substituting in equation (S9) and keeping at most $O(r)$ terms, one obtains the front radius and thickness:

$$r_F = r_0\left(\frac{8et}{3}\right)^{1/4}, \quad h_F = \left(\frac{r_0}{r_F}\right)^2, \tag{S11}$$

which well compares with the numerical solutions of the implicit relation (S9) (see figure S4($a,b$)).

The same analytical developments can be employed for the case $er \gg 1$ and $t \rightarrow \infty$:

$$h = \frac{\sqrt{3e}}{2}\frac{r}{\sqrt{t}}, \quad w = r\sqrt{4e^2r^2+1} \approx 2er^2, \tag{S12}$$

which leads to:

$$r_F = \left(\frac{8}{3}\sqrt{\frac{t}{3e}}r_0^3\right)^{1/4}, \quad h_F = \frac{4}{3}\left(\frac{r_0}{r_F}\right)^3. \tag{S13}$$

Also in this case, a good agreement is observed (see figure S4($c,d$)). The agreement improves when larger values of $e$ at larger $r$ are considered. We finally compare these theoretical results with two numerical simulations of the complete model (equation (2.5) of the manuscript). To impose the outlet condition, we consider the domain $0 < r < 3$ and employ a Sponge method to relax the thickness to zero avoiding reflections from the outlet (Högberg & Henningson 1998; Lerisson *et al.* 2020). Equation (S3) is thus modified as follows to impose the Sponge condition:

$$\frac{\partial h}{\partial t} + \frac{2erh^2}{4e^2r^2+1}\frac{\partial h}{\partial r} + \frac{4\left(2e^3r^2+e\right)h^3}{3\left(4e^2r^2+1\right)^2} = -\frac{h}{2}\left(1+\tanh\left(\eta_{sp}\left(r-r_{sp}\right)\right)\right) = -h\mathrm{Sp}(r), \tag{S14}$$

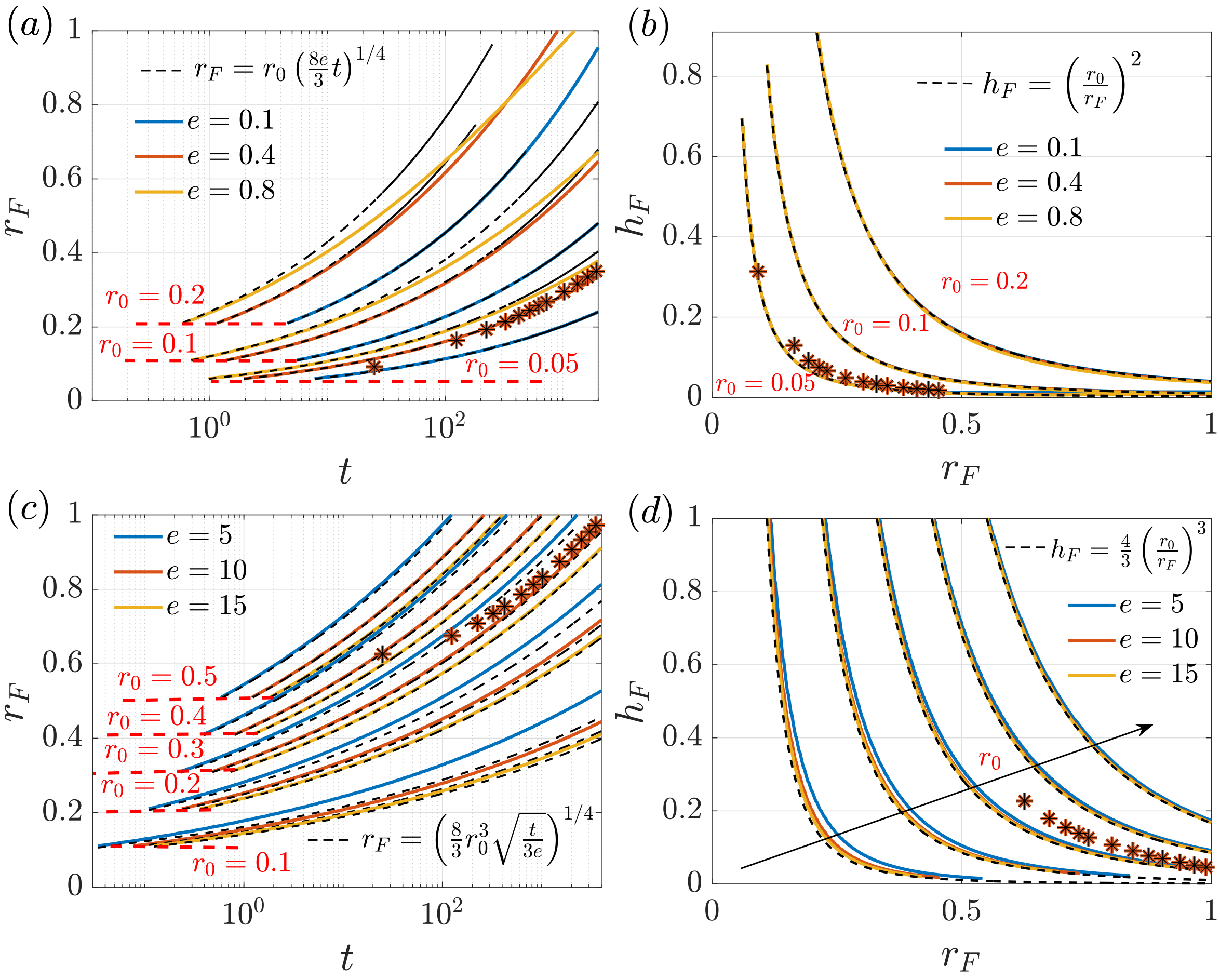

Figure S4: Spreading of an initial volume of fluid on an paraboloid. $(a, c)$ Variation of the front angle $\vartheta_F$ with time and $(b, d)$ of the thickness at the front $h_F$ with $\vartheta_F$, for different values of the initial angle $\vartheta_0$ and $e$, for the $(a, b)$ asymptotic and $(c, d)$ self-similar solution. The solid and dot-dashed lines denote the values of $\vartheta_F$ and $h_F$ on the outer and inner sides, respectively. The black dashed lines correspond to the analytical approximations of the relation $\vartheta_F(t)$ and $h_F(\vartheta_F)$, respectively, while the stars are the values recovered by a numerical simulation of the complete model with $(a, b)$ $e = 0.8$, $Bo = 5000$, $\delta = 10^{-3}$, $h_{pr} = 0.0025$ and $(c, d)$ $e = 10$, $Bo = 100$, $\delta = 10^{-2}$, $h_{pr} = 0.0025$.

where $\eta_{sp} = 3$ and $r_{sp} = 2.7$, and initial condition $h(0, t) = 1 - \text{Sp}(r)$. The numerical results show a good agreement with the prediction of the front position and thickness.